\documentclass[12pt]{article}

\usepackage{amsmath,amssymb,epsfig,subfigure,color,soul,fleqn,rotating}
\usepackage{ulem,slashed}
\usepackage{cite}
\usepackage[colorlinks,citecolor=blue,urlcolor=blue,linkcolor=blue]{hyperref}
\usepackage{xcolor}
\usepackage{ulem} 

\newcommand{\be}{\begin{eqnarray}}
\newcommand{\ee}{\end{eqnarray}}
\newcommand{\nee}{\nonumber\end{eqnarray}}

\newcommand{\drbar}{{\overline{\rm DR}}}

\newcommand{\mch}[1] {m_{\ti \x^+_{#1}}}
\newcommand{\mnt}[1] {m_{\ti \x^0_{#1}}}

\newcommand{\msg}    {m_{\ti g}}
\newcommand{\msu}[1] {m_{\ti u_{#1}}}
\newcommand{\msd}[1] {m_{\ti d_{#1}}}

\newcommand{\gsim}{\;\raisebox{-0.9ex}
           {$\textstyle\stackrel{\textstyle >}{\sim}$}\;}
\newcommand{\lsim}{\;\raisebox{-0.9ex}
          {$\textstyle\stackrel{\textstyle<}{\sim}$}\;}

\def\be            {\begin{equation}}
\def\ee            {\end{equation}}
\def\bea            {\begin{eqnarray}}
\def\eea            {\end{eqnarray}}

\definecolor{mybrown}{cmyk}{0,0.9,1.5,0.3}

\def\a              {\alpha}
\def\b               {\beta}

\def\x               {\chi}
\def\ti              {\tilde}
\def\sq              {\ti q}
\def\st              {\ti t}
\def\sc              {\ti c}
\def\sb              {\ti b}
\def\ch              {\ti \x^\pm}

\def\nt              {\ti \x^0}
\def\sg              {\ti g}
\def\stau            {\ti \tau}
\def\sneut           {\ti \nu}

\def\su              {\ti{u}}
\def\sto             {\ti{t}}
\def\sca             {\ti{c}}
\def\sd              {\ti{d}}
\def\ss              {\ti{s}}
\def\sbo             {\ti{b}}
\newcommand{\DEV}{\textnormal{DEV}}
\newcommand{\MeV}{\textnormal{MeV}}
\newcommand{\GeV}{\textnormal{GeV}}
\newcommand{\TeV}{\textnormal{TeV}}

\newcommand{\AddrGAKUGEI}{%
 \it Department of Physics, Tokyo Gakugei University, Koganei,
Tokyo 184-8501, Japan\\}

\newcommand{\AddrHEPHY}{%
 \it Institut f\"ur Hochenergiephysik der \"Osterreichischen Akademie
der Wissenschaften, A-1050 Vienna, Austria\\}

\title{\bf Impact of quark flavor violating SUSY on h(125) decays at 
future lepton colliders}

\author{Helmut~Eberl${}^{1}$, Keisho~Hidaka${}^{2}$, Elena~Ginina${}^{1}$}

\date{
\small $^1$ \AddrHEPHY
       $^2$ \AddrGAKUGEI
}

\definecolor{darkgreen}{rgb}{0,.5,0}

\begin{document}


\begingroup
\let\newpage\relax
\maketitle
\endgroup

\maketitle
\thispagestyle{empty}

\begin{abstract}
We study the CP-even neutral Higgs boson decays $h^0 \to c \bar{c}, b \bar{b}, 
b \bar{s}, \gamma \gamma, g g$ in the Minimal Supersymmetric Standard Model (MSSM) 
with general quark flavor violation (QFV) due to squark generation mixings, 
identifying the $h^0$ as the Higgs boson with a mass of 125 \GeV. 
We compute the widths of the $h^0$ decays to $c \bar c, b \bar b, b \bar s$ 
at full one-loop level. For the $h^0$ decays to $\gamma \gamma$ and $g g$ we compute 
the widths at NLO QCD level. {\it For the first time}, we perform a systematic MSSM 
parameter scan for these widths respecting all the relevant theoretical and experimental 
constraints, such as those from B-meson data, and the 125 \GeV~Higgs boson data from 
recent LHC experiments, as well as the limits on Supersymmetric (SUSY) particle 
(sparticle) masses from the LHC experiments. We also take into account the expected 
sparticle mass limits from the future HL-LHC experiment in our analysis. 
{\it In strong contrast to} the usual studies in the MSSM with Minimal Flavor 
Violation (MFV), we find that the deviations of these MSSM decay widths from 
the Standard Model (SM) values can be quite sizable and that there are significant 
correlations among these deviations. All of these sizable deviations in the $h^0$ decays 
are mainly due to large scharm-stop mixing and large sstrange-sbottom mixing. 
Such sizable deviations from the SM can be observed at high signal significance in 
future lepton colliders such as ILC, CLIC, CEPC, FCC-ee and muon collider 
{\it even after} the failure of SUSY particle discovery at the HL-LHC. In case the 
deviation pattern shown here is really observed at the lepton colliders, then it would 
strongly suggest the discovery of QFV SUSY (the MSSM with general QFV).
\end{abstract}

\newpage

\section{Introduction}
What is the Standard Model (SM)-like Higgs boson with mass of 125 \GeV~discovered at LHC 
\cite{ATLAS_Higgs, CMS_Higgs}? It can be the Higgs boson of the SM. It can be a Higgs 
boson of a New Physics (NP) theory beyond the SM. This is one of the most important 
issues in the field of present particle physics. The detailed study of the properties 
(such as mass and couplings) of the SM-like Higgs boson could shed light on this issue 
and the way to the NP theory \cite{ILC_Higgs_2017}.  
Here we study a possibility that the discovered SM-like Higgs boson is 
the lighter CP even neutral Higgs boson $h^0$ of the Minimal Supersymmetric 
Standard Model (MSSM), focusing on the decays $h^0 \to c \bar c, b \bar b, 
b \bar s, \gamma \gamma, g g$, where c, b, s, $\gamma$ and g are c-, b-, 
s-quarks, photon and gluon, respectively. 
In order to investigate such a possibility we compute the widths of these 
decays in the MSSM with general quark-flavor violation (QFV) due to 
squark generation mixing according to our previous works \cite{Bartl:h2cc, 
Eberl:h2bb, h2gagagg} and study the deviations of the MSSM widths from 
the SM widths.\\
\indent The deviations of the SM-like Higgs boson decay widths 
from their SM values are usually estimated to be rather small (typically several
\% level or less) in the MSSM with Minimal Flavor Violation (MFV) where the 
only source of quark-flavor violation is the Cabibbo-Kobayashi-Maskawa (CKM) 
quark-mixing matrix \cite{Hewett_pMSSM, Nojiri_pMSSM, Mahmoudi_pMSSM, 
Heinemeyer_pMSSM, Snowmass_Rep_BSM, Snowmass_Rep_EF}. \\
\indent In the present article, however, we show that the situation changes 
drastically yielding significant enhancement of the deviations in the widths 
once we switch-on the general QFV in the MSSM, by performing a systematic MSSM 
parameter scan respecting all the relevant theoretical and experimental constraints 
\footnote{
The decay $h^0 \to c \bar c$ was not studied in \cite{Hewett_pMSSM, Nojiri_pMSSM, 
Mahmoudi_pMSSM, Heinemeyer_pMSSM, Snowmass_Rep_BSM, Snowmass_Rep_EF}.  
The decays $h^0 \to c \bar c, \gamma \gamma, g g$ in the MSSM with QFV were 
studied in \cite{Brignole}. However, the important QFV parameters $M^2_{Q23}$, 
$M^2_{U23}$, and $M^2_{D23}$ (which are defined in Section~\ref{sec:sq.matrix}) 
were neglected and the systematic MSSM parameter scan respecting all the relevant 
theoretical and experimental constraints was not performed in \cite{Brignole}.
}.\\
\indent 
In \cite{Bejar, Curiel_1, Demir, Curiel_2, Barenboim, Heinemeyer} it was shown 
that the general QFV in the MSSM can also enhance QFV decays $h^0 \to b \bar s$ 
and $h^0 \to \bar b s$. These analyses are rather old. In the present paper we 
update them, especially by taking into account the expected mass limits 
for the superpartner particles and the heavier MSSM Higgs bosons $H^0$, $A^0$, $H^+$ 
from the future HL-LHC experiment.\\
\indent On the experimental side, the widths of these decays, $h^0 \to c \bar c, 
b \bar b, \gamma \gamma, g g$ (or corresponding effective 
couplings) can be measured precisely and {\it model-independently} at future 
lepton colliders, such as ILC, CLIC, CEPC, FCC-ee and 
muon-collider (MuC) \cite{ESU2020_Rep, Snowmass2021_Rep} 
\footnote{
The effective couplings of $g(h^0 b \bar b)$, $g(h^0 \gamma \gamma)$ and 
$g(h^0 g g)$ can be measured rather precisely but {\it model-dependently} by a 
global fit at HL-LHC \cite{ESU2020_Rep, Snowmass2021_Rep, Higgs_HL-LHC_Rep}. 
Moreover, it is difficult to measure the coupling $g(h^0 c \bar c)$ at LHC 
and HL-LHC due to the difficulty of the c-tagging and the huge hadronic QCD 
background \cite{Higgs_HL-LHC_Rep}.
}.
This enables us to clarify the possibility that the discovered SM-like 
Higgs boson is the lighter CP even neutral Higgs boson $h^0$ of the MSSM. \\

In Section ~\ref{sec:sq.matrix} we introduce the supersymmetric (SUSY) 
QFV parameters originating from the squark mass matrices. Details of 
our parameter scan are given in Section~\ref{sec:full scan}.  
In Section~\ref{sec:Deviations} we study the deviations of the 
MSSM widths from the SM widths for the decays $h^0 \to c \bar c, 
b \bar b, b \bar s, \gamma \gamma, g g$ and analyze their 
behavior in the MSSM with general QFV. The summary and conclusion 
are in Section~\ref{sec:concl}. All relevant constraints are 
listed in Appendix \ref{sec:constr} and the expected errors in the deviation 
measurements at future lepton colliders are listed in Appendix \ref{sec:error}. 
ILC sensitivity to the branching ratio $B(h^0 \to b s)$ is discussed in 
Appendix \ref{sec:ILC_sensitivity_to_BRbs}, and consistency of the MSSM 
predictions for coupling modifiers with the LHC data is discussed in 
Appendix \ref{sec:kappas}. 

\section{Squark mass matrices in the MSSM with general QFV}
\label{sec:sq.matrix}
%
In the super-CKM basis of $\sq_{0 \gamma} =
(\sq_{1 {\rm L}}, \sq_{2 {\rm L}}, \sq_{3 {\rm L}}$,
$\sq_{1 {\rm R}}, \sq_{2 {\rm R}}, \sq_{3 {\rm R}}),~\gamma = 1,...6,$  
with $(q_1, q_2, q_3)=(u, c, t),$ $(d, s, b)$, the up-type and down-type squark mass squared matrices 
${\cal M}^2_{\tilde{q}},~\tilde{q}=\tilde{u},\tilde{d}$, at the SUSY scale have the following 
most general $3\times3$ block form~\cite{Allanach:2008qq}:
\begin{equation}
    {\cal M}^2_{\tilde{q}} = \left( \begin{array}{cc}
        {\cal M}^2_{\tilde{q},LL} & {\cal M}^2_{\tilde{q},LR} \\[2mm]
        {\cal M}^2_{\tilde{q},RL} & {\cal M}^2_{\tilde{q},RR} \end{array} \right), 
        \quad \tilde{q}=\tilde{u},\tilde{d}\,.
 \label{EqMassMatrix1}
\end{equation}
Non-zero off-diagonal terms of the $3\times3$ blocks ${\cal M}^2_{\tilde{q},LL},
~{\cal M}^2_{\tilde{q},RR},~{\cal M}^2_{\tilde{q},LR}$ and ${\cal M}^2_{\tilde{q},RL}$ 
explicitly violate quark-flavor in the squark sector of the MSSM.
The left-left and right-right blocks in Eq.~(\ref{EqMassMatrix1}) are given by
\begin{eqnarray}
    & &{\cal M}^2_{\tilde{u}(\tilde{d}),LL} = M_{Q_{u(d)}}^2 + D_{\tilde{u}(\tilde{d}),LL}{\bf 1} + \hat{m}^2_{u(d)}, \nonumber \\
    & &{\cal M}^2_{\tilde{u}(\tilde{d}),RR} = M_{U(D)}^2 + D_{\tilde{u}(\tilde{d}),RR}{\bf 1} + \hat{m}^2_{u(d)},
     \label{EqM2LLRR}
\end{eqnarray}
where $M_{Q_{u}}^2=V_{\rm CKM} M_Q^2 V_{\rm CKM}^{\dag}$, $M_{Q_{d}}^2 \equiv M_Q^2$, 
$M^2_{Q,U,D}$ are the hermitian soft SUSY-breaking mass squared matrices of the squarks, 
$D_{\tilde{u}(\tilde{d}),LL}$, $D_{\tilde{u}(\tilde{d}),RR}$ are the $D$-terms, and  
$\hat{m}_{u(d)}$ are the diagonal mass matrices of the up(down)-type quarks.
$M_{Q_{u}}^2$ is related with $M_{Q_{d}}^2$
by the CKM matrix $V_{\rm CKM}$ due to the $SU(2)_{\rm L}$ symmetry.
The left-right and right-left blocks of Eq.~(\ref{EqMassMatrix1}) are given by
\begin{eqnarray}
 {\cal M}^2_{\tilde{u}(\tilde{d}),RL} = {\cal M}^{2\dag}_{\tilde{u}(\tilde{d}),LR} &=&
\frac{v_2(v_1)}{\sqrt{2}} T_{U(D)} - \mu^* \hat{m}_{u(d)}\cot\beta(\tan\beta),
\label{M2sqdef}
\end{eqnarray}
where $T_{U,D}$ are the soft SUSY-breaking trilinear 
coupling matrices of the up-type and down-type squarks entering the Lagrangian 
${\cal L}_{int} \supset -(T_{U\alpha \beta} \ti{u}^\dagger _{\a R}\ti{u}_{\b L}H^0_2 $ 
$+ T_{D\alpha \beta} \ti{d}^\dagger _{\a R}\ti{d}_{\b L}H^0_1)$,
$\mu$ is the higgsino mass parameter, and 
$\tan\beta = v_2/v_1$ with $v_{1,2}=\sqrt{2} \left\langle H^0_{1,2} \right\rangle$.
The squark mass squared matrices are diagonalized by the $6\times6$ unitary matrices $U^{\tilde{q}}$,
$\tilde{q}=\tilde{u},\tilde{d}$, such that
\begin{eqnarray}
&&U^{\tilde{q}} {\cal M}^2_{\tilde{q}} (U^{\tilde{q} })^{\dag} = {\rm diag}(m_{\tilde{q}_1}^2,\dots,m_{\tilde{q}_6}^2)\,,
\label{Umatr}
\end{eqnarray}
with $m_{\tilde{q}_1} < \dots < m_{\tilde{q}_6}$.
The physical mass eigenstates
$\sq_i, i=1,...,6$ are given by $\sq_i =  U^{\sq}_{i \alpha} \sq_{0\alpha} $.

In this paper we focus on the $\ti{c}_L - \ti{t}_L$, $\ti{c}_R - \ti{t}_R$, $\ti{c}_R - \ti{t}_L$, 
$\ti{c}_L - \ti{t}_R$, $\ti{s}_L - \ti{b}_L$, $\ti{s}_R - \ti{b}_R$, $\ti{s}_R - \ti{b}_L$, and 
$\ti{s}_L - \ti{b}_R$ mixing which is described by the QFV parameters $M^2_{Q_{u}23} \simeq M^2_{Q23}$, 
$M^2_{U23}$, $T_{U23}$, $T_{U32}$, $M^2_{Q23}$, $M^2_{D23}$, $T_{D23}$ and $T_{D32}$, respectively. 
We will also often refer to the quark-flavor conserving (QFC) parameters $T_{U33}$ and $T_{D33}$ 
which induce the $\ti{t}_L - \ti{t}_R$ and $\ti{b}_L - \ti{b}_R$ mixing, respectively,  
and play an important role in this study.\\
The slepton parameters are defined analogously to the squark ones. 
In our analysis we assume that there is no SUSY lepton-flavor violation.
We also assume that R-parity is conserved and that the lightest neutralino 
$\nt_1$ is the lightest SUSY particle (LSP). All the parameters in this 
study are assumed to be real, except the CKM matrix $V_{CKM}$.

\section{Parameter scan}
\label{sec:full scan}

In our MSSM parameter scan we take into account all the relevant constraints, 
i.~e., theoretical constraints from vacuum stability conditions and experimental 
constraints, such as those from $K$- and $B$-meson data, electroweak precision 
data, and the $H^0$ mass and coupling data from recent LHC experiments, as well 
as the SUSY particle (sparticle) mass limits from current LHC experiments (see 
Appendix \ref{sec:constr}). 
Here $H^0$ is the discovered SM-like Higgs boson which we 
identify as the lightest $CP$ even neutral Higgs boson $h^0$ in the MSSM. 
Concerning squark generation mixings, we only consider the 
mixing between the second and third generation of squarks. The mixing between 
the first and the second generation squarks is strongly constrained by the 
$K$- and $D$-meson data ~\cite{Gabbiani:1996hi, PDG2020}. 
The experimental constraints on the mixing of the first and third generation squarks 
are not so strong \cite{Dedes}, but we do not consider this mixing since its 
effect is essentially similar to that of the mixing of the second and third 
generation squarks. We generate the input parameter points by using random numbers 
in the ranges shown in Table~\ref{table1}, where some parameters are fixed as given in 
the last box. All input parameters are $\drbar$ parameters defined at scale $Q = 1~\TeV$, 
except $m_{A^0}(pole)$ which is the pole mass of the $CP$ odd Higgs boson $A^0$.
The parameters that are not shown explicitly are taken to be zero. 
The entire scan range lies in the decoupling Higgs limit, i.~e., in the scenarios 
with large $\tan\beta \geq 10$ and large $m_{A^0} \geq 1350$ \GeV~(see Table~\ref{table1}), 
respecting the fact that the discovered Higgs boson is SM-like. It is well known that the 
lightest MSSM Higgs boson $h^0$ is SM-like (including its couplings) in this limit.
We do not assume the GUT relation for the gaugino masses $M_1$, $M_2$, $M_3$. 
The masses and mixing matrices of the SUSY particles and the Higgs bosons 
are renormalized basically at one-loop level by using the public code 
{\tt SPheno}-v3.3.8~\cite{SPheno1, SPheno2}
\footnote{
This version SPheno-v3.3.8 implements full flavor (generation) mixings 
in the sfermion (squarks and sleptons) sector as described in Section 2 
and calculates the masses and mixings of the SUSY particles and the MSSM 
Higgs bosons $h^0, H^0, A^0, H^\pm$ taking into accounts the full flavor 
mixings in the sfermion sector \cite{SPheno_HP}. 
}
based on the technique given in \cite{Pierce}. 
From 377180 input points generated in the scan 3208 points survived all constraints. 
These are 0.85\% of the generated points. We show these survival points 
in all scatter plots in this article.

\begin{table}[h!]
\footnotesize{
\caption{
Scanned ranges and fixed values of the MSSM parameters (in units of GeV or GeV$^2$, 
except for $\tan\beta$). The parameters that are not shown explicitly are 
taken to be zero. $M_{1,2,3}$ are the U(1), SU(2), SU(3) gaugino mass parameters.}
\begin{center}
\begin{tabular}{|c|c|c|c|c|c|}
    \hline
\vspace*{-0.3cm}
& & & & &\\
\vspace*{-0.3cm}
     $\tan\beta$ & $M_1$ &  $M_2$ & $M_3$ & $\mu$ &  $m_{A^0}(pole)$\\ 
& & & & &\\
    \hline
\vspace*{-0.3cm}
& & & & &\\
\vspace*{-0.3cm}
     10 $\div$ 80 & $100 \div 2500$ & $100 \div 2500$  & $2500 \div 5000$ & $100 \div 2500$ & $1350 \div 6000$\\
& & & & &\\
    \hline
    \hline
\vspace*{-0.3cm}
& & & & &\\
\vspace*{-0.3cm}
      $ M^2_{Q 22}$ & $ M^2_{Q 33}$ &  $|M^2_{Q 23}| $ & $ M^2_{U 22} $ & $ M^2_{U 33} $ &  $|M^2_{U 23}| $\\ 
& & & & &\\
     \hline
\vspace*{-0.3cm}
& & & & &\\
\vspace*{-0.3cm}
      $2500^2 \div 4000^2$ & $2500^2 \div 4000^2$ & $< 1000^2$  & $1000^2 \div 4000^2$ & $600^2 \div 3000^2$& $ < 2000^2$\\
& & & & &\\
    \hline
    \hline
\vspace*{-0.3cm}    
& & & & &\\
\vspace*{-0.3cm}      
      $ M^2_{D 22} $ & $ M^2_{D 33}$ &  $ |M^2_{D 23}|$ & $|T_{U 23}|  $ & $|T_{U 32}|  $ &  $|T_{U 33}|$\\ 
& & & & &\\
    \hline
\vspace*{-0.3cm}      
& & & & &\\
\vspace*{-0.3cm}  
       $ 2500^2 \div 4000^2$ & $1000^2 \div 3000^2 $ & $ < 2000^2$  & $< 4000 $ & $ < 4000$& $< 5000 $\\
& & & & &\\
 \hline 
\multicolumn{6}{c}{}\\[-3.6mm]  
\cline{1-4}
\vspace*{-0.3cm}      
     & & & \\
\vspace*{-0.3cm}      
     $ |T_{D 23}| $ & $|T_{D 32}|  $ &  $|T_{D 33}|$ &$|T_{E 33}| $\\ 
     & & & \\
    \cline{1-4}
\vspace*{-0.3cm}      
     & & & \\
\vspace*{-0.3cm}      
     $< 3000 $ & $< 3000 $& $ < 4000$& $ < 500$\\
     & & & \\
    \cline{1-4}
\end{tabular}\\[3mm]
\begin{tabular}{|c|c|c|c|c|c|c|c|c|}
    \hline
\vspace*{-0.3cm}      
    & & & & & & & &\\
\vspace*{-0.3cm}      
    $M^2_{Q 11}$ & $M^2_{U 11} $ &  $M^2_{D 11} $ & $M^2_{L 11}$ & $M^2_{L 22} $ &  $M^2_{L 33}$ & $M^2_{E 11}$&$M^2_{E 22}$ & $M^2_{E 33} $\\ 
    & & & & & & & &\\
    \hline
\vspace*{-0.3cm}      
    & & & & & & & &\\
\vspace*{-0.3cm}      
    $4500^2$ & $4500^2$ & $4500^2$  & $1500^2$ & $1500^2$ & $1500^2$& $1500^2$& $1500^2$&$1500^2$\\
    & & & & & & & &\\
    \hline
\end{tabular}
\end{center}
\label{table1}
}
\end{table}
%

\section{125 GeV Higgs boson decays in the MSSM with general QFV}
\label{sec:Deviations}
%
We compute the decay widths $\Gamma(h^0 \to c \bar c)$, 
$\Gamma(h^0 \to b \bar b)$ and $\Gamma(h^0 \to b \bar s / \bar b s)$ 
at full 1-loop level in the $\overline{DR}$ renormalization scheme in the 
MSSM with general QFV ~\cite{Bartl:h2cc, Eberl:h2bb} and study the 
deviation of the MSSM predictions from the SM ones.
We also compute the decay widths $\Gamma(h^0 \to g g)$ and 
$\Gamma(h^0 \to \gamma \gamma)$ at the next-to-leading order (NLO) QCD level in the 
$\overline{DR}$ renormalization scheme in the MSSM with general QFV ~\cite{h2gagagg} and 
study the deviation of the MSSM widths from the SM ones, where g is a gluon and 
$\gamma$ is a photon. As the $h^0$ decays to $g g$ and $\gamma \gamma$ are 
loop-induced decays, these decays are sensitive to New Physics. 

Here, we remark the differences between our previous works 
~\cite{Bartl:h2cc, Eberl:h2bb, h2gagagg} and the present work:
In the present work, we update the constraints on the MSSM 
parameters significantly including the expected sparticle mass 
limits from the future HL-LHC experiments, and study also the branching 
ratio of the explicitly QFV decay $B(h^0 \to b \bar s / \bar b s)$ 
and the correlations among the deviations of the MSSM widths of the 
various decay modes from the corresponding SM ones.\\

\subsection{Expectations}
\label{subsec:Expectations}
%
We find that large squark trilinear couplings $T_{U23,32,33}$, 
$T_{D23,32,33}$, large $M^2_{Q23}$, $M^2_{U23}$, $M^2_{D23}$, 
large bottom Yukawa coupling $Y_b$ for large $\tan\beta$, and 
large top Yukawa coupling $Y_t$ can lead to large MSSM 1-loop 
contributions to these decay widths, resulting in sizable deviation 
of the MSSM widths from the SM values.
This is mainly due to the following reasons:

The lighter up-type squarks $\su_{1,2,3}$ are strong $\sc_{L,R}$ - 
$\st_{L,R}$ mixtures for large $M^2_{Q23}$, $M^2_{U23}$, 
$T_{U23,32,33}$. The lighter down-type squarks $\sd_{1,2,3}$ 
are strong $\ss_{L,R}$ - $\sb_{L,R}$ mixtures for large 
$M^2_{Q23}$, $M^2_{D23}$, $T_{D23,32,33}$.
Here note that $|T_{U23,32,33}|$ the size of which are controlled 
by $Y_t$ due to the vacuum stability conditions can be large 
because of large $Y_t$ and that $|T_{D23,32,33}|$ the size of 
which are controlled by $Y_b$ due to the vacuum stability 
conditions can be large thanks to large $Y_b$ for large 
$\tan\beta$ (see Eqs. (\ref{eq:CCBfcU} - \ref{eq:CCBfvD}) in Appendix \ref{sec:constr}). 
In the following we assume these setups.
%
\subsubsection{Expectations for fermionic decays}
\label{subsubsec:Expect_fermionic_decays}
\begin{figure*}[t!]
\centering
\hspace*{-0.6cm}
  \subfigure[]{
  {\mbox{\resizebox{4.8cm}{!}{\includegraphics{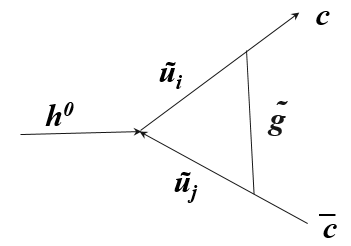}}}}
  \label{h02cc_gluino_loop}}
\hspace*{-4mm}  
  \subfigure[]{
  {\mbox{\resizebox{5.1cm}{!}{\includegraphics{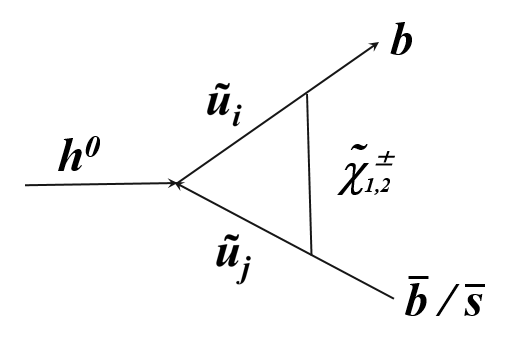}}}}
  \label{h02bb_chargino_loop}}
\hspace*{-4mm}   
  \subfigure[]{
  {\mbox{\resizebox{4.8cm}{!}{\includegraphics{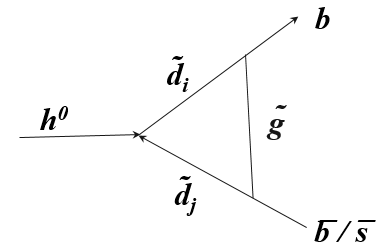}}}}
  \label{h02bb_gluino_loop}}
  \hspace*{-4mm} 
\caption{
(a) The $\su_i$-$\sg$ loop corrections to $\Gamma(h^0 \to c \bar c)$, 
(b) the $\su_i$-$\ch_{1,2}$ loop and (c) the $\sd_i$-$\sg$ loop corrections 
to $\Gamma(h^0 \to b \, \, \bar b / \bar s)$.
}
\label{1-loop_diag_to_h0_decay}
\end{figure*}

%
The main MSSM 1-loop corrections to $\Gamma(h^0 \to c \, \bar c)$ stem 
from the lighter up-type squarks ($\su_{1,2,3}$) - gluino ($\sg$) 
loops at the decay vertex which have $h^0-\su_i-\su_j$ couplings 
containing $H^0_2-\sc_R-\st_L$, $H^0_2-\sc_L-\st_R$, 
$H^0_2-\st_L-\st_R$ couplings, i.~e., $T_{U23,32,33}$ 
(see Fig.~\ref{h02cc_gluino_loop}). 
Note that $h^0$ is a mixture of $Re(H_1^0)$ (which couples to 
the down-type squarks $\sd_i$) and $Re(H_2^0)$ (which couples to 
the up-type squarks $\su_i$ ), i.~e., 
$h^0 = -\sin\alpha (\sqrt{2} Re(H_1^0) - v_1) + \cos\alpha (\sqrt{2} Re(H_2^0) - v_2)$ and that
$h^0$ is dominated by $Re(H^0_2)$ component in our decoupling 
Higgs scenario with large $m_{A^0} \, (> 1350 \, \GeV)$ and large 
$\tan \beta \, (> 10)$ (see Table \ref{table1}). 
Hence, the large $Re(H^0_2)$ component of $h^0$ and the large 
trilinear couplings $T_{U23,32,33}$ can enhance the 
$h^0-\su_i-\su_j$ couplings, which together with the large QCD 
couplings involved can result in a strong enhancement of the 
$\su_i$-$\sg$ loop corrections to $\Gamma(h^0 \to c \, \bar c)$, 
leading to a large deviation of the MSSM width 
$\Gamma(h^0 \to c \bar c)$ from its SM value.\\ 
Here note that $\su_i$ - neutralino ($\nt_{i}$) loops and 
$\sd_i$ - chargino ($\ch_{1,2}$) loops at the decay vertex are 
not so important by the following reason with $\nt_{i}$ and 
$\ch_{1,2}$ being mixtures of photino $\ti \gamma$, zino $\ti Z$, 
and neutral higgsinos $\ti H_{1,2}^0$ and mixtures of charged 
wino $\ti W^\pm$ and charged higgsino $\ti H^\pm$, respectively: 
(i) The former loops where $h^0$ directly couples to $\su_i-\su_j$ 
are suppressed by the relatively small electroweak-Yukawa couplings of 
the neutralino to c and $\su_i$ compared with the QCD couplings of 
gluino to c and $\su_i$ in the $\su_i$-$\sg$ loops. 
(ii) The former loops where $h^0$ directly couples to the neutralino 
are also suppressed by the relatively small couplings of the neutralino 
and cannot be enhanced by the large trilinear couplings $T_{U23,32,33}$.
(iii) The latter loops where $h^0$ directly couples to $\sd_i-\sd_j$ 
are suppressed by the small $Re(H^0_1)$ component of $h^0$ though 
they can be enhanced by the large trilinear couplings $T_{D23,32,33}$. 
They are further suppressed by the relatively small electroweak-Yukawa 
couplings of the chargino to c and $\sd_i$ compared with the QCD couplings.
(iv) The latter loops where $h^0$ directly couples to the 
chargino are also suppressed by the relatively small electroweak-Yukawa 
couplings of the chargino to c and $\sd_i$ and cannot be enhanced by 
the large trilinear couplings $T_{D23,32,33}$.
\\
%

The main MSSM 1-loop corrections to $\Gamma(h^0 \to b \bar b)$ and 
$\Gamma(h^0 \to b \bar{s}/\bar{b} s)$ stem from 
(i) $\su_{1,2,3}$ - chargino ($\ch_{1,2}$) loops at the decay 
vertex which have $h^0-\su_i-\su_j$ couplings to be enhanced by  
large $T_{U23,32,33}$ (see Fig.~\ref{h02bb_chargino_loop}) 
\footnote{
Note that the $\su_{1,2,3}$ - $\ch_{1,2}$ loops where $h^0$ couples directly 
to $\ch_{1,2}$ cannot be enhanced by the large $T_{U23,32,33}$ and hence 
they are not so important. 
} and
(ii) $\sd_{1,2,3}$ - $\sg$ loops at the decay vertex which have 
$h^0-\sd_i-\sd_j$ couplings containing $H^0_1-\ss_R-\sb_L$, 
$H^0_1-\ss_L-\sb_R$, $H^0_1-\sb_L-\sb_R$ couplings, 
i.~e., $T_{D23,32,33}$ (see Fig.~\ref{h02bb_gluino_loop})
\footnote{
Here note that the $h^0-\sd_i-\sd_j$ couplings are suppressed 
by the small $Re(H^0_1)$ component of $h^0$, but can be enhanced 
by large $T_{D23,32,33}$.
Note also that $\sd_{1,2,3}$ - $\nt_{i}$ loops at the decay vertex 
are not so important compared with $\sd_{1,2,3}$ - $\sg$ loops at 
the vertex by a reason similar to the reason why the 
$\su_i$ - $\nt_{i}$ loop corrections to $\Gamma(h^0 \to c \, \bar c)$ 
are suppressed compared with the $\su_i$ - $\sg$ loop corrections. 
}.
Hence large trilinear couplings $T_{U23,32,33}$ 
and $T_{D23,32,33}$ can enhance the MSSM 1-loop corrections 
to $\Gamma(h^0 \to b \bar b)$ and $\Gamma(h^0 \to b \bar{s}/\bar{b} s)$  
due to the $\su_i$ - $\ch_{1,2}$ and $\sd_i$ - $\sg$ loops, 
leading to large deviation of the MSSM widths $\Gamma(h^0 \to b \bar b)$ 
and $\Gamma(h^0 \to b \bar{s}/\bar{b} s)$ from their SM values. \\

Note that the wave function corrections for the external $h^0$ in the 
decays $h^0 \to c \bar c/b \bar b$ which have the $h^0-\su_i-\su_j$ and 
$h^0-\sd_i-\sd_j$ couplings (for $\su_i$ and $\sd_i$ loops, respectively) 
can also be enhanced by the large trilinear couplings $T_{U23,32,33}$ and 
$T_{D23,32,33}$, resulting in a further enhancement of the MSSM 1-loop 
corrections to the widths $\Gamma(h^0 \to c \bar c / b \bar b)$.\\

Here we remark that the MSSM one-loop corrections to the decay amplitude 
for $h^0 \to c \, \bar c$ are expected to be significantly larger than 
those for $h^0 \to b \, \bar b$ due to the following reasons:\\
(i) The main MSSM 1-loop corrections to $h^0 \to c \, \bar c$ stem 
from $\su_i$ - $\sg$ loops (Fig.~\ref{h02cc_gluino_loop}). 
The main MSSM 1-loop corrections to $h^0 \to b \, \bar b$ stem from 
$\su_i$ - $\ch_j$ loops (Fig.~\ref{h02bb_chargino_loop}) and 
$\sd_i$ - $\sg$ loops (Fig.~\ref{h02bb_gluino_loop}).\\
(ii) The $\su_i$ - $\ch_j$ loops of Fig.~\ref{h02bb_chargino_loop} 
are suppressed by the relatively small electroweak-Yukawa couplings 
of $\ch_j$ compared with the $\su_i$ - $\sg$ loops of 
Fig.~\ref{h02cc_gluino_loop} having the large QCD couplings of $\sg$.\\
(iii) The $\sd_i$ - $\sg$ loops of Fig.~\ref{h02bb_gluino_loop} 
are suppressed by the relatively small $h^0-\sd_i-\sd_j$ couplings due to 
the small $Re(H^0_1)$ component of $h^0$ compared with the $\su_i$ - $\sg$ 
loops of Fig.~\ref{h02cc_gluino_loop}.\\
(iv) Hence the MSSM one-loop corrections to the decay amplitude 
for $h^0 \to c \, \bar c$ (Fig.~\ref{h02cc_gluino_loop}) are expected to 
be significantly larger than those for $h^0 \to b \, \bar b$
(Fig.~\ref{h02bb_chargino_loop} + Fig.~\ref{h02bb_gluino_loop}).\\
\indent On the other hand, the SM width $\Gamma(h^0 \to c \, \bar c)_{SM}$ 
is much smaller than $\Gamma(h^0 \to b \, \bar b)_{SM}$ mainly due to the much 
smaller charm Yukawa coupling than the bottom Yukawa one, which together with 
the item (iv) results in {\it much larger} relative deviation of the MSSM 
width from the SM width for the decay $h^0 \to c \, \bar c$ than that for 
$h^0 \to b \,\bar b$ (see Eq.(\ref{DEVX})) {\it in strong contrast to usual 
expectations}. We will see this tendency explicitly in the plots shown below 
(e.g. see Fig.~\ref{DEVc_DEVb}).\\

\subsubsection{Expectations for bosonic decays}
\label{subsubsec:Expect_bosonic_decays}
\begin{figure*}[t!]
\centering
  \subfigure[]{
  {\mbox{\resizebox{5cm}{!}{\includegraphics{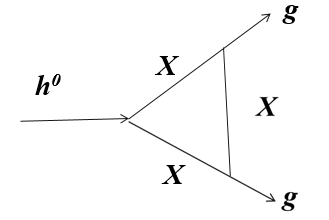}}}}
  \label{h02gg_loops}}
  \hspace{5mm}
  \subfigure[]{
  {\mbox{\resizebox{5cm}{!}{\includegraphics{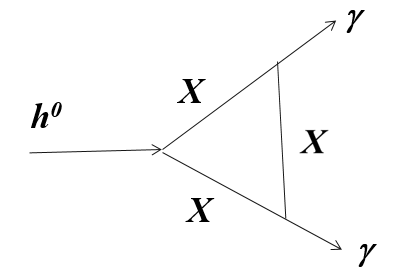}}}}
  \label{h02gamgam_loops}}
\caption{
(a) The SM (X = top quark) and MSSM (X = $\su_{1,2,3}$) loop contributions  
to $\Gamma(h^0 \to g g)$, and 
(b) the SM (X = $W^+$ boson, top quark) and MSSM (X = $\su_{1,2,3}$) loop 
contributions to $\Gamma(h^0 \to \gamma \gamma)$.
The NLO QCD correction diagrams are not shown in these figures. 
For (a) $h^0 \to g g$, the MSSM ($\su_{i}$) loop diagrams with the contact 
interactions of $\su_{i}$-$\su_{i}$-gluon-gluon are not shown. 
For (b) $h^0 \to \gamma \gamma$, the SM ($W^+$) loop diagram with that 
of W-W-$\gamma$-$\gamma$ and the MSSM ($\su_{i}$) loop diagrams with those 
of $\su_{i}$-$\su_{i}$-$\gamma$-$\gamma$ are not shown.
}
\label{1-loop_diag_to_h02gagagg}
\end{figure*}

%
Similar arguments hold for the loop-induced decays $h^0 \to g g, \gamma \gamma$. 
The main SM 1-loop contribution to $\Gamma(h^0 \to g g)$ stems from the top-quark loop. 
The bottom-quark loop contribution to this width is much suppressed by the small 
$Re(H_1^0)$ component of $h^0$ in our decoupling Higgs scenario. 
The main MSSM 1-loop contributions to $\Gamma(h^0 \to g g)$  stem from 
the lighter up-type squark ($\su_{1,2,3}$) loops which have $h^0-\su_i-\su_i$ 
couplings (see Fig.~\ref{h02gg_loops}). The large trilinear couplings 
$T_{U23,32,33}$ can enhance the $h^0-\su_i-\su_i$ couplings and hence the 
$\su_i$ loops, resulting in sizable deviation of the MSSM width 
$\Gamma(h^0 \to g g)$ from its SM value. The lighter down-type squark 
($\sd_{1,2,3}$) loop contributions to this width are suppressed by the 
small $Re(H_1^0)$ component of $h^0$.\\
The main SM 1-loop contributions to $\Gamma(h^0 \to \gamma \gamma)$ stem from 
$W^+$ boson and top-quark loops. The bottom-quark and tau-lepton loop 
contributions to this width are suppressed by the small $Re(H_1^0)$ component 
of $h^0$. The main MSSM 1-loop contributions to $\Gamma(h^0 \to \gamma \gamma)$  
stem from the lighter up-type squark ($\su_{1,2,3}$) loops which have 
$h^0-\su_i-\su_i$ couplings (see Fig.~\ref{h02gamgam_loops}). The large 
trilinear couplings $T_{U23,32,33}$ can enhance the $h^0-\su_i-\su_i$ couplings 
and hence the $\su_i$ loops, leading to sizable deviation of the MSSM width 
$\Gamma(h^0 \to \gamma \gamma)$ from its SM value. 
The lighter down-type squark ($\sd_{1,2,3}$) and the charged slepton 
loop contributions to this width are suppressed by the small $Re(H_1^0)$ 
component of $h^0$. The chargino loops can also contribute to this width,  
but they can not be enhanced by the large trilinear couplings $T_{U23,32,33}$. 
The $H^+$ boson loop contribution to this width is strongly suppressed 
by its large mass $m_{H^+}$ ($\simeq m_{A^0}$) in our decoupling 
Higgs scenario with large $m_{A^0} \, (> 1350 \, \GeV)$ and large 
$\tan \beta \, (> 10)$ (see Table \ref{table1}). 
Here note that the deviation of the MSSM width $\Gamma(h^0 \to \gamma \gamma)$ 
from its SM value is not so large since the $W^+$ boson loop contribution 
dominates this width. \\

\subsection{Scatter plot analysis}
\label{subsec:Scatter_plot_analysis}
%
We compute the decay widths 
$\Gamma(h^0 \to X \bar X)$ ($X=c,b$) and $\Gamma(h^0 \to b \, s) \equiv 
\Gamma(h^0 \to b \, \bar s) + \Gamma(h^0 \to \bar b \, s)$ at full 1-loop 
level in the $\overline{DR}$ renormalization scheme in the MSSM 
\newpage 
\noindent with general QFV using Fortran codes developed by us \cite{Bartl:h2cc, Eberl:h2bb}.
\footnote{
The SM widths $\Gamma(h^0 \to X \bar X)_{SM}$ ($X=c,b$) are computed in 
Refs. \cite{CERN_YR4, Almeida2014}, but we do not use these SM widths. 
Instead, we compute the SM widths $\Gamma(h^0 \to X \bar X)_{SM}$ ($X=c,b$) by taking 
the decoupling SUSY/Higgs limit of the MSSM width $\Gamma(h^0 \to X \bar X)_{MSSM}$, 
i.~e., the limit of large SUSY mass scale $M_{SUSY}$, large $m_{A^0}(pole)$, 
large $\tan \beta$ and no SUSY QFV (no squark generation mixing) in order to 
calculate the relative deviation of the MSSM width $\Gamma(h^0 \to X \bar X)_{MSSM}$ 
from the SM width $\Gamma(h^0 \to X \bar X)_{SM}$ at full 1-loop level {\it consistently}. 
We have obtained $\Gamma(h^0 \to c \bar c)_{SM} = 0.128 \MeV$ and 
$\Gamma(h^0 \to b \bar b)_{SM} = 2.89 \MeV$. 
} 
We compute the decay widths $\Gamma(h^0 \to X X)$ ($X=g,\gamma$) 
at the NLO QCD level in the $\overline{DR}$ renormalization scheme in the MSSM 
with general QFV using Fortran codes developed by us \cite{h2gagagg}. 
\footnote{
The SM widths $\Gamma(h^0 \to X X)_{SM}$ ($X=g,\gamma$) are computed in 
Refs. \cite{CERN_YR4, Almeida2014}, but we do not use these SM widths. 
Instead, we compute the SM widths $\Gamma(h^0 \to X X)_{SM}$ ($X=g,\gamma$) 
at the NLO QCD level by ourselves in order to calculate the relative deviation 
of the MSSM width $\Gamma(h^0 \to X X)_{MSSM}$ from the SM width 
$\Gamma(h^0 \to X X)_{SM}$ at the NLO QCD level {\it consistently}.
We also compute the SM widths $\Gamma(h^0 \to X X)_{SM}$ ($X=g,\gamma$) by taking 
the decoupling SUSY/Higgs limit of the MSSM width $\Gamma(h^0 \to X X)_{MSSM}$ 
at the NLO QCD level. We have found that the former SM widths agree with the latter 
SM widths very well. We have obtained $\Gamma(h^0 \to g g)_{SM} = 0.262 \MeV$ and 
$\Gamma(h^0 \to \gamma \gamma)_{SM} = 0.0111 \MeV$.  
}
The details of the computation of the decay widths $\Gamma(h^0 \to X X)$ ($X=g,\gamma$) 
at the NLO QCD level in the MSSM with general QFV are explained in Section 3 of 
\cite{h2gagagg} (see especially footnote 1 in Section 3 of \cite{h2gagagg}).\\
In the following we will show scatter plots in various planes related 
with these decay widths obtained from the MSSM parameter scan described 
above (see Table \ref{table1}), respecting all the relevant constraints 
(see Appendix \ref{sec:constr}).\\

\subsubsection{Definition of relative deviations from SM predictions}
\label{subsubsec:Definition_of_DEV}
%
We define the relative deviation of the decay width $\Gamma(X) 
(\equiv \Gamma(h^0 \to X \bar X))$ from the SM width as follows:
\be
\DEV(X) \equiv \frac{\Gamma(X)}{\Gamma(X)_{SM}} - 1 \, \, (X=c,b,g,\gamma).
  \label{DEVX}
\ee

\noindent Here $\Gamma(X)_{SM}$ is the SM prediction for the decay width $\Gamma(X)$. 

\noindent According to Ref. \cite{Snowmass2021_Rep} we define the effective $h^0 X X$ 
coupling $g(h^0 X X)$ as follows: 
\be
g(h^0 X X)^2 \equiv \frac{\Gamma(X)}{\Gamma(X)_{SM}}.
  \label{Eff_Coupling}
\ee

As the SM effective coupling $g(h^0 X X)_{SM} = 1$ by definition, 
the so-called coupling modifier $\kappa_{X} (\equiv g(h^0 X X)/g(h^0 X X)_{SM})$ 
is equal to $g(h^0 X X)$.
The relative deviation DEV(X) is related with the effective coupling $g(h^0 X X)$ 
and the coupling modifier $\kappa_X$ as follows:
\be
\DEV(X) = g(h^0 X X)^2 - 1 = \kappa_X^2 - 1.
  \label{DEV_Coup_kappa}
\ee

\noindent We define the relative deviation of the width ratio $\Gamma(X)/\Gamma(Y)$ 
from its SM prediction as follows:
\be
\DEV(X/Y) \equiv \frac{\Gamma(X)/\Gamma(Y)}{\Gamma(X)_{SM}/\Gamma(Y)_{SM}} - 1.
  \label{DEVRXY}
\ee
Note that we have the following approximation: 
\be
\DEV(X/Y) \simeq \DEV(X) - \DEV(Y) \, (\mbox{for} \,|\DEV(Y)| \ll 1).
  \label{DEVRXY_APPROX}
\ee
\noindent It is important to notice that a significant (substantial) part of the  
experimental systematic and statistical errors of the measured widths $\Gamma(X)$ 
and $\Gamma(Y)$ cancel out in the width ratio $\Gamma(X)/\Gamma(Y)$, which results 
in a relatively small experimental error of the measured width ratio. The theoretical 
errors of the MSSM widths $\Gamma(X)_{MSSM}$ and $\Gamma(Y)_{MSSM}$ also cancel out 
significantly in the width ratio $\Gamma(X)_{MSSM}/\Gamma(Y)_{MSSM}$, which leads 
to a relatively small theoretical error of the MSSM width ratio; e.g., the phase-space 
factor proportional to $1/m_{h^0}$ cancels out in the MSSM width ratio, where we impose 
the constraint $m_{h^0} = 125.09 \pm 3.48$~GeV (see Table \ref{TabConstraints}).  
Therefore, the experimental measurement errors as well as the MSSM prediction 
uncertainties tend to cancel out significantly in the width ratios, making the 
measurement of these width ratios a very sensitive probe of virtual SUSY loop 
effects in these $h^0$ decays at future lepton colliders. 
Moreover, as we expect from Eq. (\ref{DEVRXY_APPROX}), the  
deviation of the MSSM width ratio from the SM prediction can be significantly 
enhanced compared with that of a single MSSM width from the SM; e.g., we will see 
below that DEV($\gamma$/g) can be as large as about +9\% (see Fig.~\ref{DEVgam2g_TU32M2U23}). 
Furthermore, there can be significant correlations 
between the deviation of the single MSSM width from the SM value and that of 
the MSSM width ratio from the SM; e.g. there is a very strong correlation 
between DEV(c) and DEV(b/c) as shown in Fig.~\ref{DEVb2c_DEVc} below. 

\subsubsection{Scatter plots for fermionic decays}
\label{subsubsec:Scatter_plots_fermionic_decays}

{\bf - Scatter plots for DEVs of fermionic decays}\\

\begin{figure*}[h!]
\centering
 \renewcommand{\subfiglabelskip}{-7mm}
 \subfigure[]{
 {\mbox{\resizebox{7.0cm}{!}{\includegraphics{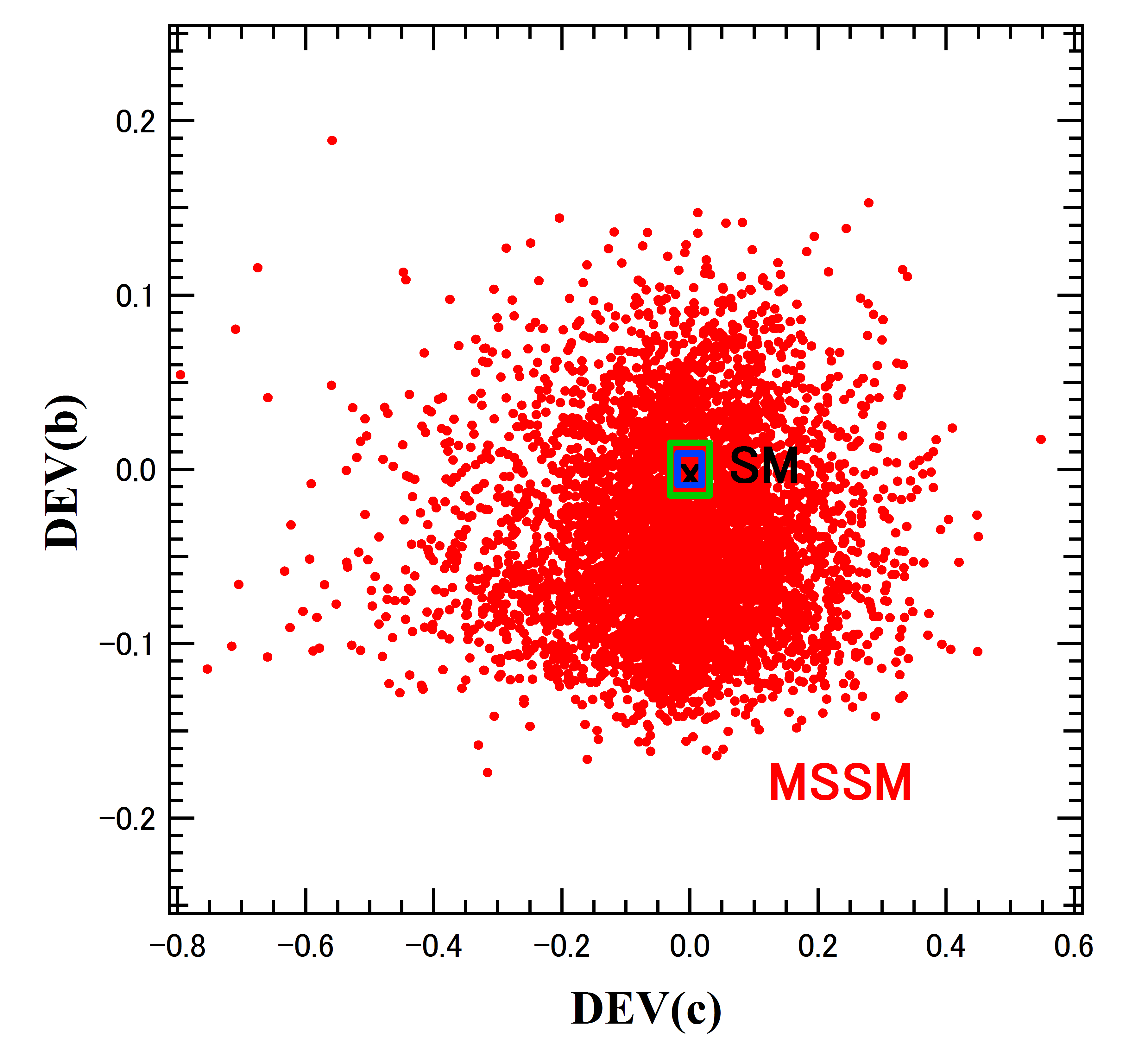}}}}
 \label{DEVc2DEVb}}
 \subfigure[]{
 {\mbox{\resizebox{7.0cm}{!}{\includegraphics{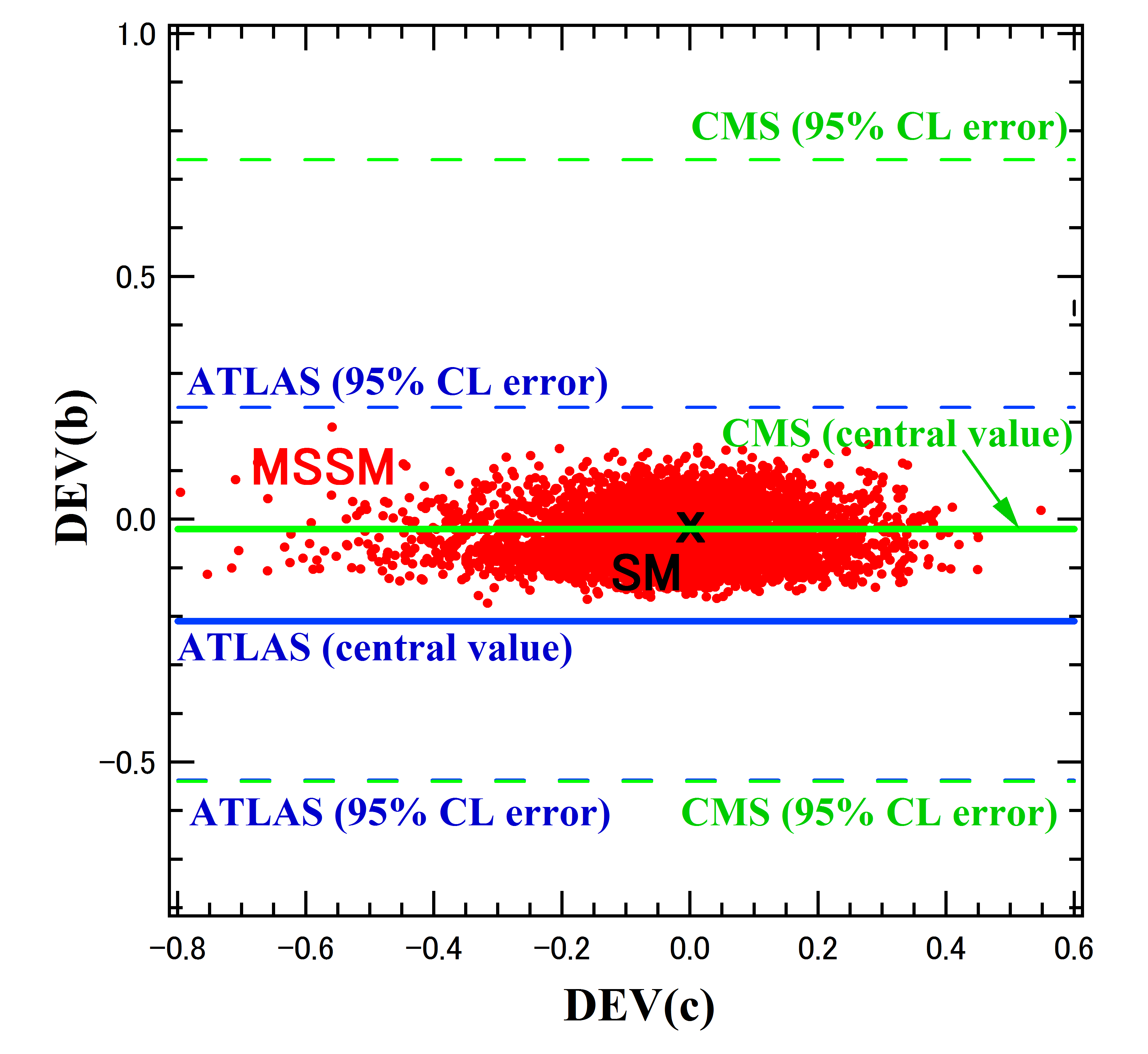}}}}
 \label{DEVc2DEVb_LHC}}
\caption{
Shown in (a) is the scatter plot in the DEV(c)-DEV(b) plane obtained 
from the MSSM parameter scan described in Section \ref{sec:full scan}. 
"X" marks the SM point. 
The green and blue boxes indicate the expected $1\sigma$ errors 
at ILC250 and ILC250+500, respectively (see Table \ref{table_DEVerror_LC}).
Though in principle the expected $1\sigma$ error should be shown by an error ellipse, 
here it is shown by an error box as an approximation since such $1\sigma$ error ellipse 
in the $\kappa_c$-$\kappa_b$ plane (and hence, in the DEV(c)-DEV(b) plane) 
is not given in \cite{ESU2020_Rep, Snowmass2021_Rep}. 
The expected absolute $1\sigma$ errors at the other lepton colliders are similar 
to those at ILC (see Table \ref{table_DEVerror_LC}). 
In (b) we show also the ATLAS and CMS data of DEV(b) obtained 
from the recent $\kappa_b$ data \cite{kappa_bgamg_ATLAS, kappa_bgamg_CMS}
by using the DEV(b)-$\kappa_b$ relation DEV(b) = $\kappa_b^2$ - 1.
}
\label{DEVc_DEVb}
\end{figure*}

\begin{figure*}[t!]
\centering
\renewcommand{\subfiglabelskip}{-7mm}
 \subfigure[]{
 {\mbox{\resizebox{7.0cm}{!}{\includegraphics{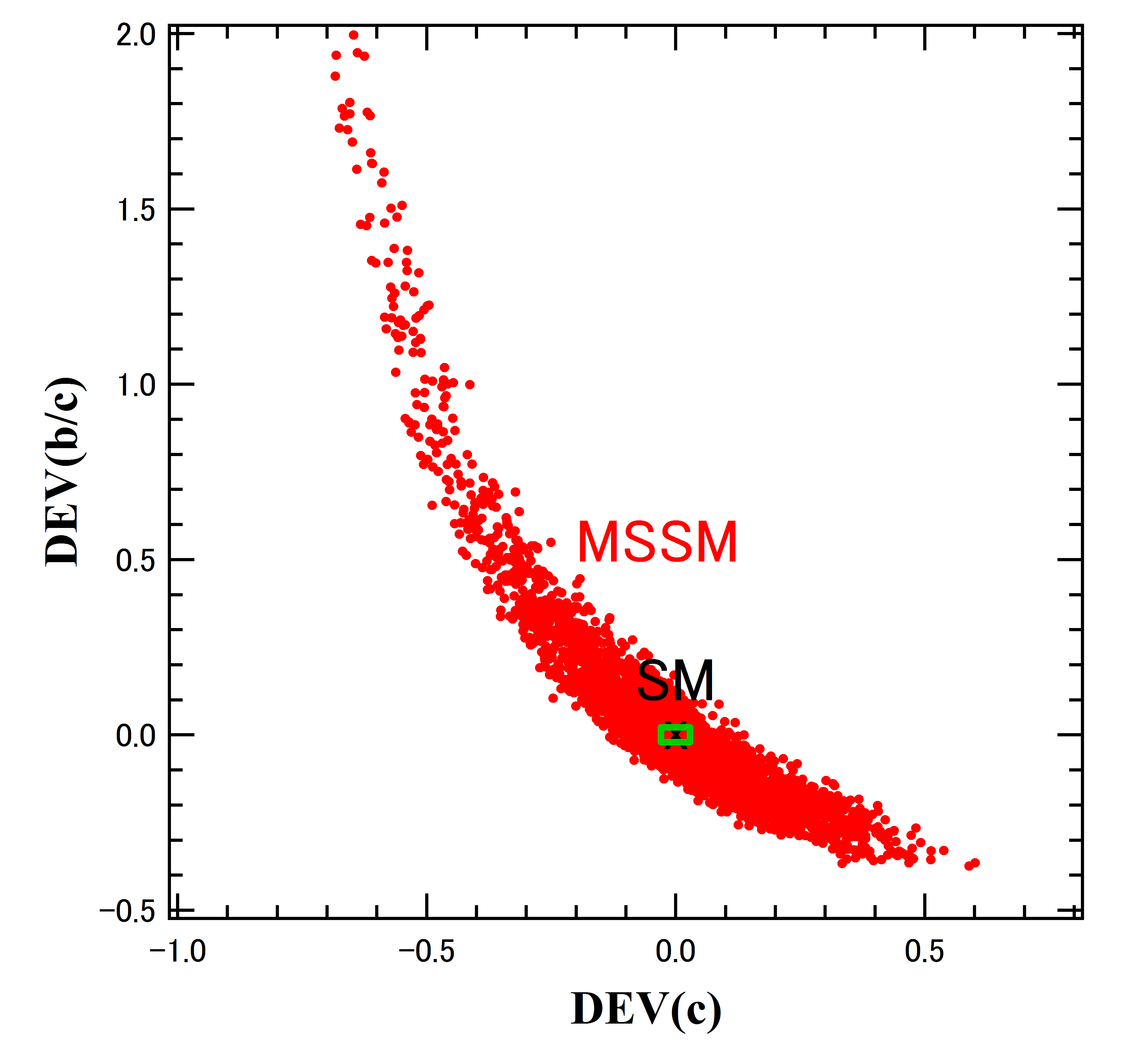}}}}
 \label{DEVb2c_DEVc}}
 \subfigure[]{
 {\mbox{\resizebox{7.0cm}{!}{\includegraphics{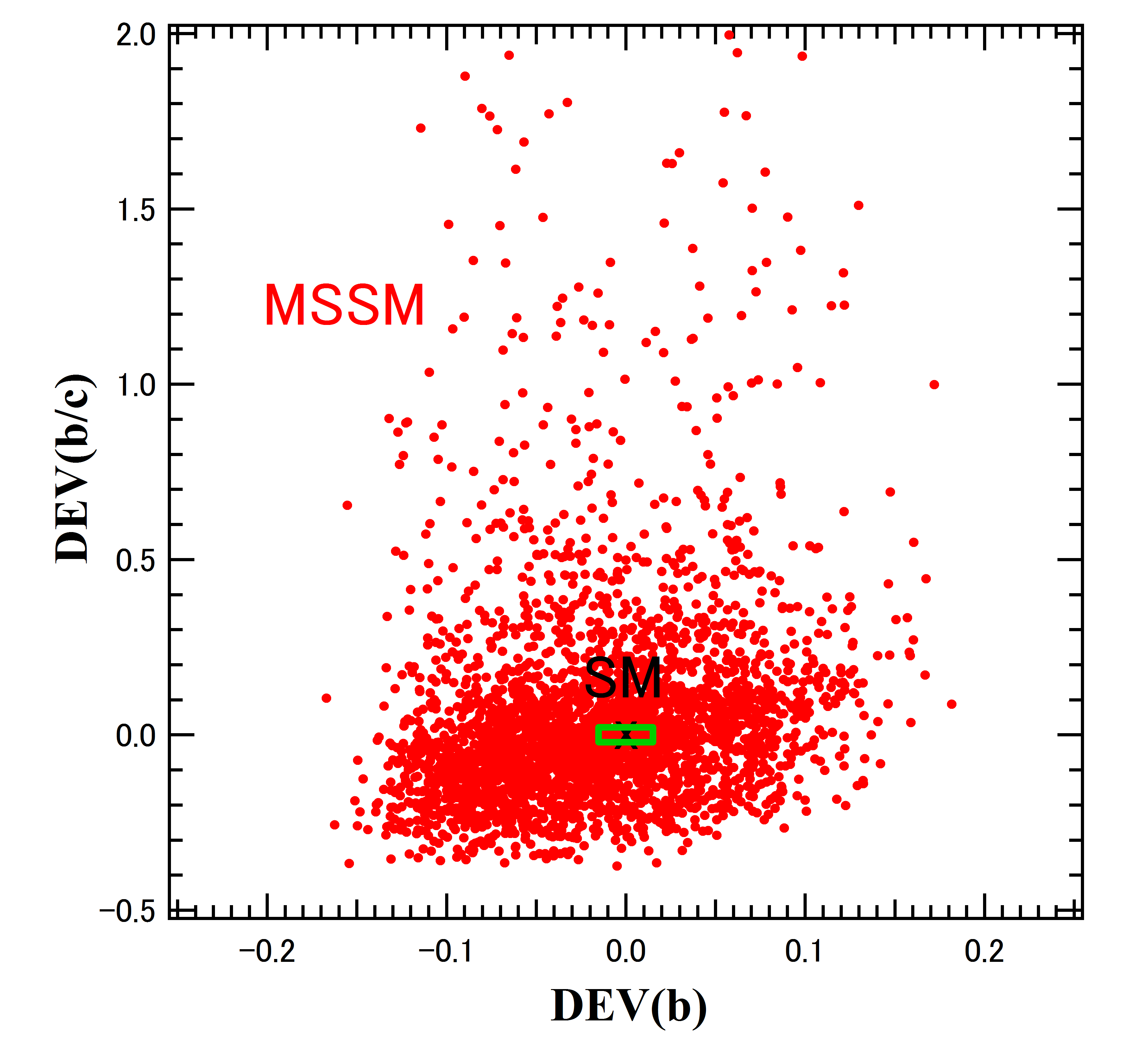}}}}
 \label{DEVb2c_DEVb}}
\caption{
The scatter plot in the (a) $\DEV(c)$-$\DEV(b/c)$ and (b) $\DEV(b)$-$\DEV(b/c)$ 
planes obtained from the MSSM parameter scan described in Section \ref{sec:full scan}.  
"X" marks the SM point. 
The expected absolute $1\sigma$ errors at ILC250 shown by the green box are given 
by ($\Delta \DEV(c)$, $\Delta \DEV(b)$, $\Delta \DEV(b/c)$)=(3.6\%, 1.7\%, 3.1\%) 
(see Table \ref{table_DEVerror_LC}). Though, in principle, the expected $1\sigma$ 
error should be shown by an error ellipse, here it is shown by an error box as 
an approximation. The expected absolute $1\sigma$ errors at the other 
lepton colliders are similar to those at ILC (see Table \ref{table_DEVerror_LC}). 
}
\label{DEVb2c_DEVbc}
\end{figure*}

\vspace{-0.4cm}
\noindent In Fig.~\ref{DEVc_DEVb} we show the scatter plot in the DEV(c)-DEV(b) 
plane obtained from the MSSM parameter scan. 
DEV(c) and DEV(b) can be quite large simultaneously since 
large trilinear couplings $T_{U23}$, $T_{U32}$ and $T_{U33}$ can enhance 
both DEV(c) and DEV(b) as explained above. 
From Fig.~\ref{DEVc2DEVb}, indeed we see that DEV(c) and DEV(b) can be quite 
large simultaneously: DEV(c) can be as large as $\sim\pm 60 \%$ and DEV(b) 
can be as large as $\sim\pm 20 \%$.  
Future lepton colliders together with HL-LHC can observe such large 
deviations from SM at very high significance (see Appendix \ref{sec:error}); e.g., 
from Appendix \ref{sec:error} we see that the expected absolute 1$\sigma$ error of DEV(c) is 
sufficiently small $\Delta$\DEV(c) = (3.6\%, 2.4\%, 1.8\%) and that of DEV(b) is 
also sufficiently small $\Delta$\DEV(b) = (1.7\%, 1.1\%, 0.9\%) at 
(ILC250, ILC250+500, ILC250+500+1000) together with HL-LHC, respectively 
(see Fig.~\ref{DEVc2DEVb}). The expected absolute 1$\sigma$ errors of DEV(c) and 
DEV(b) at the other lepton colliders are similar to those at ILC (see Appendix \ref{sec:error}). \\
In Fig.~\ref{DEVc2DEVb_LHC}, we show also the ATLAS and CMS data of 
DEV(b) at 95\% CL obtained from the recent $\kappa_b$ data \cite{kappa_bgamg_ATLAS, kappa_bgamg_CMS} 
shown in Table~\ref{TabConstraints} by using the relation $\DEV(b)=\kappa_b^2 -1$:
DEV(b) = -0.21 $^{+0.44}_{-0.33}$ (95\% CL) (ATLAS) and 
DEV(b) = -0.02 $^{+0.76}_{-0.52}$ (95\% CL) (CMS). 
Here note that the current LHC data of $\kappa_c$ (and hence, that of DEV(c) also) has very 
large uncertainties (errors) \cite{kappa_bgamg_ATLAS, kappa_bgamg_CMS}.
We see that both the SM and the MSSM are consistent with the recent ATLAS and CMS data, 
and that the errors of the recent ATLAS and CMS data are relatively very large.
\\
\indent In Fig.~\ref{DEVb2c_DEVc}, we show the scatter plot in the 
$\DEV(b/c)$-$\DEV(c)$ plane obtained from the MSSM parameter scan 
described in Section \ref{sec:full scan}. We see 
that there is a strong correlation between $\DEV(b/c)$ and $\DEV(c)$, 
and that $\DEV(b/c)$ can be quite large for large $| \DEV(c)|$: $\DEV(b/c)$ 
can exceed +100\% for $\DEV(c) \lsim -0.5$. This strong correlation 
stems from the fact that the two DEVs have a common origin of enhancement, i.~e., 
the large trilinear couplings $T_{U23,32,33}$. This behavior is consistent 
with the expectation from the argument above.\\  
In Fig.~\ref{DEVb2c_DEVb}, we show the scatter plot in the 
$\DEV(b/c)$-$\DEV(b)$ plane obtained from the MSSM parameter scan.
We see that there is an appreciable correlation 
between $\DEV(b/c)$ and $\DEV(b)$, which comes also from the fact that 
the two DEVs have a common origin of enhancement, i.~e., the 
large trilinear couplings $T_{U23,32,33}$ and $T_{D23,32,33}$. \\


\noindent 
{\bf - Scatter plots for QFV decay $h^0 \to b \, s$}\\

\vspace{-0.4cm}
\noindent  
As for the explicitly QFV decay $h^0 \to b \, s$, Refs. \cite{Bejar, Curiel_1, 
Demir, Curiel_2, Barenboim} computed $B(h^0 \to b \, s)$ in the MSSM with 
general QFV. However, they neglected LR and RL squark flavor (generation) 
mixings which we have found very important. Moreover, the constraints 
used in \cite{Bejar, Curiel_1, Demir, Curiel_2, Barenboim} are incomplete 
and/or old (obsolete).\\
Ref. \cite{Heinemeyer} computed $B(h^0 \to b \, s)$ at full 1-loop level 
in the MSSM with general QFV including LL, RR, LR, RL squark flavor (generation) 
mixings respecting relevant constraints as we do here. 
However, some of the constraints used in \cite{Heinemeyer} (including those 
from B meson data) are already obsolete; e.g., 
they performed a $B(h^0 \to b \, s)$ contour plot analysis in the squark flavor mixing 
parameter planes in six benchmark scenarios (S1-S6). All of the benchmark scenarios 
except S5 are already excluded by the recent gluino mass limit from LHC, 
$m_{\sg} > 2.35$ TeV for $m_{\nt_1} < 1.55$ TeV (see Appendix \ref{sec:constr}).\\

In the present work, we update these constraints (including the B meson data) 
and perform a systematic MSSM parameter scan respecting the updated constraints. 
Furthermore, we take into account also the expected mass limits for the SUSY 
particles and the heavier MSSM Higgs bosons $H^0$, $A^0$, $H^+$ from the 
future HL-LHC experiment.\\
We compute $B(h^0 \to b \, s)$ approximately from the full 1-loop level width 
$\Gamma(h^0 \to b \, s)$ in the MSSM with general QFV by dividing it by the LO 
total width of the $h^0$ decay obtained from the public code 
{\tt SPheno}-v3.3.8~\cite{SPheno1, SPheno2}.\\

In Fig.~\ref{BRbs_DEVb} we show the scatter plot in the 
$B(h^0 \to b s)$-$\DEV(b)$ plane obtained from the MSSM parameter 
scan described in Section \ref{sec:full scan}. 
We see that $B(h^0 \to b s)$ can be as large as $\sim$0.1\% 
and that $B(h^0 \to b s)$ and $\DEV(b)$ can be sizable simultaneously.
This is due to the fact that $B(h^0 \to b s)$ and $\DEV(b)$ have a common 
origin of enhancement, i.~e., large trilinear couplings $T_{U23,32,33}$ and 
$T_{D23,32,33}$ (see Fig.~\ref{h02bb_chargino_loop} and Fig.~\ref{h02bb_gluino_loop}).\\
On the other hand, from our contour plot analysis of the branching ratio $B(h^0 \to b \, s)$ 
shown below in Fig.~\ref{BRbs_TU32tanb}, we find that it can be as large as $\sim 0.15\%$ 
respecting all the updated constraints including the expected mass limits for the SUSY particles 
and the heavier MSSM Higgs bosons from the future HL-LHC experiment. 
Here we remark that \cite{Heinemeyer} found that $B(h^0 \to b \, s)$ can be as large as 
$\sim O(0.1\%)$, however, respecting already outdated constraints in the MSSM with general QFV. 
It is very small ($B(h^0 \to b \, s) \lsim 10^{-7}$) in the SM \cite{Bejar, Benitez, Kamenik}. 
The ILC250+500+1000 sensitivity to this branching ratio $B(h^0 \to b \, s)$ could be 
$\sim 0.1\%$ at 4$\sigma$ signal significance \cite{Tian} (see also \cite{Barducci}) 
(see Appendix \ref{sec:ILC_sensitivity_to_BRbs}). 
Hence, such QFV decay $h^0 \to b \, s$ in the MSSM with general QFV 
can be observed at ILC with very high signal significance. 
\footnote{
The expected upper bound on $B(h^0 \to b s)$ at FCC-ee 
is \cite{Selvaggi_FCC_Meeting_2024}:
$B(h^0 \to b s) < 4.5 \cdot 10^{-4}$~(95\%~CL).
The expected upper bound on $B(h^0 \to b s)$ at CEPC 
is \cite{Manqi_Higgs2023}:
$B(h^0 \to b s) < 2.2 \cdot 10^{-4}$~(95\%~CL).
} 
Note that the LHC and HL-LHC sensitivity to this QFV decay branching ratio 
should not be so good due to huge QCD background \cite{Barducci}.\\

Here we comment on the effect of resummation of the bottom 
Yukawa coupling for $\Gamma(h^0 \to b \, \bar b)$ and 
$\Gamma(h^0 \to b \, s)$ in the MSSM. We have studied the effect 
of the resummation of the bottom Yukawa coupling for large 
$\tan\beta$ \cite{Carena}. It turns out that in 
our case with large $m_{A^0}$ close to the decoupling Higgs limit 
(see Table \ref{table1}), the resummation effect (the 
so-called $\Delta_b$ effect) is very small ($<$ 0.1\%) 
\cite{Eberl:h2bb, EPS-HEP2015}. \\

\newcommand{\len}{7cm}
\begin{figure*}[!t]
\centering
 {\mbox{\resizebox{\len}{!}{\includegraphics{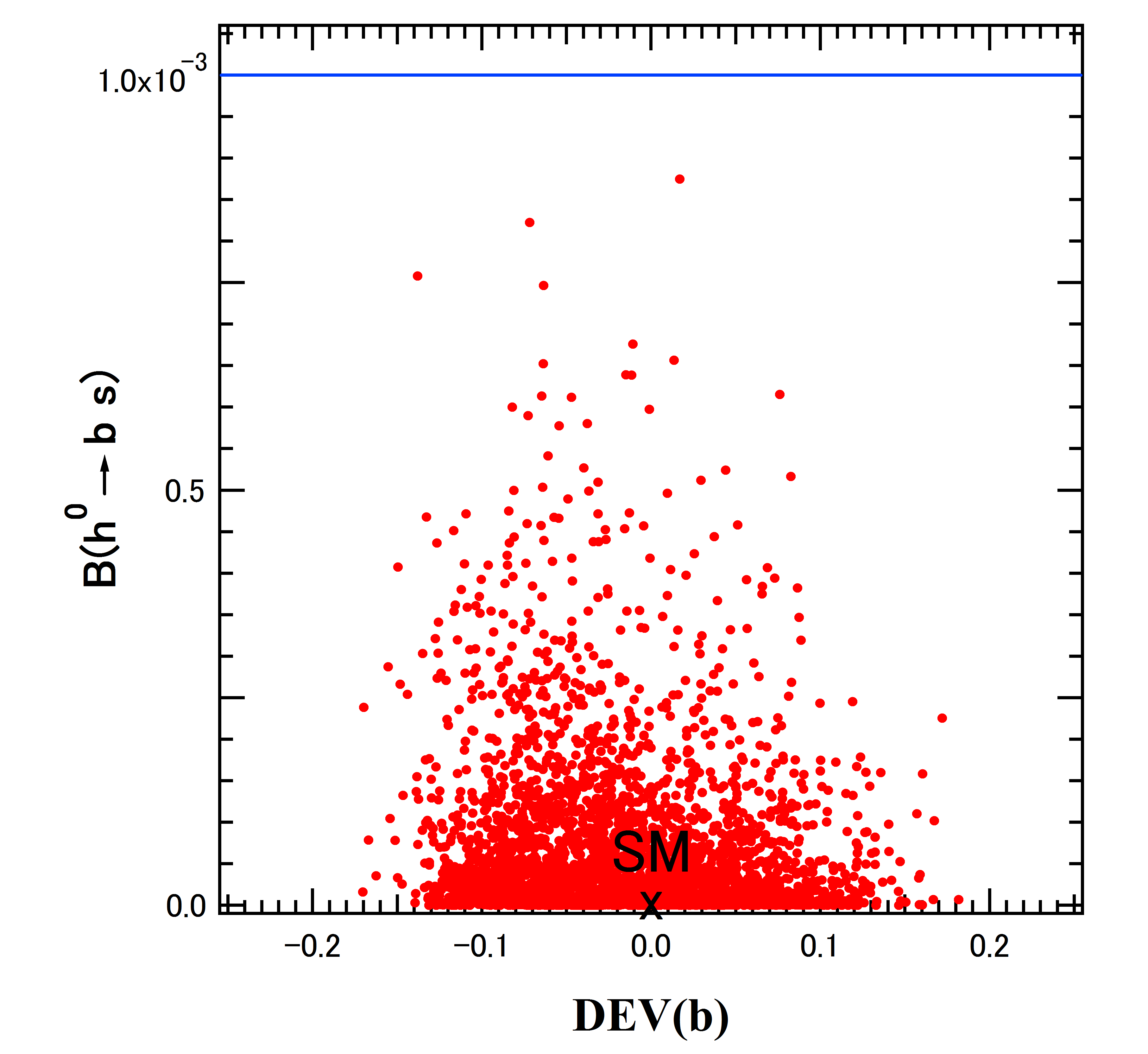}}}}
%
\caption{
The scatter plot in the $B(h^0 \to b s)$-$\DEV(b)$ plane obtained from the 
MSSM parameter scan described in Section \ref{sec:full scan}. 
The blue horizontal line indicates the ILC250+500+1000 sensitivity to 
$B(h^0 \to b s)$ of $\sim$0.1\% at 4$\sigma$ signal significance. 
"X" marks to the SM point. 
}
\label{BRbs_DEVb}
\end{figure*}

\subsubsection{Scatter plots for bosonic decays}
\label{subsubsec:Scatter_plots_bosonic_decays}

{\bf - Scatter plot in DEV($\gamma$)-DEV(g) plane}\\

\vspace{-0.4cm}

\begin{figure*}[!t]
\centering
\renewcommand{\subfiglabelskip}{-7mm}
 \subfigure[]{
 {\mbox{\resizebox{7.0cm}{!}{\includegraphics{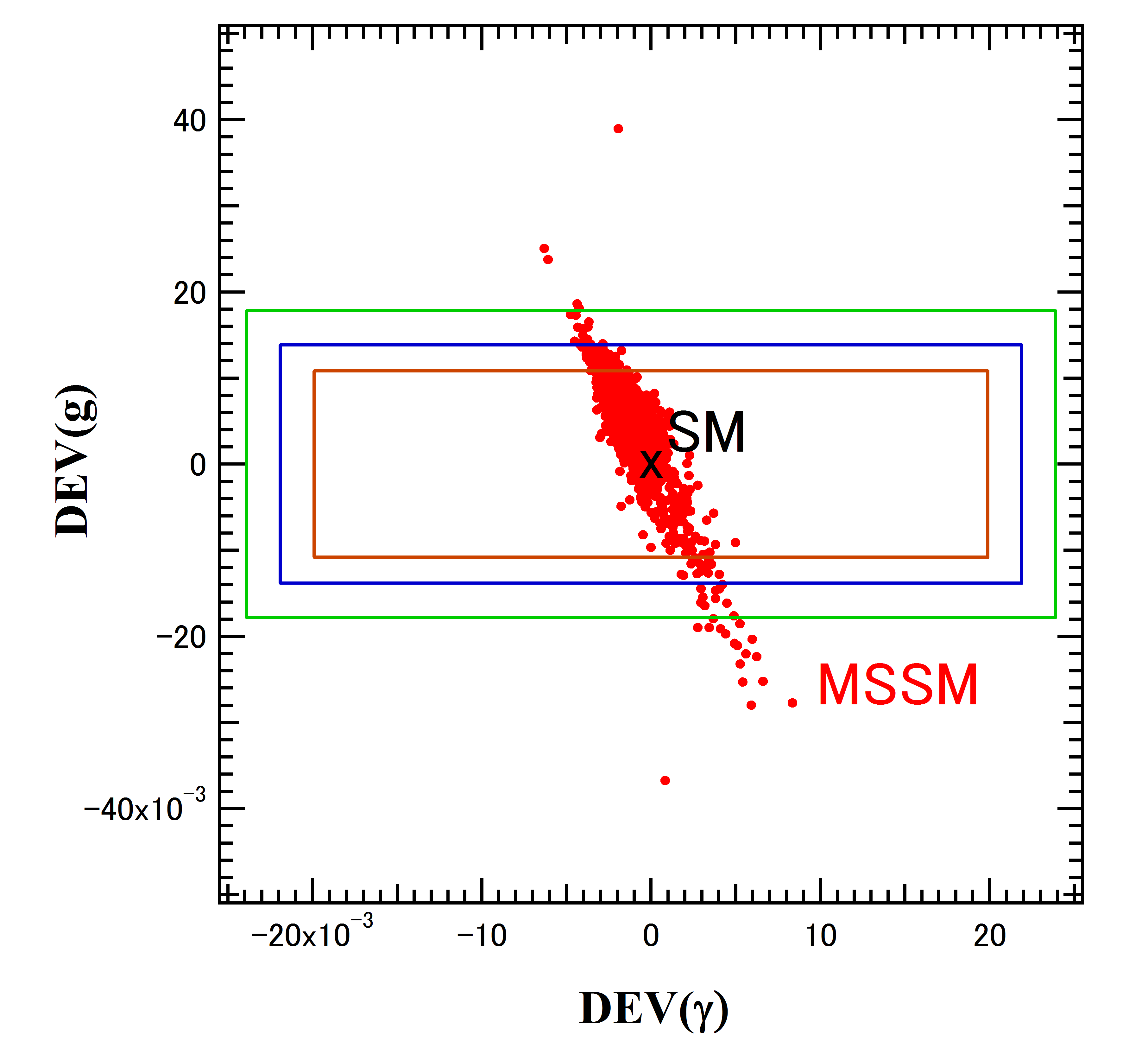}}}}
 \label{DEVgam2DEVg}}
 \subfigure[]{
 {\mbox{\resizebox{7.0cm}{!}{\includegraphics{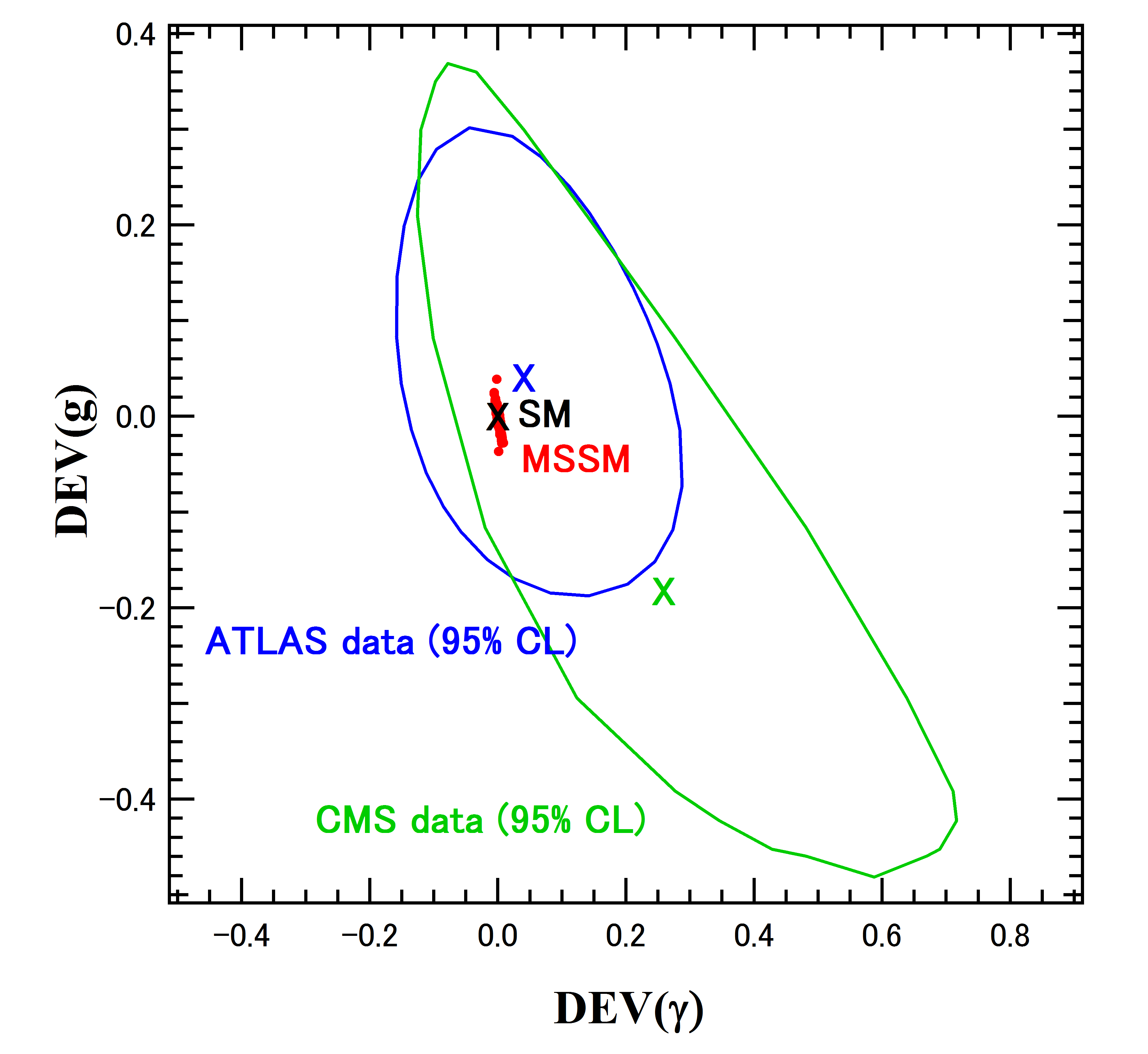}}}}
 \label{DEVgam2DEVg_LHC}}
\caption{
In (a), we show the scatter plot in the DEV($\gamma$)-DEV(g) plane obtained 
from the MSSM parameter scan described in Section \ref{sec:full scan}. 
"X" marks the SM point. 
The green, blue and brown boxes indicate the expected $1\sigma$ errors at ILC250, 
ILC250+500 and ILC250+500+1000, respectively (see Table \ref{table_DEVerror_LC}). 
Though, in principle, the expected $1\sigma$ error should be shown by an error ellipse, 
here it is shown by an error box as an approximation since such $1\sigma$ error ellipse 
in the $\kappa_\gamma$-$\kappa_g$ plane (and hence, in the DEV($\gamma$)-DEV(g) plane)
is not given in \cite{ESU2020_Rep, Snowmass2021_Rep}. 
The expected absolute $1\sigma$ errors at the other lepton colliders are similar 
to those at ILC (see Table \ref{table_DEVerror_LC}). 
We have confirmed that all the constraints in Appendix \ref{sec:constr} are 
satisfied at the MSSM parameter points corresponding to the two isolated points 
with $\DEV(g) \sim \pm4\%$ in (a). 
In (b), we show also the ATLAS and CMS data of DEV($\gamma$) and DEV(g) at 95\% CL 
obtained from the recent ATLAS and CMS data of $\kappa_\gamma$ and $\kappa_g$ 
\cite{kappa_bgamg_ATLAS, kappa_bgamg_CMS} by using the relation DEV(X) = $\kappa_X^2$ - 1. 
The black, blue and green "X" marks the SM point, and the central values of the 
ATLAS and CMS data, respectively.  
}
\label{DEVgam_DEVg}
\end{figure*}

\noindent In Fig.~\ref{DEVgam_DEVg} we show the scatter plot in the DEV($\gamma$)-DEV(g) 
plane obtained from the MSSM parameter scan. We see that there is a strong 
correlation between DEV($\gamma$) and DEV(g). This correlation is due to the fact 
that the lighter up-type squark ($\su_{1,2,3}$) loop (stop-scharm mixture loop) 
contributions dominate the two DEVs. From Fig.~\ref{DEVgam2DEVg} we see that 
DEV($\gamma$) and DEV(g) can be sizable simultaneously: DEV($\gamma$) can be as large 
as $\sim\pm 1 \%$ and DEV(g) can be as large as $\sim\pm 4 \%$. ILC together with 
HL-LHC can observe such sizable deviation DEV(g) at fairly high significance though 
they can not observe such moderate deviation DEV($\gamma$) significantly (see 
Appendix \ref{sec:error}):
The expected absolute 1$\sigma$ error of DEV($\gamma$) is $\Delta$\DEV($\gamma$) = 
(2.4\%, 2.2\%, 2.0\%) and that of DEV(g) is $\Delta$\DEV(g) = 
(1.8\%, 1.4\%, 1.1\%) at (ILC250, ILC250+500, ILC250+500+1000)  
together with HL-LHC, respectively (see Fig.~\ref{DEVgam2DEVg}). 
The expected absolute 1$\sigma$ errors of DEV($\gamma$) and DEV(g) at 
the other lepton colliders are similar to those at ILC (see Appendix \ref{sec:error}). 
It is important to note that DEV(g) can be as large as $\sim -7\%$ as 
can be seen in the contour plot of Fig.~\ref{DEVg_TU32M2U23} in the 
following, which corresponds to more than $5\sigma$ deviation from the SM 
at ILC250+500 and ILC250+500+1000. We would see this fact if we generate 
much more MSSM parameter points in our parameter scan. \\
In Fig.~\ref{DEVgam2DEVg_LHC}, we show also the ATLAS and CMS data of 
DEV($\gamma$) and DEV(g) at 95\% CL obtained from the recent ATLAS and CMS data 
of $\kappa_\gamma$ and $\kappa_g$ \cite{kappa_bgamg_ATLAS, kappa_bgamg_CMS}  
by using the relation DEV(X) = $\kappa_X^2$ - 1. 
We see that both the SM and the MSSM are allowed by the recent ATLAS 
and CMS data and that the errors of the ATLAS and CMS data are too large 
to distinguish the MSSM from the SM. \\
%

\subsection{Contour plot analysis}
\label{subsec:Contour_plot_analysis}
%
\subsubsection{Benchmark scenario}
\label{subsubsec:Benchmark scenario}
%
In order to see the relevant MSSM parameter dependence of DEV(b), DEV(c), 
DEV(b/c), $B(h^0 \to b s)$, DEV($\gamma$), DEV(g) and DEV($\gamma/g$) 
in more detail, we take a reference scenario P1 where we have sizable 
DEV(b), DEV(c), DEV(b/c), $B(h^0 \to b s)$, DEV($\gamma$), DEV(g) and 
DEV($\gamma/g$) and then variate the relevant MSSM parameters around 
this point P1. All MSSM input parameters for P1 are shown in 
Table \ref{table2}, where one has DEV(c) = -0.11, DEV(b) = -0.15, 
DEV(b/c) = -0.042, $B(h^0 \to b s)$ = 0.040\%, DEV($\gamma$) = 0.011, 
DEV(g) = -0.045 and DEV($\gamma/g$) = 0.059. 
The scenario P1 satisfies all the present experimental and theoretical 
constraints shown in Appendix \ref{sec:constr}. 
The resulting physical masses of the particles are shown in 
Table~\ref{physmasses}. The flavor decompositions of the lighter squarks 
$\su_{1,2,3}$ and $\sd_{1,2,3}$ are shown in Table~\ref{flavourdecomp}. 
For the calculation of the masses and the 
mixings of the sparticles and the MSSM Higgs bosons, 
as well as for the low-energy observables, especially those in the B and K 
meson sectors (see Table~\ref{TabConstraints}), we use the public code 
{\tt SPheno} v3.3.8~\cite{SPheno1, SPheno2}. 
As for the effective mixing angle $\alpha$ in the CP even neutral 
Higgs boson sector, we obtain $\alpha = -0.0303$ with 
$H^0 = \cos\alpha (\sqrt{2} Re(H_1^0) - v_1) + \sin\alpha (\sqrt{2} Re(H_2^0) - v_2)$ and 
$h^0 = -\sin\alpha (\sqrt{2} Re(H_1^0) - v_1) + \cos\alpha (\sqrt{2} Re(H_2^0) - v_2)$. \\
We compute the coupling modifier $\kappa_X$ by using $\DEV(X) = \kappa_X^2 - 1$ (see Eq. (\ref{DEV_Coup_kappa})).
We obtain $\kappa_b = 0.922$ ($\DEV(b) = -0.15$), $\kappa_g = 0.977$ ($\DEV(g) = -0.045$) and
$\kappa_\gamma = 1.005$ ($\DEV(\gamma) = 0.011$) in the scenario P1, which satisfy the LHC data 
in Table~\ref{TabConstraints}.
For the other coupling modifiers, and the invisible decay branching ratio $B_{inv}$ 
and the undetected decay branching ratio $B_{und}$, see Appendix~\ref{sec:kappas}.\\
Furthermore, we have confirmed that in all contour plots shown below
the red hatched regions (which include always our reference point P1) 
satisfy all the constraints in Appendix~\ref{sec:constr} and also all 
the expected sparticle mass limits 
and ($m_{A^0}$, $\tan\beta$) limit 
from the negative search for the sparticles and the MSSM Higgs bosons 
$H^0$, $A^0$, $H^\pm$ 
in the future HL-LHC experiments \cite{ESUpgrade, 
SnowmassRep, HL-LHC_ATLAS, HL-LHC_EPS-HEP2023, snu_HL-LHC, MSSM_Higgs@HL-LHC, 
MSSM_Higgs@HL-LHC_EPJC}. 
In this confirmation, we have used contour plots of all the sparticle 
masses in the individual parameter plane. \\
Here, concerning the constraint from the negative searches for the MSSM 
Higgs bosons at LHC, we remark the following point: 
In all contour plots, $m_{A^0}$ is fixed to be $\simeq 5.3$ TeV 
and hence $m_{H^0,A^0,H^+} \simeq 5.3$ TeV for which $H^0$, $A^0$, $H^+$ 
are obviously too heavy to be produced at LHC(13TeV) and HL-LHC(14TeV). 
Here note that $m_{H^0} \simeq m_{A^0} \simeq m_{H^+}$ in the decoupling 
Higgs limit in the MSSM. Hence, the entire contour plot planes are allowed 
by the negative search for the MSSM Higgs bosons $H^0$, $A^0$, $H^\pm$ 
at LHC(13TeV) and HL-LHC(14TeV). 

\begin{table}[h!]
\footnotesize{
\caption{
The MSSM parameters for the reference point P1 (in units 
of GeV or GeV$^2$ except for $\tan\beta$). All parameters are 
defined at scale Q = 1 TeV, except $m_{A^0}(pole)$. 
The parameters that are not shown here are taken to be zero.
}
\begin{center}
\begin{tabular}{|c|c|c|c|c|c|}
    \hline
\vspace*{-0.3cm}
& & & & &\\
\vspace*{-0.3cm}
     $\tan\beta$ & $M_1$ &  $M_2$ & $M_3$ &  $\mu$ &  $m_{A^0}(pole)$\\ 
& & & & &\\
    \hline
\vspace*{-0.3cm}
& & & & &\\
\vspace*{-0.3cm}
    33 & 1660 & 765 & 4615 & 870 & 5325\\
& & & & &\\
    \hline
    \hline
\vspace*{-0.3cm}
& & & & &\\
\vspace*{-0.3cm}
      $M^2_{Q 22}$ & $ M^2_{Q 33}$ &  $M^2_{Q 23}$ & $ M^2_{U 22} $ & $ M^2_{U 33} $ &  $M^2_{U 23} $\\ 
& & & & &\\
     \hline
\vspace*{-0.3cm}
& & & & &\\
\vspace*{-0.3cm}
     3975$^2$ & 3160$^2$ & 920$^2$ & 3465$^2$ & 1300$^2$ & 795$^2$\\
& & & & &\\
    \hline
    \hline
\vspace*{-0.3cm}    
& & & & &\\
\vspace*{-0.3cm}      
      $ M^2_{D 22} $ & $ M^2_{D 33}$ &  $ M^2_{D 23}$ & $T_{U 23}  $ & $T_{U 32}  $ &  $T_{U 33}$\\ 
& & & & &\\
    \hline
\vspace*{-0.3cm}      
& & & & &\\
\vspace*{-0.3cm}  
      2620$^2$ & 2425$^2$ & -1625$^2$ & -2040 & -1880 & -4945\\
& & & & &\\
 \hline 
\multicolumn{6}{c}{}\\[-3.6mm]  
\cline{1-4}
\vspace*{-0.3cm}      
     & & & \\
\vspace*{-0.3cm}      
     $ T_{D 23} $ & $T_{D 32}  $ &  $ T_{D 33}$ &$T_{E 33} $\\ 
     & & & \\
    \cline{1-4}
\vspace*{-0.3cm}      
     & & & \\
\vspace*{-0.3cm}      
     -2360 & 1670 & -2395 & -300\\
     & & & \\
    \cline{1-4}
\end{tabular}\\[3mm]
\begin{tabular}{|c|c|c|c|c|c|c|c|c|}
    \hline
\vspace*{-0.3cm}      
    & & & & & & & &\\
\vspace*{-0.3cm}      
    $M^2_{Q 11}$ & $M^2_{U 11}$ &  $M^2_{D 11}$ & $M^2_{L 11}$ & $M^2_{L 22}$ & $M^2_{L 33}$ & $M^2_{E 11}$ & $M^2_{E 22}$ & $M^2_{E 33} $\\ 
    & & & & & & & &\\
    \hline
\vspace*{-0.3cm}      
    & & & & & & & &\\
\vspace*{-0.3cm}      
    $4500^2$ & $4500^2$ & $4500^2$  & $1500^2$ & $1500^2$ & $1500^2$ & $1500^2$ & $1500^2$ & $1500^2$\\
    & & & & & & & &\\
    \hline
\end{tabular}
\end{center}
\label{table2}
}
\end{table}

\begin{table}
\caption{Physical masses in GeV of the particles for the scenario of Table~\ref{table2}.}
\begin{center}
\begin{tabular}{|c|c|c|c|c|c|}
  \hline
  $\mnt{1}$ & $\mnt{2}$ & $\mnt{3}$ & $\mnt{4}$ & $\mch{1}$ & $\mch{2}$ \\
  \hline \hline
  $781$ & $882$ & $911$ & $1669$ & $782$ & $914$ \\
  \hline
\end{tabular}
\vskip 0.4cm
\begin{tabular}{|c|c|c|c|c|}
  \hline
  $m_{h^0}$ & $m_{H^0}$ & $m_{A^0}$ & $m_{H^+}$ \\
  \hline \hline
  $124$  & $5325$ & $5325$ & $5359$ \\
  \hline
\end{tabular}
\vskip 0.4cm
\begin{tabular}{|c|c|c|c|c|c|c|}
  \hline
  $\msg$ & $\msu{1}$ & $\msu{2}$ & $\msu{3}$ & $\msu{4}$ & $\msu{5}$ & $\msu{6}$ \\
  \hline \hline
  $4424$ & $868$ & $3011$ & $3331$ & $3877$ & $4402$ & $4402$ \\
  \hline
\end{tabular}
\vskip 0.4cm
\begin{tabular}{|c|c|c|c|c|c|}
  \hline
 $\msd{1}$ & $\msd{2}$ & $\msd{3}$ & $\msd{4}$ & $\msd{5}$ & $\msd{6}$ \\
  \hline \hline
  $1705$ & $2833$ & $3010$ & $3877$ & $4397$ & $4403$ \\
  \hline
\end{tabular}
\vskip 0.4cm
\begin{tabular}{|c|c|c|c|c|c|c|c|c|}
  \hline
  $m_{\sneut_1}$ &  $m_{\sneut_2}$ &  $m_{\sneut_3}$ &  $m_{\ti l_1}$ &  $m_{\ti l_2}$ 
                        &  $m_{\ti l_3}$ &  $m_{\ti l_4}$ &  $m_{\ti l_5}$ &  $m_{\ti l_6}$ \\
  \hline \hline
   $1509$ &  $1509$ &  $1528$ &  $1489$ &  $1489$ &  $1509$ &  $1512$ &  $1512$ &  $1545$ \\
  \hline
\end{tabular}
\end{center}
\label{physmasses}
\end{table}
%

\begin{table}[h!]
\caption{Flavor decompositions of the mass eigenstates $\su_{1,2,3}$ and $\sd_{1,2,3}$ for the scenario 
         of Table~\ref{table2}. Shown are the expansion coefficients of the mass eigenstates in terms 
         of the flavor eigenstates. Imaginary parts of the coefficients are negligibly small.} 
\begin{center}
\begin{tabular}{|c|c|c|c|c|c|c|c|}
  \hline
  & $\su_L$ & $\sca_L$ & $\sto_L$ & $\su_R$ & $\sca_R$ & $\sto_R$ \\
  \hline
  $\su_1$  & $0.0001$ &  $-0.0190$ & $-0.0959$ & $0$ & $0.0767$ & $-0.9922$ \\
  \hline 
  $\su_2$  &  $-0.0283$ &  $-0.0736$ & $0.9740$ & $0$ & $0.1978$ & $-0.0775$ \\
  \hline 
  $\su_3$  & $0.0088$ & $0.0288$ & $-0.1886$ & $0$ & $0.9771$ & $0.0932$ \\
  \hline  \hline
  & $\sd_L$ & $\ss_L$ & $\sbo_L$ & $\sd_R$ & $\ss_R$ & $\sbo_R$ \\
  \hline
  $\sd_1$  & $0$ & $-0.0012$ &  $0.0099$ & $0$ & $0.6585$ & $0.7525$ \\
  \hline 
  $\sd_2$  & $0$ &  $-0.0058$ & $0.0448$ & $0$ & $-0.7521$ & $0.6575$ \\
  \hline 
  $\sd_3$  & $-0.0018$ & $-0.1183$ & $0.9919$ & $0$ & $0.0274$ & $-0.0373$ \\
  \hline
\end{tabular}
\end{center}
\label{flavourdecomp}
\end{table}

\subsubsection{Contour plots for fermionic decays}
\label{subsubsec:Cont_plots_fermionic}
In Fig. \ref{fig_DEVb} we show contours of $\DEV(b)$ around 
the benchmark point P1 in various parameter planes. 
%
Fig.~\ref{DEVb_TU23TU32} shows contours of $\DEV(b)$ in the $T_{U23}$-$T_{U32}$ plane. 
We see that $|\DEV(b)|$ increases quickly with the increase of $|T_{U23}|$ and 
$|T_{U32}|$ as expected. We also see that 
it is large ($-0.18 \lsim \DEV(b) \lsim -0.13$) respecting all the constraints 
in a significant part of this parameter plane. 
From Fig.~\ref{DEVb_TU23TU33}, we see that $\DEV(b)$ is also sensitive to $T_{U33}$
and can be as large as $\sim -0.18$. 
As can be seen in Fig.~\ref{DEVb_TU32M2U23}, $\DEV(b)$ is fairly sensitive to $M^2_{U 23}$, 
especially for large $|T_{U32}| \gsim 3$ TeV, as expected, and is large 
($-0.18 \lsim \DEV(b) \lsim -0.15$) respecting all the constraints in a 
significant part of this parameter plane. \\
As for the contours of $\DEV(b)$ around P1 in the planes of down-type squark 
parameters $T_{D23}$, $T_{D32}$, $T_{D33}$ and $M^2_{D23}$, we have found 
that $\DEV(b)$ is rather insensitive to these parameters as expected: 
the weaker dependence of DEV(b) on the down-type squark parameters such as $T_{D23}$ 
than that on the up-type squark parameters such as $T_{U23}$ stems from the fact 
that the $\sd_{1,2,3}$ - $\sg$ loops (Fig.~\ref{h02bb_gluino_loop}) are less 
important than the $\su_{1,2,3}$ - $\ch_{1,2}$ loops (Fig.~\ref{h02bb_chargino_loop}) 
mainly due to the small $Re(H_1^0)$ component of $h^0$ in our decoupling Higgs scenario, 
which suppresses the $h^0-\sd_i-\sd_j$ couplings. \\
As shown in Table \ref{table_DEVerror_LC}, the expected absolute 1$\sigma$ error 
of $\DEV(b)$ measured at ILC is given by $\Delta \DEV(b)$ = (1.7\%, 1.1\%, 0.9\%) 
at (ILC250, ILC250+500, ILC250+500+1000) and similar results are obtained for the 
future lepton colliders other than ILC. 
Therefore, such large deviation $\DEV(b)$ ($\sim$ -15\% to -20\%) in the sizable 
region allowed by all the constraints (including the expected sparticle 
and MSSM Higgs boson mass limits from HL-LHC) as shown in Fig.~\ref{fig_DEVb} can be 
observed with very high significance at the future lepton colliders such as ILC. \\

Just to show how the allowed parameter region is affected by the change 
of the total theoretical error of $m_{h^0}$, we have drawn $m_{h^0}$ bound lines 
in the contour plots of DEVs and $B(h^0 \to b s)$ in the hypothetical case that 
the total theoretical error is $\pm 2$ GeV instead of $\pm 3$ GeV. As can be seen in 
Figs.~\ref{fig_DEVb}~-~\ref{fig_DEVc} and Figs.~\ref{fig_BRbs}~-~\ref{fig_DEVgam2g}, 
the allowed parameter region is only a little affected by the change of the total 
theoretical error. \\

In Fig.~\ref{fig_DEVc} we show contour plots of $\DEV(c)$ 
around the benchmark point P1 in various parameter planes. 
Fig.~\ref{DEVc_TU23TU32} shows contours of $\DEV(c)$ in the 
$T_{U23}$-$T_{U32}$ plane. We see that $\DEV(c)$ is sensitive to $T_{U23}$ and $T_{U32}$: 
$|\DEV(c)|$ quickly increases with the increase of $T_{U23}$ and $T_{U32}$ as expected 
(see Fig.~\ref{h02cc_gluino_loop} and the related argument above). 
We find also that $\DEV(c)$ is sizable ($-0.15 \lsim \DEV(c) \lsim -0.05$) 
respecting all the constraints in a significant part of this parameter plane. 
From Fig.~\ref{DEVc_TU23TU33}, we see that $\DEV(c)$ is sensitive also 
to $T_{U33}$, quickly increases with the increase of $|T_{U33}|$ as expected, 
and it is large ($-0.20 \lsim \DEV(c) \lsim -0.05$) respecting all the constraints 
in a sizable part of this parameter plane.
As can be seen in Fig.~\ref{DEVc_TU32M2U23}, $\DEV(c)$ is sensitive to 
$T_{U32}$ and $M^2_{U 23}$ increasing with the increase of $T_{U32}$ and 
$M^2_{U 23}$ as expected and is quite large ($-0.25 \lsim \DEV(c) \lsim 0.1$) 
respecting all the constraints in a significant part of this parameter plane. 
Note that $\DEV(c)$ is especially large for the large product 
$T_{U32} \cdot M^2_{U 23} (< 0)$. 
This is due to the following reason:\\
(i) The gluino loop contribution (Fig.~\ref{h02cc_gluino_loop}) to the 
decay amplitude for $h^0 \to c \, \bar c$ can be very large positive or 
negative for a large value of the product $T_{U32} \cdot M^2_{U23}$ since the 
leading gluino loop contribution is proportional to the product 
$T_{U32} \cdot M^2_{U23}$ in terms of the mass-insertion (MI) approximation 
(see Fig.~\ref{h02cc_gluino_loop_MI}) \cite{Eberl:h2bb}. \\
(ii) Hence, the interference term between the tree diagram and the gluino loop 
contribution of Fig.~\ref{h02cc_gluino_loop} can be very large 
(positive or negative) for a large value of the product $T_{U32} \cdot M^2_{U23}$, which 
can lead to even {\it negative} width $\Gamma(h^0 \to c \, \bar c)$ at one-loop level. 
In this case the perturbation theory breaks down.\\
(iii) Therefore, in principle, the deviation of $\Gamma(h^0 \to c \, \bar c)$ 
from the SM value ($\DEV(c)$) can be very large for a large value of the product 
$T_{U32} \cdot M^2_{U23}$. \\
\indent As for the contours of $\DEV(c)$ around P1 in the planes of down-type squark 
parameters $T_{D23}$, $T_{D32}$, $T_{D33}$ and $M^2_{D23}$, we have found 
that $\DEV(c)$ is insensitive to these parameters as expected:
the main MSSM 1-loop corrections to $\Gamma(h^0 \to c \bar c)$ stem from 
$\su_{1,2,3}$ - $\sg$ loops (see Fig.~\ref{h02cc_gluino_loop}), 
which do not involve the down-type squark parameters. \\
As shown in Table \ref{table_DEVerror_LC}, the expected absolute 1$\sigma$ error 
of $\DEV(c)$ measured at ILC is given by $\Delta \DEV(c)$ = (3.6\%, 2.4\%, 1.8\%) 
at (ILC250, ILC250+500, ILC250+500+1000) and similar results are obtained for the 
future lepton colliders other than ILC. 
Hence, such large deviation $\DEV(c)$ ($\sim$ +10\% to -25\%) in the sizable 
region allowed by all the constraints (including the expected sparticle mass limits 
from HL-LHC) as shown in Fig.~\ref{fig_DEVc} can be observed with very high 
significance at the future lepton colliders such as ILC. \\

\renewcommand{\len}{5.3cm}
\begin{figure*}[h!]
\centering
\renewcommand{\subfiglabelskip}{-2mm} 
\subfigure[]{
   {\mbox{\hspace*{-0.5cm} \resizebox{\len}{!}{\includegraphics{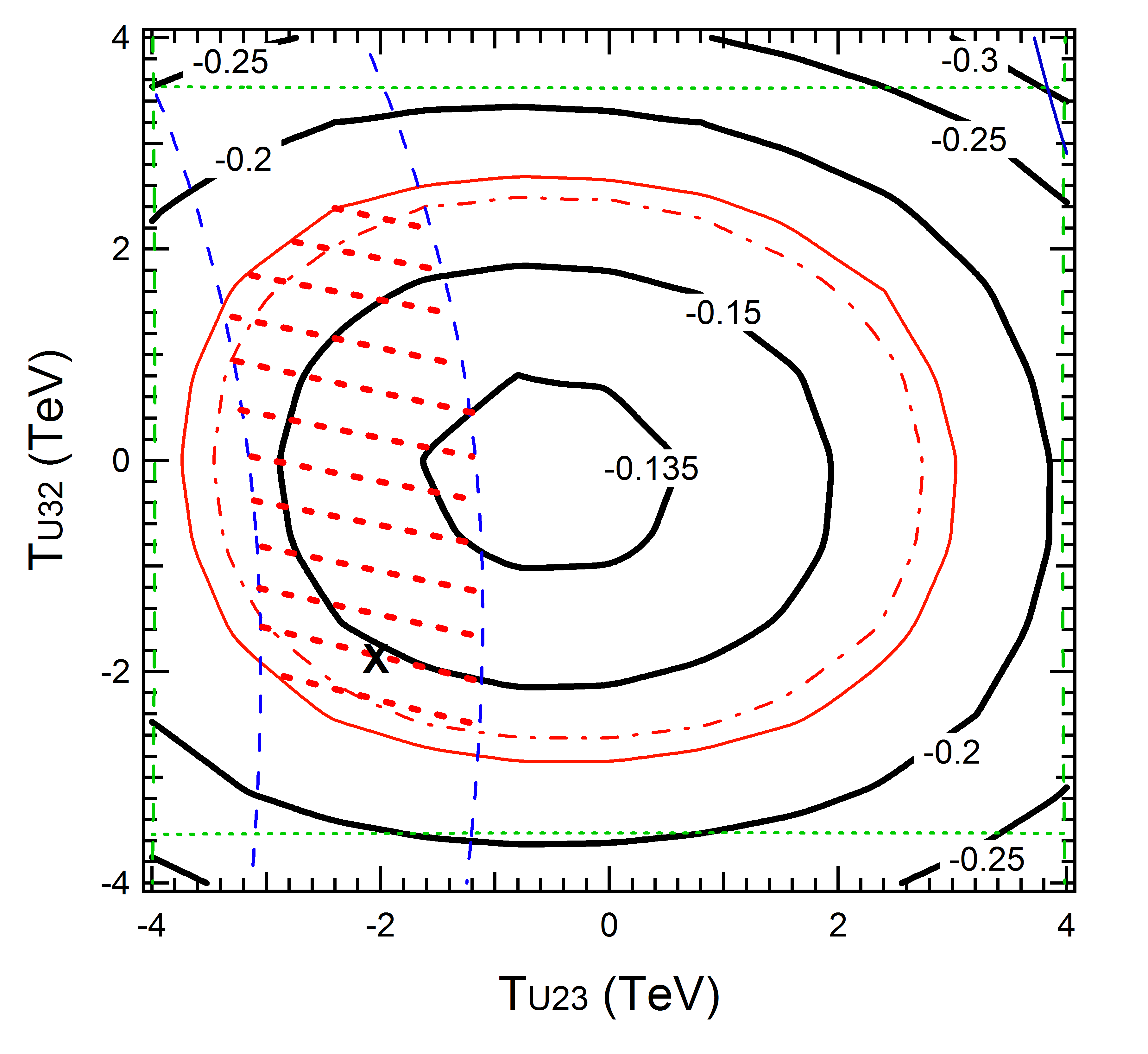}} \hspace*{-0.2cm}}}
   \label{DEVb_TU23TU32}} 
\renewcommand{\subfiglabelskip}{-7mm} 
\subfigure[]{
   {\mbox{\hspace*{-0.5cm} \resizebox{\len}{!}{\includegraphics{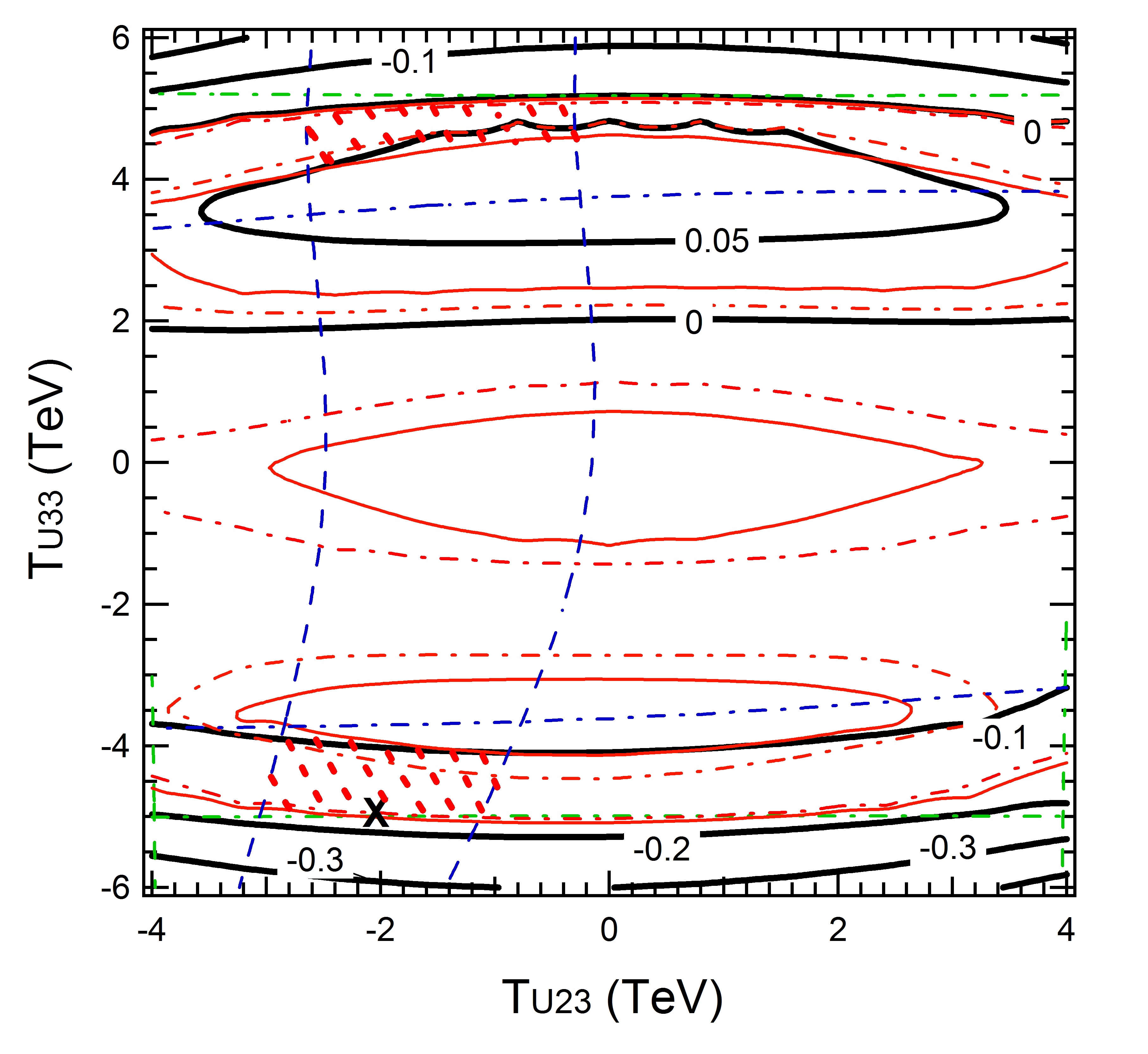}} \hspace*{-0.7cm}}}
   \label{DEVb_TU23TU33}}
\renewcommand{\subfiglabelskip}{-11mm}
\subfigure[]{
   {\mbox{\hspace*{0cm} \resizebox{\len}{!}{\includegraphics{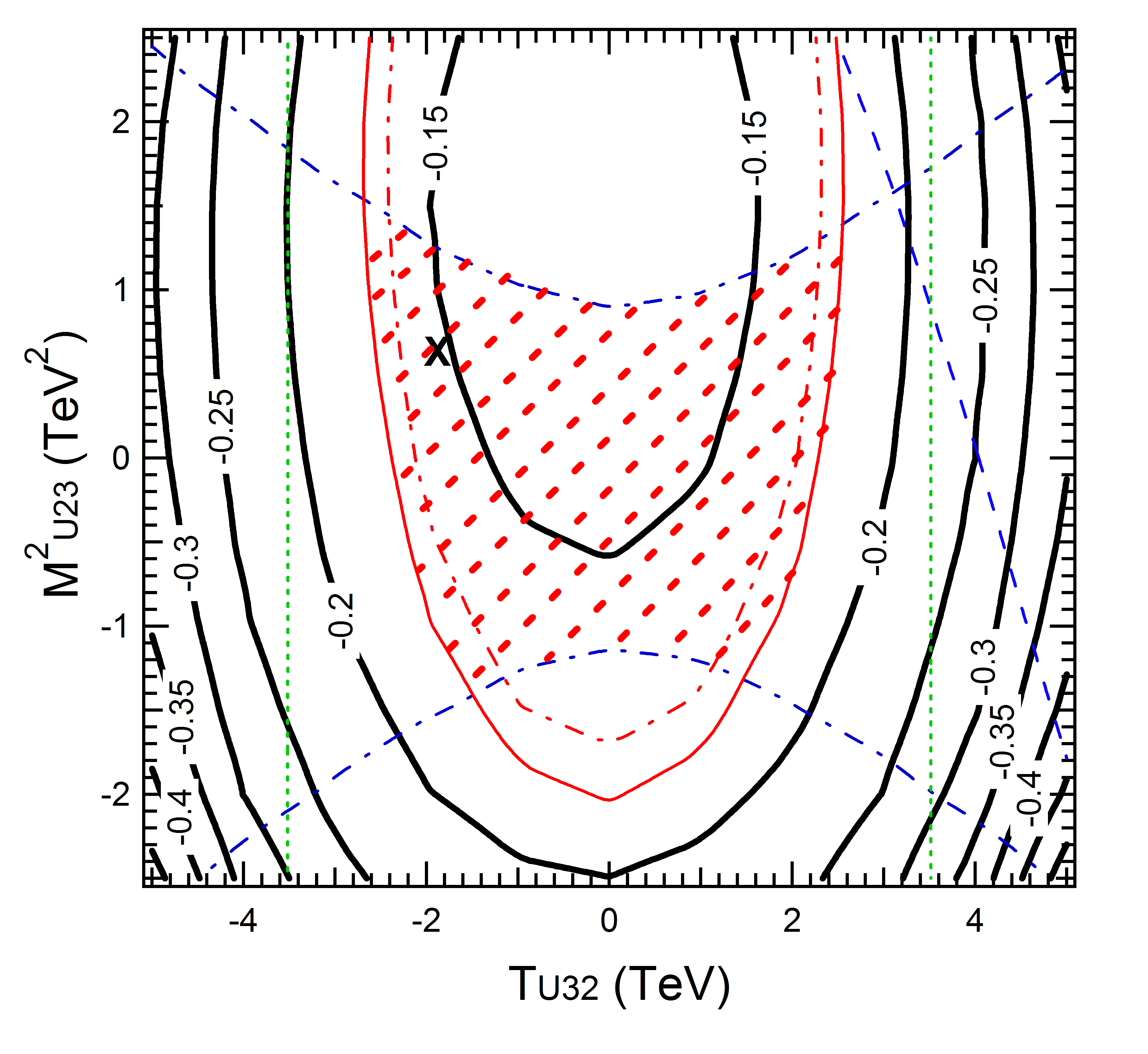}} \hspace*{-0.6cm}}}
  \label{DEVb_TU32M2U23}}
\caption{
     Contour plots of $\DEV(b)$ around the benchmark point P1 in the parameter planes of
     (a) $T_{U 23}$ - $T_{U 32}$, (b) $T_{U 23}$ - $T_{U 33}$, and (c) $T_{U 32}$ - $M^2_{U 23}$. 
     The parameters other than the shown ones in each plane are fixed as in Table~\ref{table2}.
     The "X" marks P1 in the plots. 
     The red hatched region satisfies all the constraints in Appendix \ref{sec:constr} 
     and also all the expected sparticle mass limits and ($m_{A^0}$, $\tan\beta$) limit 
     at 95\% CL from negative search for sparticles and the heavier MSSM Higgs bosons $H^0$, 
     $A^0$, $H^\pm$ in the future HL-LHC experiments \cite{ESUpgrade, SnowmassRep, 
     HL-LHC_ATLAS, HL-LHC_EPS-HEP2023, snu_HL-LHC, MSSM_Higgs@HL-LHC, MSSM_Higgs@HL-LHC_EPJC}. 
     The red solid lines, 
     green dashed lines, green dotted lines, green dash-dotted lines, 
     blue solid lines, blue dashed lines, and blue dash-dotted lines 
     show the $m_{h^0}$ bound, vacuum stability bound for $T_{U23}$, $T_{U32}$, $T_{U33}$, 
     $B(b \to s \gamma)$ bound, $B(B_s \to \mu^+ \mu^-)$ bound, 
     and $m_{\su_1}$ bound, respectively. 
     Just for reference, the red dash-dotted lines show the $m_{h^0}$ bounds for a hypothetical 
     case that the total theoretical error is $\pm 2$ GeV instead of $\pm 3$ GeV; i.e. the contours of 
     $m_{h^0}$ (GeV) = 125.09 + (0.48 + 2) = 127.57 and $m_{h^0}$ (GeV) = 125.09 - (0.48 + 2) = 122.61. 
     We see that only a small part of the red hatched region is excluded by this hypothetical 
     $m_{h^0}$ bounds.
     }
\label{fig_DEVb}
\end{figure*}    

\renewcommand{\len}{5.3cm}
\begin{figure*}[h!]
\centering
\renewcommand{\subfiglabelskip}{-2mm} 
 \subfigure[]{
   {\mbox{\hspace*{-0.7cm} \resizebox{\len}{!}{\includegraphics{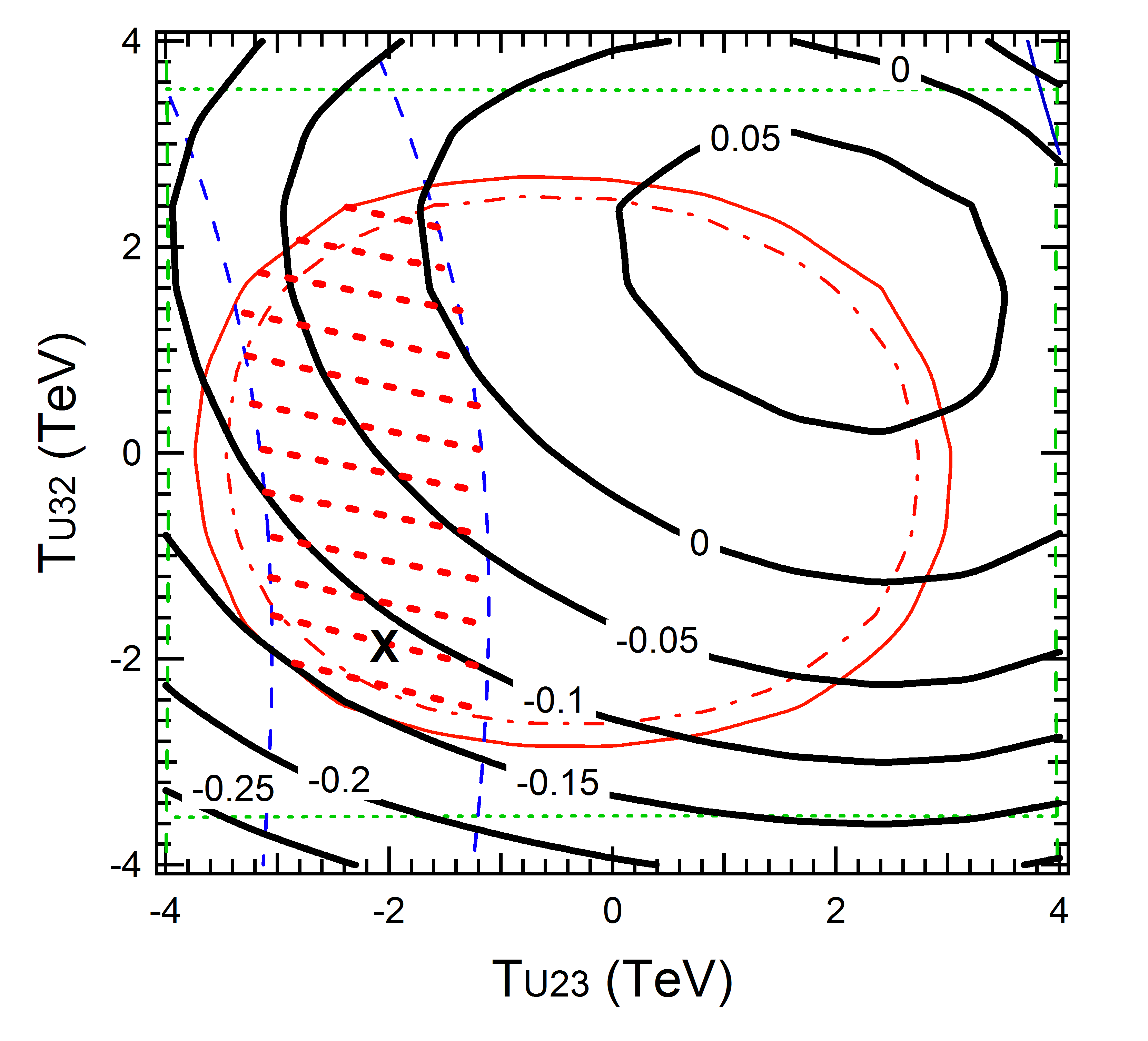}} \hspace*{-0.3cm}}}
   \label{DEVc_TU23TU32}} 
\renewcommand{\subfiglabelskip}{-7mm}
 \subfigure[]{
   {\mbox{\hspace*{-0.5cm} \resizebox{\len}{!}{\includegraphics{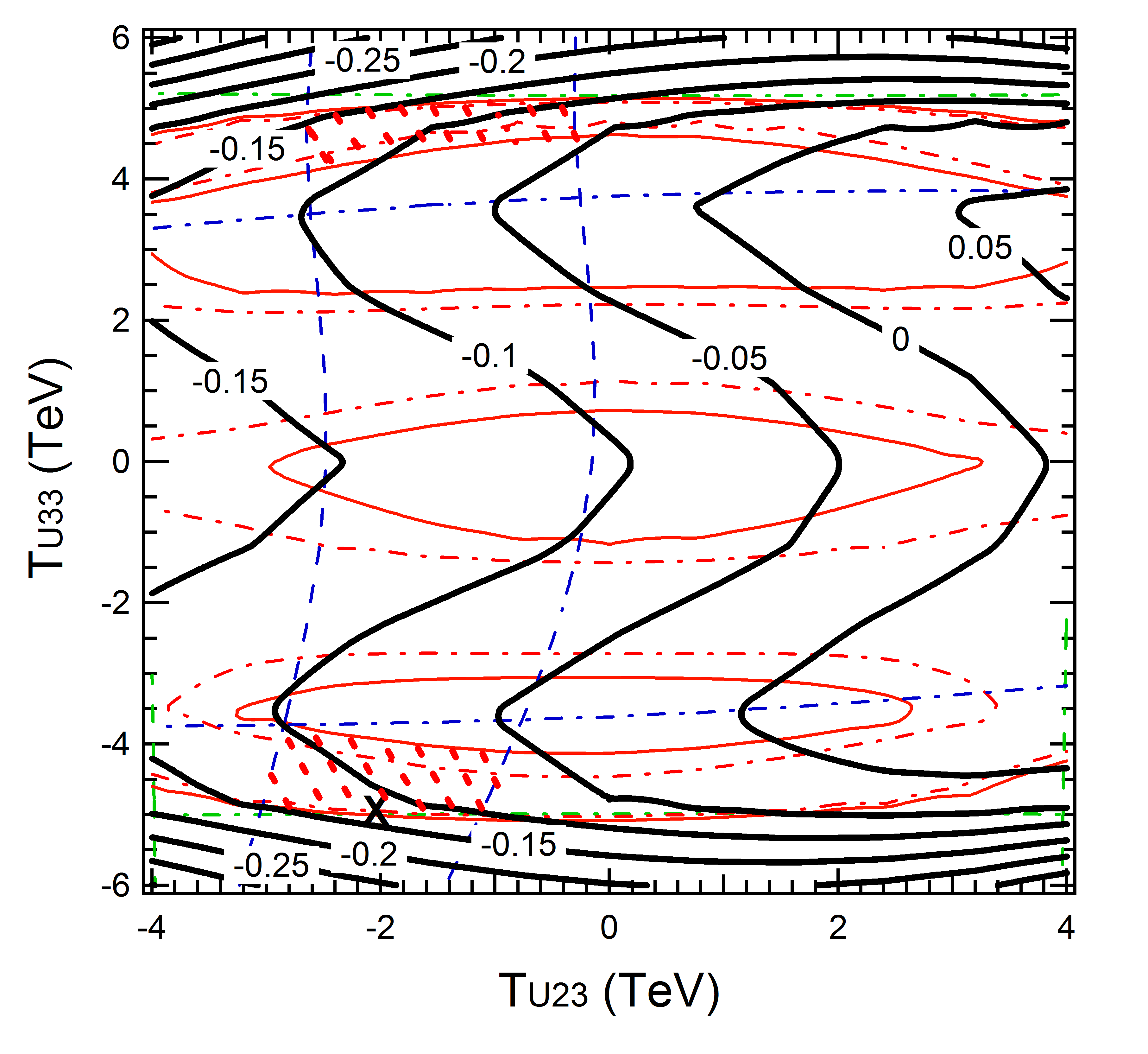}} \hspace*{-0.7cm}}}
   \label{DEVc_TU23TU33}}
\renewcommand{\subfiglabelskip}{-13mm}
 \subfigure[]{
   {\mbox{\hspace*{0cm} \resizebox{\len}{!}{\includegraphics{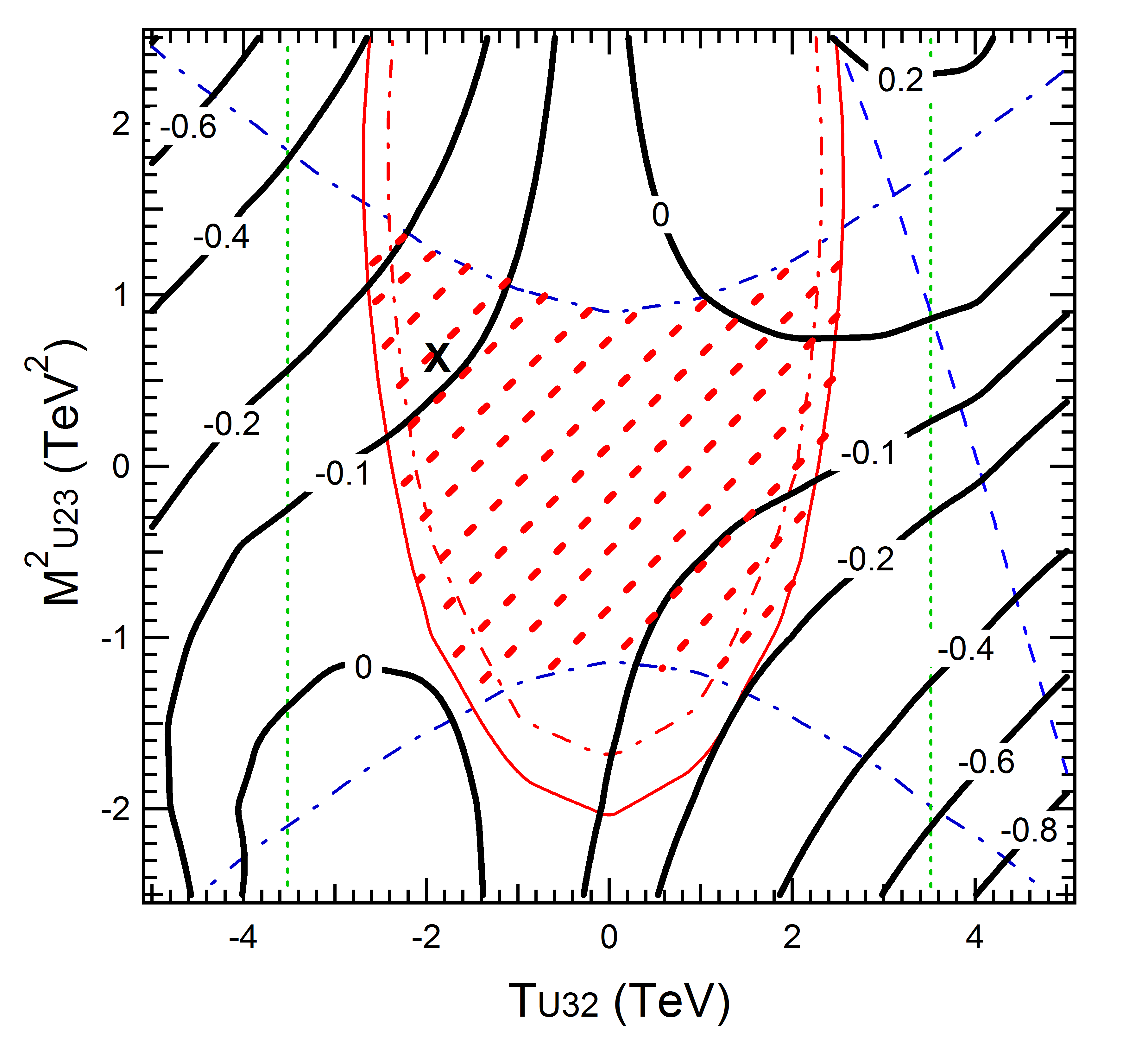}} \hspace*{-0.8cm}}}
   \label{DEVc_TU32M2U23}}
\caption{
     Contour plots of $\DEV(c)$ around the benchmark point P1 in the parameter planes of
     (a) $T_{U 23}$ - $T_{U 32}$, (b) $T_{U 23}$ - $T_{U 33}$ and (c) $T_{U 32}$ - $M^2_{U 23}$.
     The parameters other than the shown ones in each plane are fixed as in Table~\ref{table2}.
     The "X" marks P1 in the plots. 
     The definitions of the red hatched regions and the bound lines are 
     the same as those in Fig.~\ref{fig_DEVb}.
     }
\label{fig_DEVc}
\end{figure*} 

\begin{figure*}[t!]
\centering
  {\mbox{\resizebox{8.0cm}{!}{\includegraphics{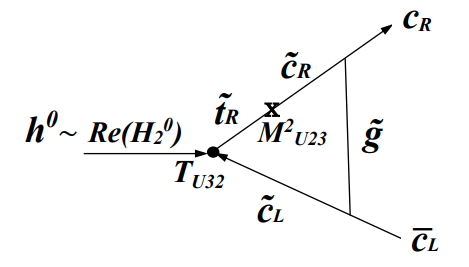}}}}
\caption{
The leading gluino loop contribution to the decay amplitude for $h^0 \to c \, \bar c$ 
in terms of the MI approximation.
}
\label{h02cc_gluino_loop_MI}
\end{figure*}

In Fig. \ref{fig_BRbs} we show contours of $B(h^0 \to b s)$ 
around the benchmark point P1 in various parameter planes. 
Fig.~\ref{BRbs_TU23TU32} shows contours of $B(h^0 \to b s)$ in the $T_{U23}$-$T_{U32}$ plane. 
We see that $B(h^0 \to b s)$ is sensitive to both $T_{U23}$ and $T_{U32}$, increases 
with the increase of $|T_{U23}|$ and $|T_{U32}|$, as expected 
(see Fig.~\ref{h02bb_chargino_loop}), and can be as large as about 0.00045 
in the allowed region. 
From Fig.~\ref{BRbs_TU23TU33}, we see that $B(h^0 \to b s)$ is also sensitive 
to $T_{U33}$, increases with the increase of $|T_{U33}|$, as expected, and 
can be as large as about 0.0006 in the allowed region. 
From Fig.~\ref{BRbs_TU32tanb}, we find that $B(h^0 \to b s)$ is also sensitive to 
$tan\b$ as expected and can be as large as about 0.0015 in the allowed region.
We also see that it is sizable ($0.00025 \lsim B(h^0 \to b s) \lsim 0.0015$) 
respecting all the constraints in a significant part of this parameter plane.
From Fig.~\ref{BRbs_TD23TD32} we see that $B(h^0 \to b s)$ is sensitive to 
both $T_{D23}$ and $T_{D32}$ (being the $\ss_R$ - $\sbo_L$ and $\ss_L$ - $\sbo_R$ 
mixing parameter, respectively), increases with the increase of $|T_{D23}|$ and 
$|T_{D32}|$ (with $T_{D23} \cdot T_{D32} < 0$), as expected, and can reach 
0.001 in the allowed region.

The 4$\sigma$ signal significance sensitivity to $B(h^0 \to b s)$ of 
ILC250+500+1000 is $B(h^0 \to b s)$ = 0.001 as mentioned above. 
Hence, such large $B(h^0 \to b s)$ ($\sim$ 0.001 to 0.0015) 
in the region allowed by all the constraints (including the expected 
sparticle mass limits from HL-LHC) as shown in Figs. \ref{fig_BRbs} 
can be observed at ILC with high significance.\\

\renewcommand{\len}{5.3cm}
\begin{figure*}[h!]
\centering
 \subfigure[]{
   {\mbox{\hspace*{-0.6cm} \resizebox{\len}{!}{\includegraphics{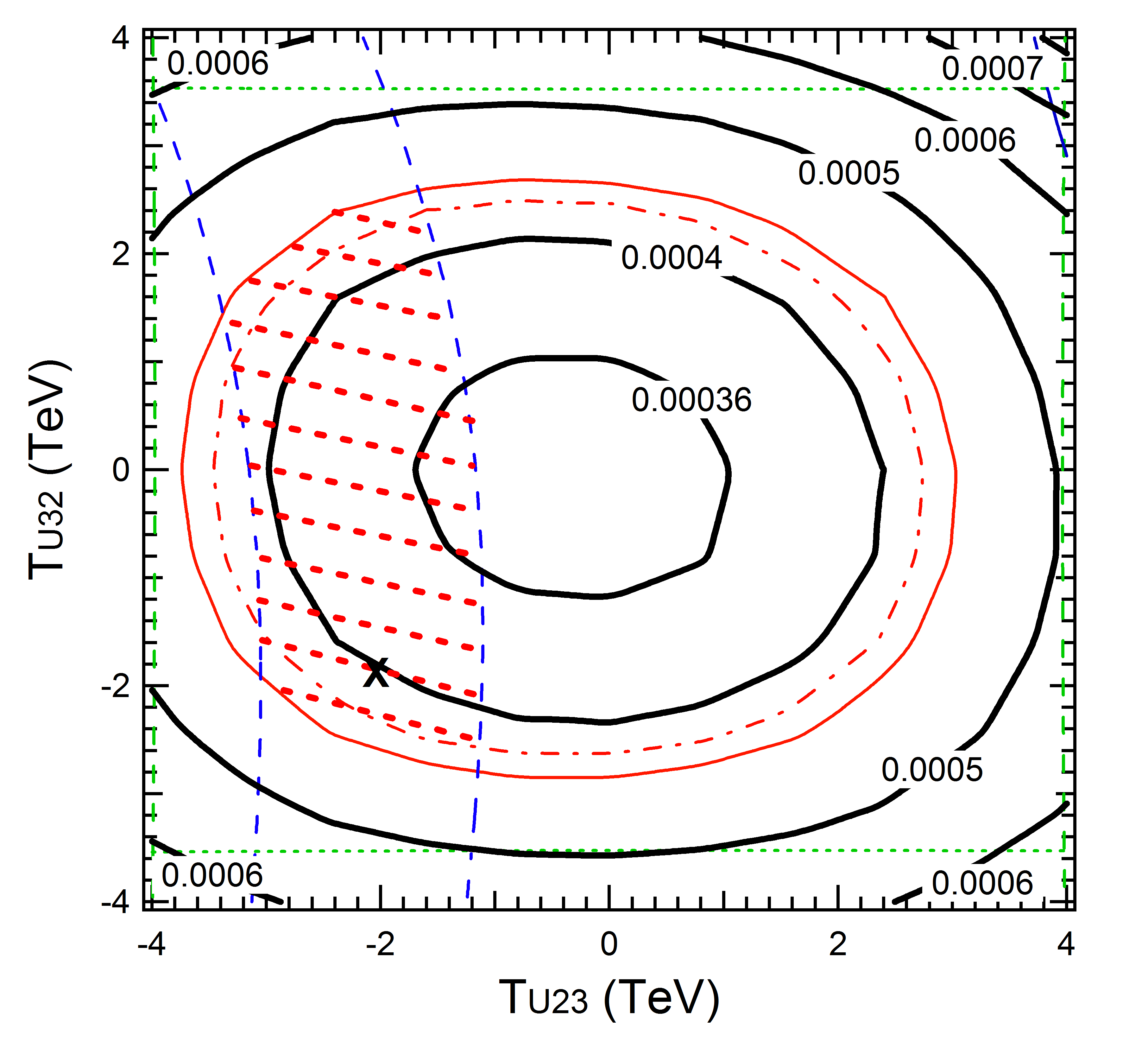}} \hspace*{0cm}}}
   \label{BRbs_TU23TU32}} 
 \subfigure[]{
   {\mbox{\hspace*{-0.6cm} \resizebox{\len}{!}{\includegraphics{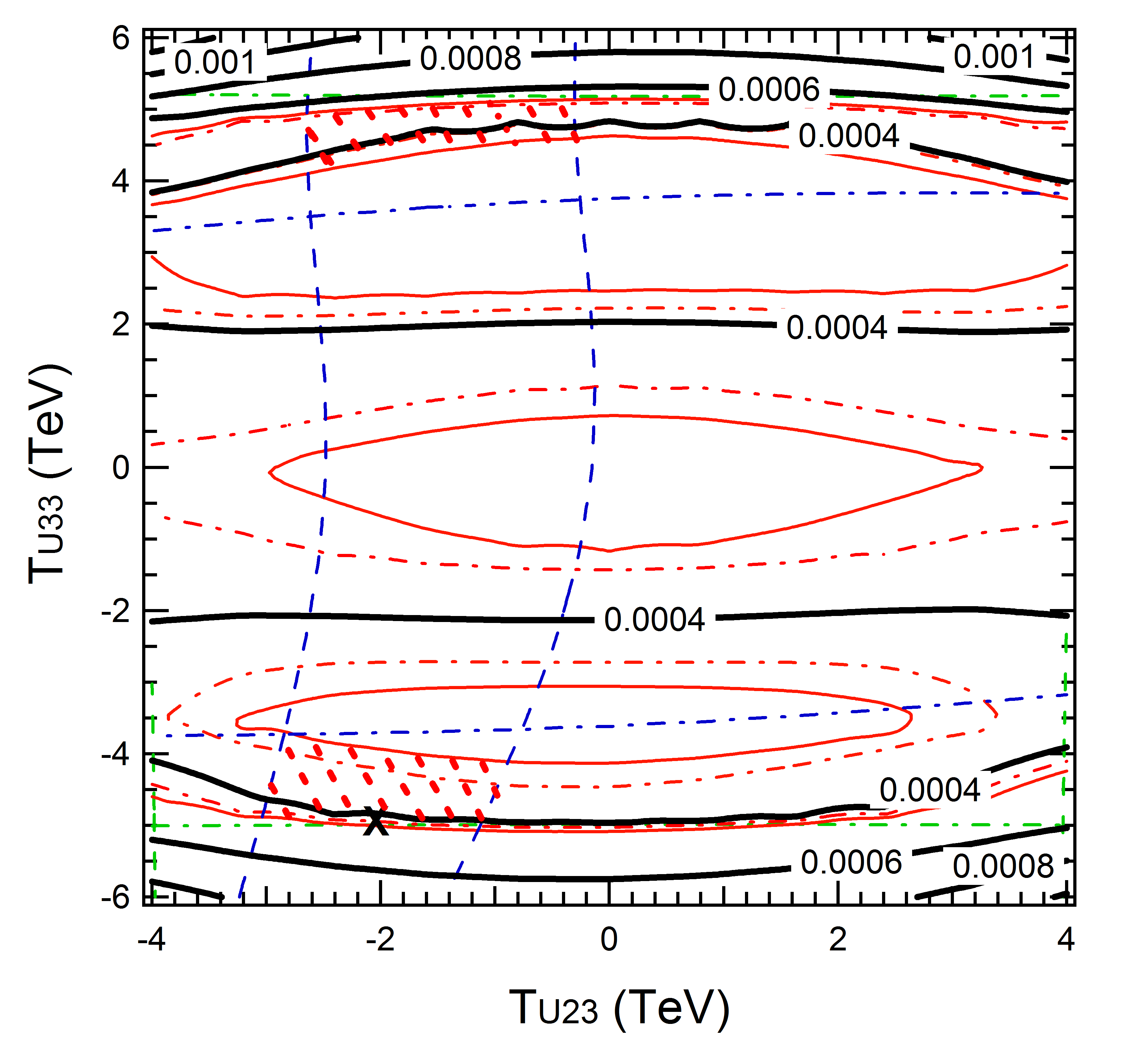}} \hspace*{0cm}}}
   \label{BRbs_TU23TU33}}\\
 \subfigure[]{
   {\mbox{\hspace*{-0.6cm} \resizebox{\len}{!}{\includegraphics{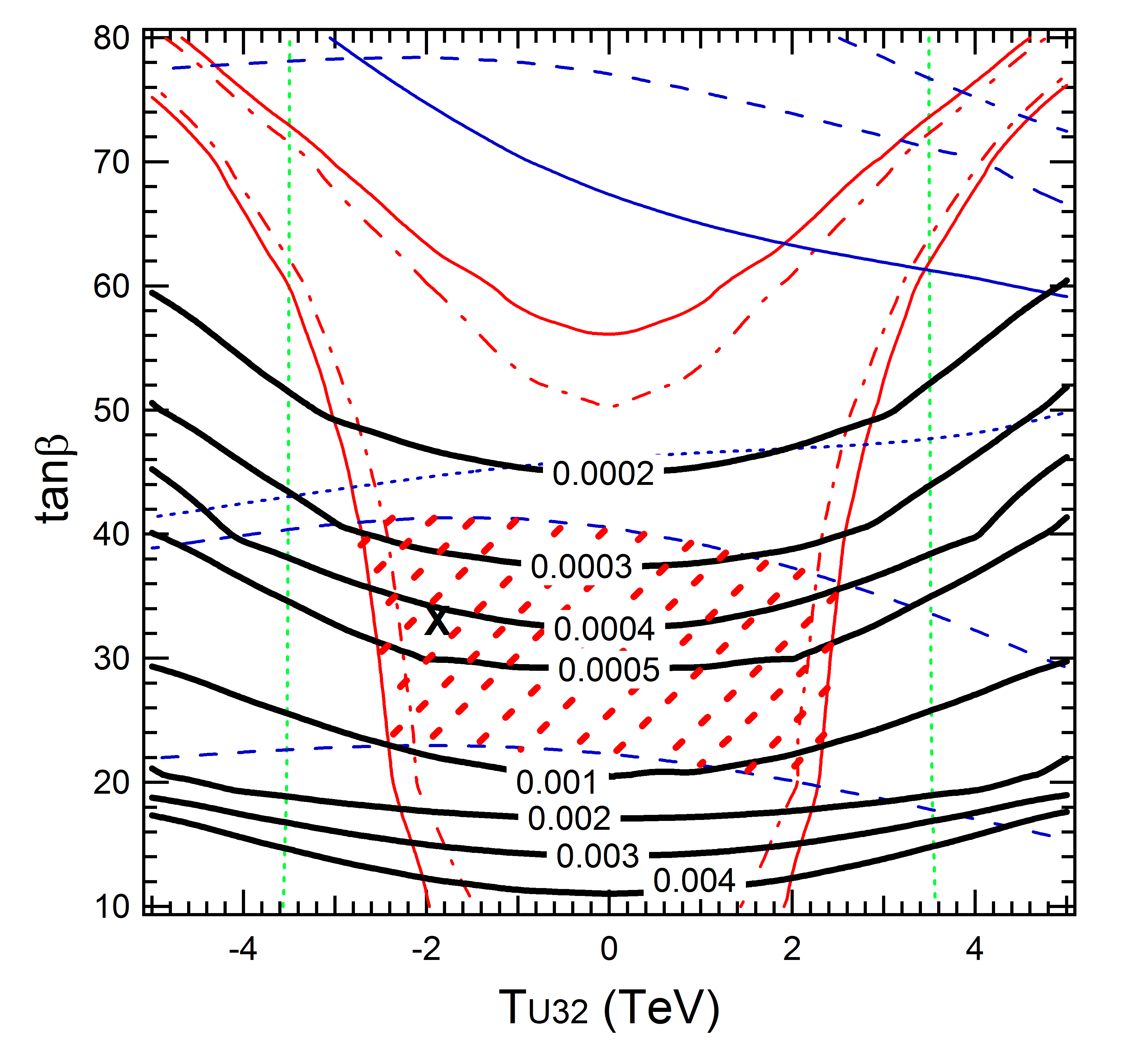}} \hspace*{0cm}}}
  \label{BRbs_TU32tanb}}
 \subfigure[]{
   {\mbox{\hspace*{-0.6cm} \resizebox{\len}{!}{\includegraphics{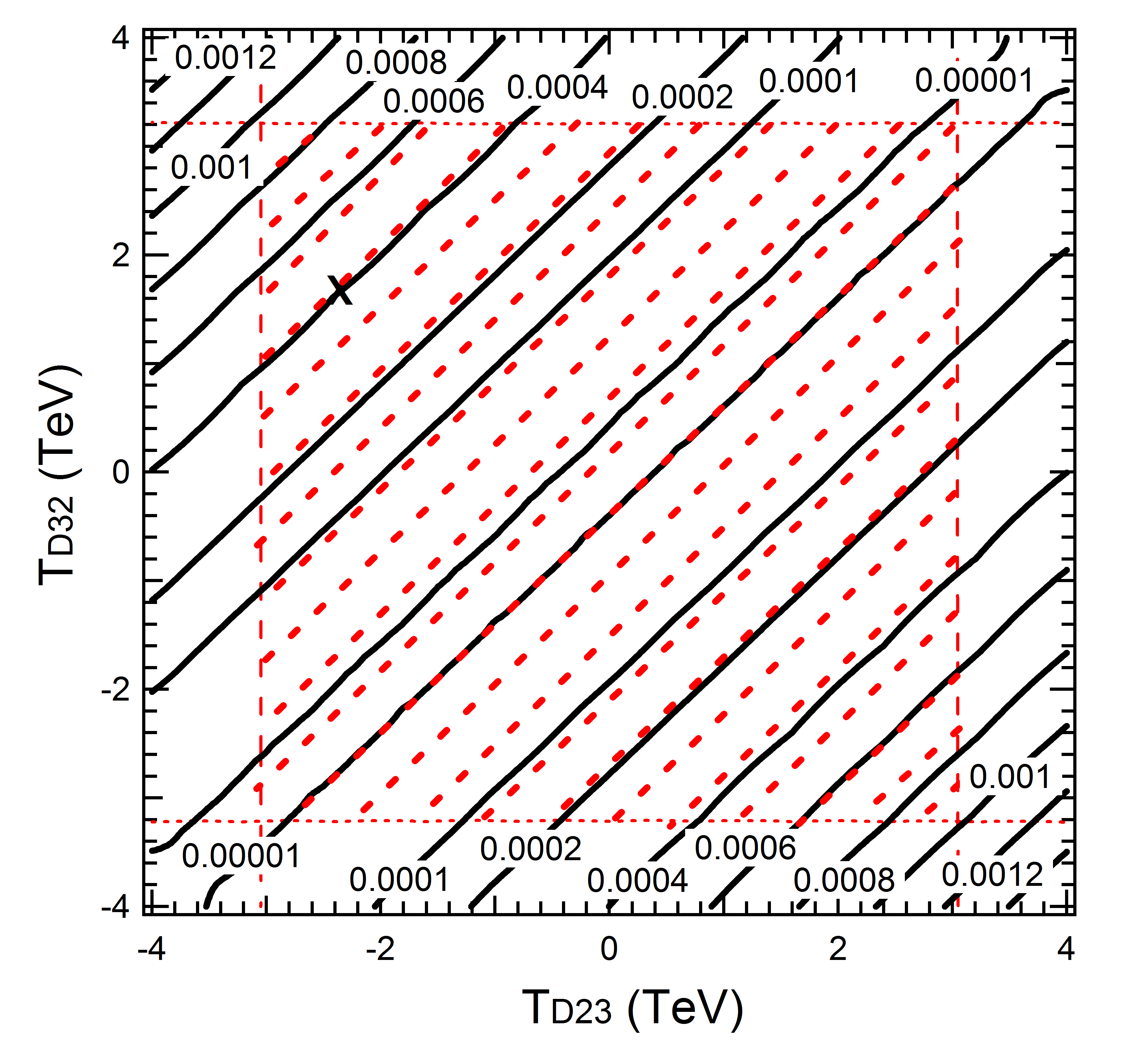}} \hspace*{0cm}}}
  \label{BRbs_TD23TD32}}
\caption{
     Contour plots of $B(h^0 \to b s)$ around the benchmark point P1 in the parameter planes of
     (a) $T_{U 23}$ - $T_{U 32}$, (b) $T_{U 23}$ - $T_{U 33}$, (c) $T_{U 32}$ - $tan\beta$,
     and (d) $T_{D 23}$ - $T_{D 32}$. 
     The parameters other than the shown ones in each plane are fixed as in Table~\ref{table2}.
     The "X" marks P1 in the plots. 
     Note that 4$\sigma$ signal significance sensitivity to $B(h^0 \to b s)$ of ILC250+500+1000 
     is $B(h^0 \to b s) = 0.001$. 
     The definitions of the red hatched regions and the bound lines are 
     the same as those in Fig.~\ref{fig_DEVb}. 
     In addition to these the blue dotted lines, the red dashed lines and red dotted lines show 
     the $\Delta M_{B_s}$ bound, and the vacuum stability bound for $T_{D23}$ and $T_{D32}$, respectively. 
     In Figs. 10(a)-10(c), we see that only a small part of the red hatched region is excluded 
     by the hypothetical $m_{h^0}$ bounds mentioned in Fig. \ref{fig_DEVb}. 
     The entire region of Fig. 10(d) is allowed by the hypothetical $m_{h^0}$ bounds. 
     }
\label{fig_BRbs}
\end{figure*}

\subsubsection{Contour plots for bosonic decays}
\label{subsubsec:Cont_plots_bosonic}
In Fig.~\ref{fig_DEVg} we show contour plots of $\DEV(g)$ 
around the benchmark point P1 in various parameter planes. 
Fig.~\ref{DEVg_TU23TU32} shows contours of $\DEV(g)$ in the $T_{U23}$-$T_{U32}$ plane. 
We see that $\DEV(g)$ is fairly sensitive to $T_{U23}$ and $T_{U32}$ as expected 
(see Fig.~\ref{h02gg_loops} and the related argument above).  
It is sizable ($-0.05 \lsim \DEV(g) \lsim -0.044$) in the allowed region. 
From Fig.~\ref{DEVg_TU23TU33}, we see that $\DEV(g)$ is also sensitive  
to $T_{U33}$ increasing quickly with the increase of $|T_{U33}|$ as expected. 
As can be seen in Fig.~\ref{DEVg_TU32M2U23}, $\DEV(g)$ is sensitive to 
$M^2_{U 23}$ increasing with the increase of $|M^2_{U 23}|$ as expected 
from decreasing $m_{\su_1}$ for increasing $|M^2_{U 23}|$, and it is sizable 
($-0.07 \lsim \DEV(g) \lsim -0.045$) in the allowed region. \\
As for the contours of $\DEV(g)$ around P1 in the planes of the down-type squark 
parameters $T_{D23}$, $T_{D32}$, $T_{D33}$ and $M^2_{D23}$, we have found 
that $\DEV(g)$ is insensitive to these parameters as expected. \\

As shown in Table \ref{table_DEVerror_LC}, the expected absolute 1$\sigma$ error 
of $\DEV(g)$ measured at ILC is given by $\Delta \DEV(g)$ = (1.8\%, 1.4\%, 1.1\%)
at (ILC250, ILC250+500, ILC250+500+1000), and similar results are obtained for the 
future lepton colliders other than ILC. 
Hence, such large deviation $\DEV(g)$ ($\sim$ -4\% to -7\%) in the sizable 
region allowed by all the constraints (including the expected sparticle mass 
limits from HL-LHC) as shown in Fig.~\ref{fig_DEVg} can be observed with high 
significance at the future lepton colliders such as ILC. \\

\renewcommand{\len}{5.3cm}
\begin{figure*}[h!]
%
\renewcommand{\subfiglabelskip}{-3mm} 
 \subfigure[]{
   {\mbox{\hspace*{-0.4cm} \resizebox{\len}{!}{\includegraphics{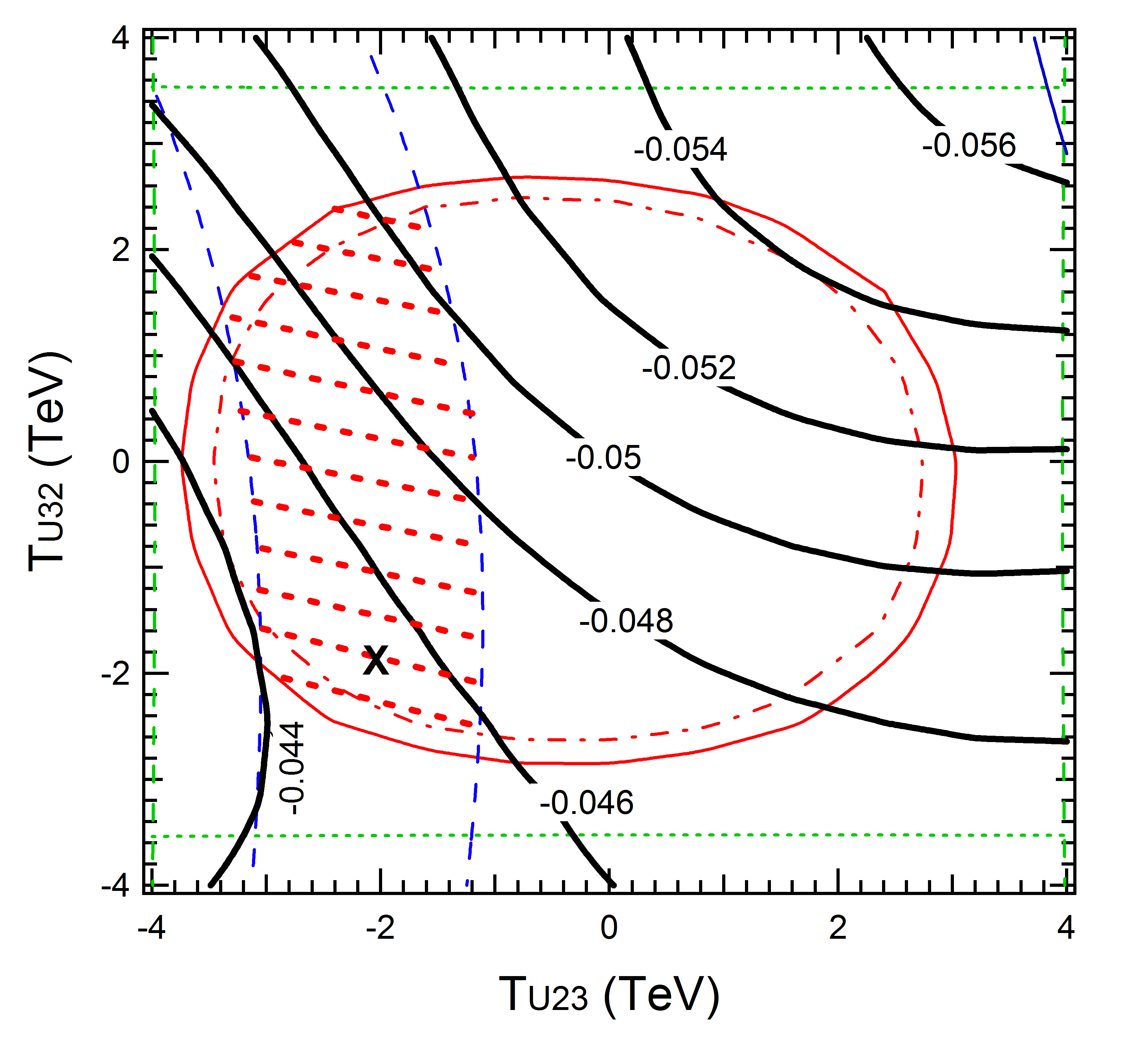}} \hspace*{-0.2cm}}}
   \label{DEVg_TU23TU32}} 
\renewcommand{\subfiglabelskip}{-7mm}
 \subfigure[]{
   {\mbox{\hspace*{-0.6cm} \resizebox{\len}{!}{\includegraphics{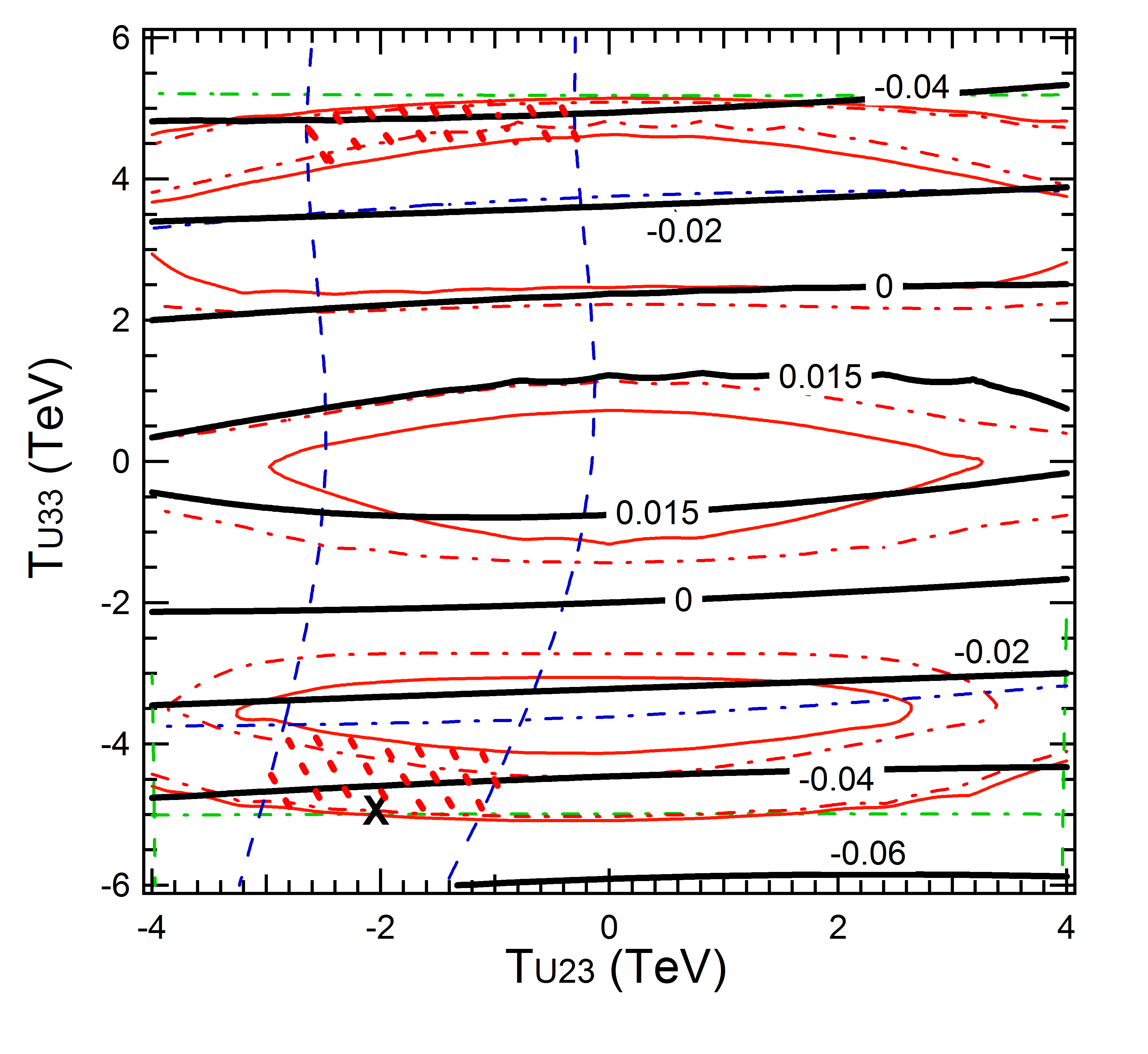}} \hspace*{-0.7cm}}}
   \label{DEVg_TU23TU33}}
\renewcommand{\subfiglabelskip}{-13mm}
 \subfigure[]{
   {\mbox{\hspace*{0cm} \resizebox{\len}{!}{\includegraphics{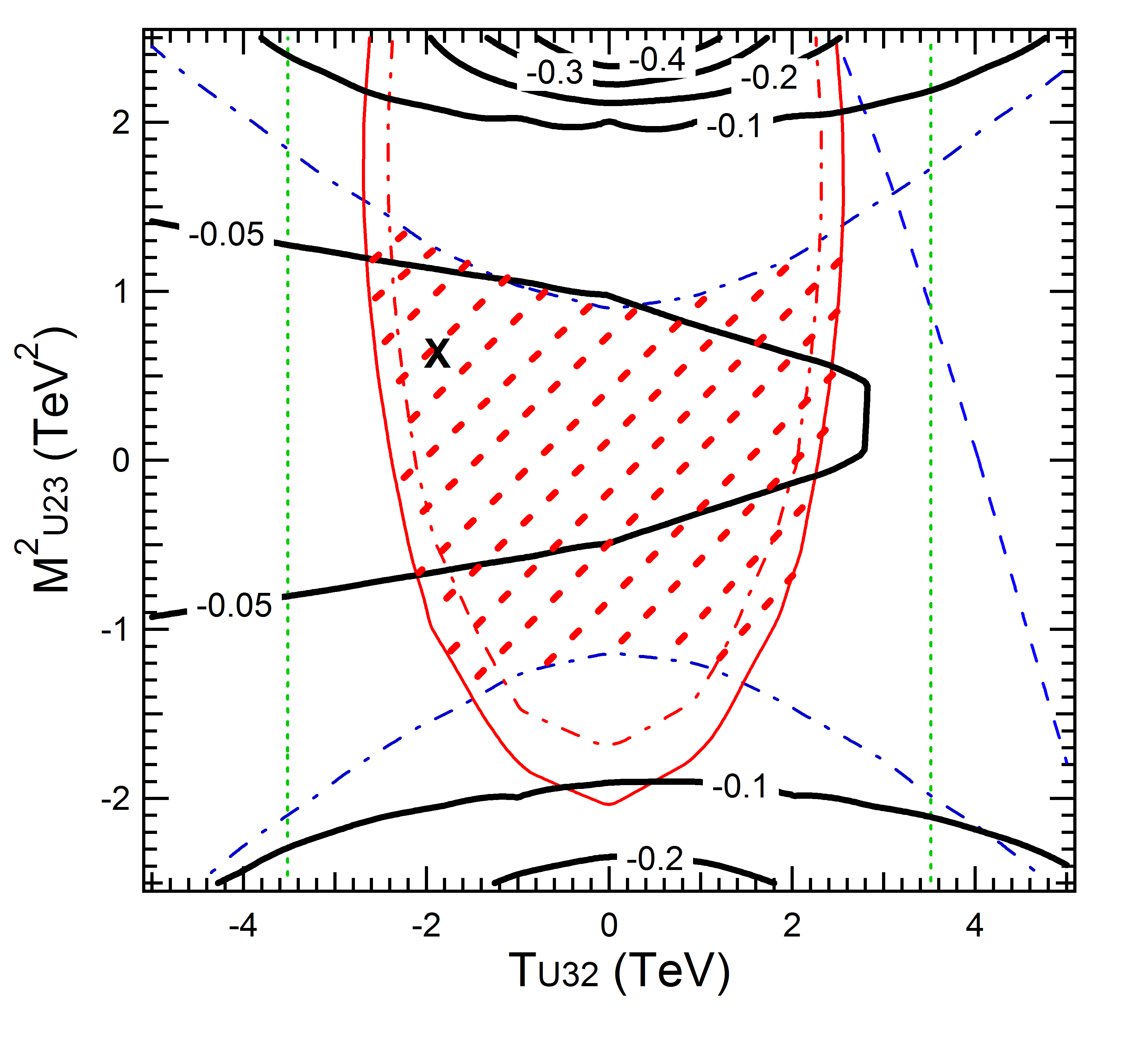}} \hspace*{-0.7cm}}}
   \label{DEVg_TU32M2U23}}  
\caption{
     Contour plots of $\DEV(g)$ around the benchmark point P1 in the parameter planes of
     (a) $T_{U 23}$ - $T_{U 32}$, (b) $T_{U 23}$ - $T_{U 33}$ and (c) $T_{U 32}$ - $M^2_{U 23}$.
     The parameters other than the shown ones in each plane are fixed as in Table~\ref{table2}.
     The "X" marks P1 in the plots. 
     The definitions of the red hatched regions and the bound lines are 
     the same as those in Fig.~\ref{fig_DEVb}.
     }
\label{fig_DEVg}
\end{figure*}

As for contour plots of $\DEV(\gamma)$ around the benchmark point P1, 
we have found that the tendency of them is similar to that of $\DEV(g)$, 
except the overall normalization and sign. 
This is due to the fact that, as can be seen in Fig. \ref{DEVgam2DEVg}, 
for any MSSM parameter point, approximately we have (see \cite{h2gagagg} also)
\be
\DEV(\gamma) \sim - \frac{1}{4}\DEV(g). 
  \label{DEVgamDEVg_Corr}
\ee

%

\begin{figure*}[!htb]
\renewcommand{\len}{5.2cm}
%
\renewcommand{\subfiglabelskip}{-4mm} 
 \subfigure[]{
   {\mbox{\hspace*{-0.4cm} \resizebox{\len}{!}{\includegraphics{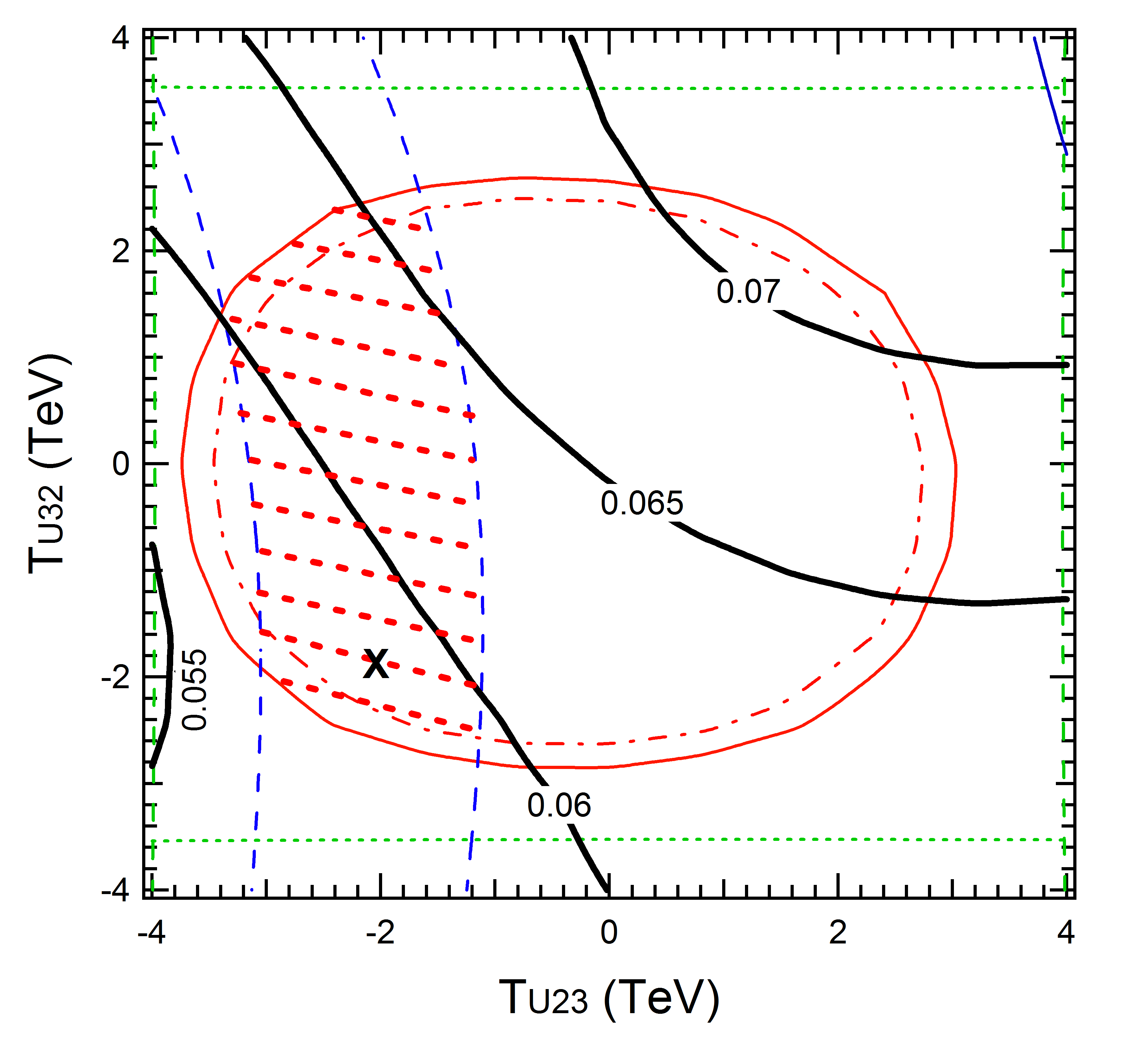}} \hspace*{-0.3cm}}}
   \label{DEVgam2g_TU23TU32}} 
\renewcommand{\subfiglabelskip}{-8mm}
 \subfigure[]{
   {\mbox{\hspace*{-0.45cm} \resizebox{\len}{!}{\includegraphics{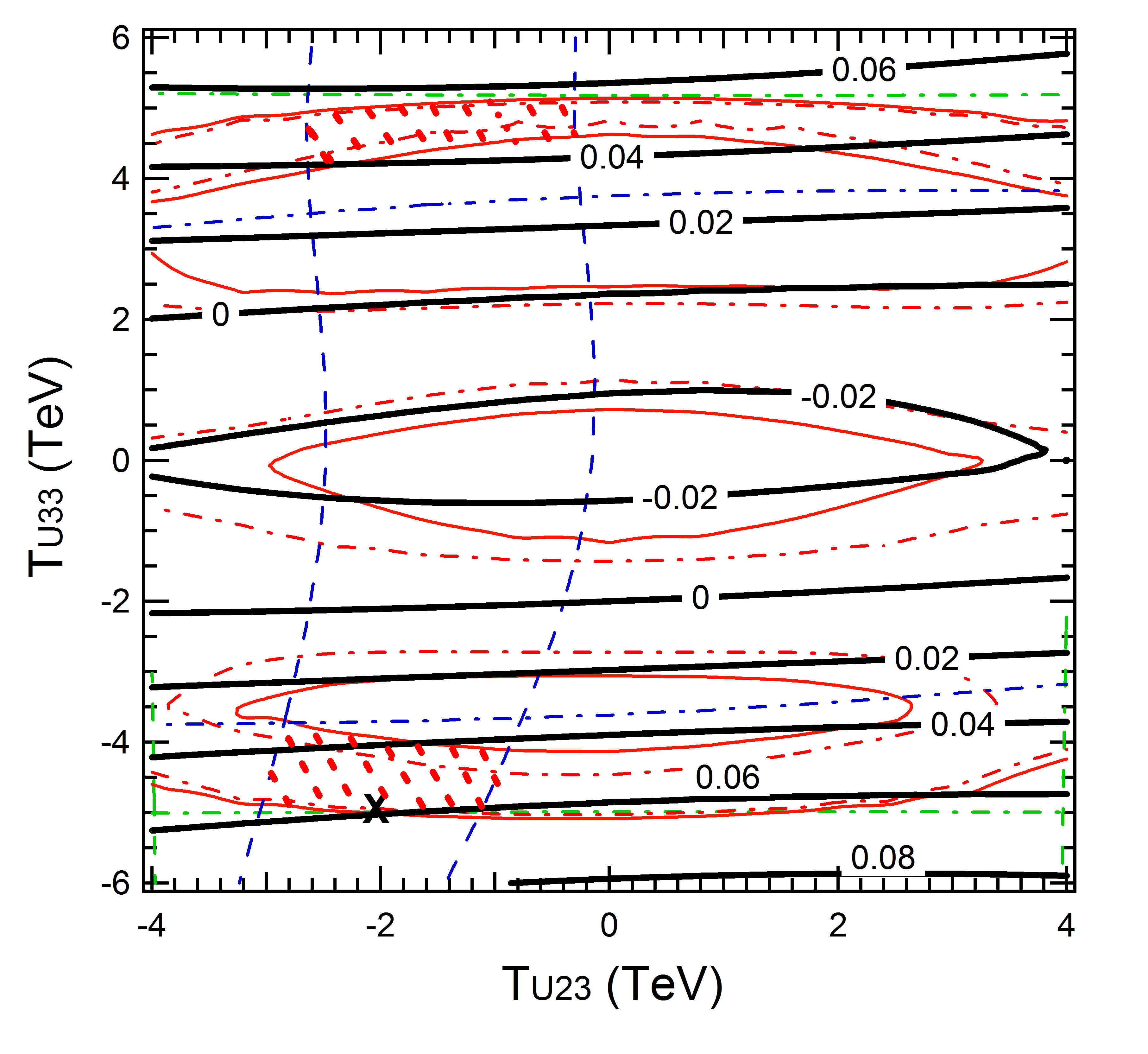}} \hspace*{-0.6cm}}}
   \label{DEVgam2g_TU23TU33}}
\renewcommand{\subfiglabelskip}{-14mm}
 \subfigure[]{
   {\mbox{\hspace*{-0.1cm} \resizebox{\len}{!}{\includegraphics{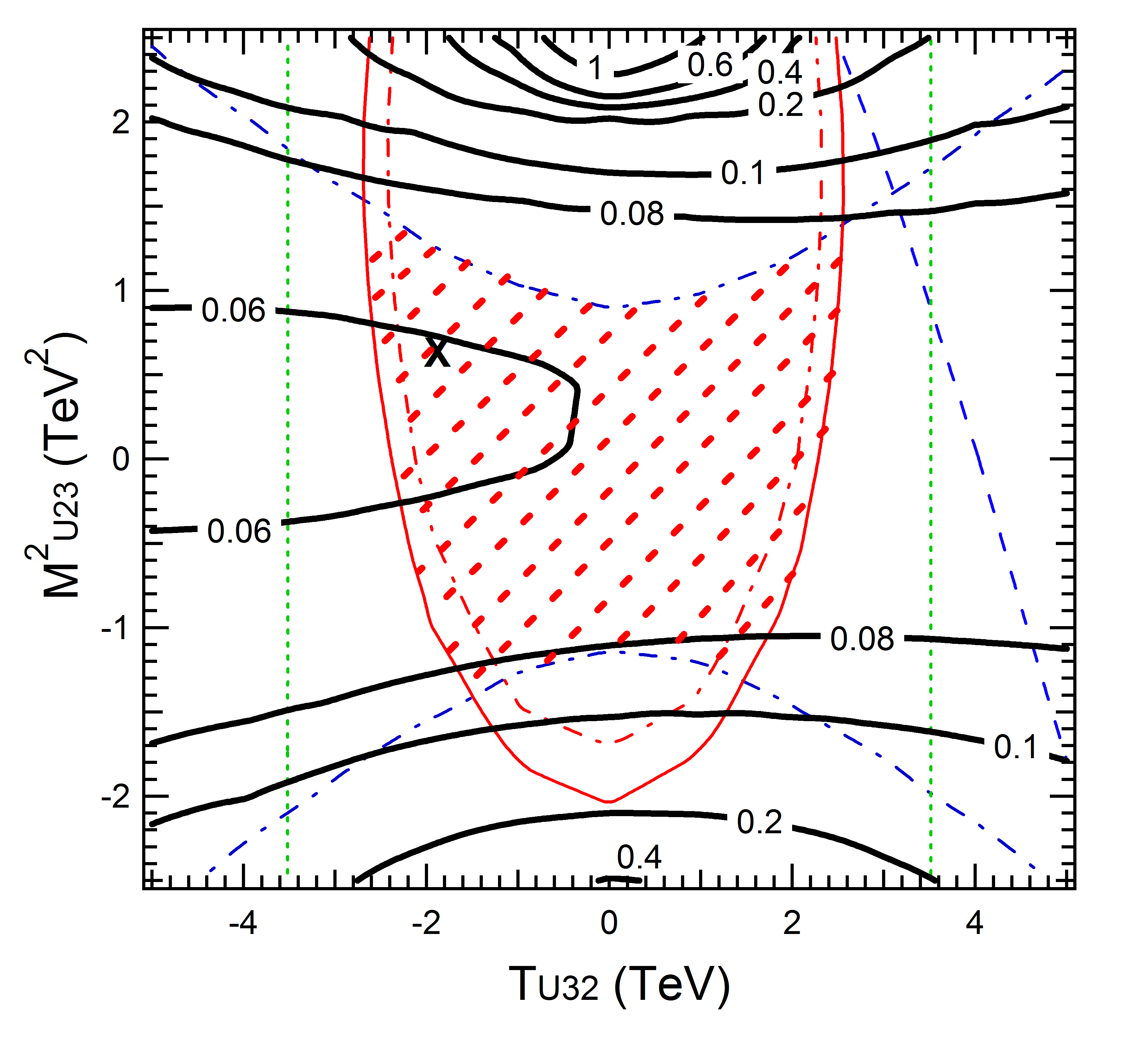}} \hspace*{-1cm}}}
   \label{DEVgam2g_TU32M2U23}}
\caption{
     Contour plots of $\DEV(\gamma/g)$ around the benchmark point P1 in the parameter planes of
     (a) $T_{U 23}$ - $T_{U 32}$, (b) $T_{U 23}$ - $T_{U 33}$ and (c) $T_{U 32}$ - $M^2_{U 23}$.  
     The parameters other than the shown ones in each plane are fixed as in Table~\ref{table2}.
     The "X" marks P1 in the plots. 
     The definitions of the red hatched regions and the bound lines are 
     the same as those in Fig.~\ref{fig_DEVb}.
     } 
\label{fig_DEVgam2g}
\end{figure*}

In Fig.~\ref{fig_DEVgam2g}, we show contour plots of $\DEV(\gamma/g)$ 
around the benchmark point P1 in various parameter planes.  
Fig.~\ref{DEVgam2g_TU23TU32} shows contours of $\DEV(\gamma/g)$ in the 
$T_{U23}$-$T_{U32}$ plane. We see that $\DEV(\gamma/g)$ is fairly sensitive to 
$T_{U23}$ and $T_{U32}$ as expected (see Fig.~\ref{1-loop_diag_to_h02gagagg} 
and the related argument above). It is sizable ($0.06 \lsim \DEV(\gamma/g) 
\lsim 0.07$) in the allowed region. 
From Fig.~\ref{DEVgam2g_TU23TU33}, we see that $\DEV(\gamma/g)$ is also sensitive 
to $T_{U33}$, quickly increases with increase of $|T_{U33}|$ as expected. 
As can be seen in Fig.~\ref{DEVgam2g_TU32M2U23}, $\DEV(\gamma/g)$ is also 
sensitive to $M^2_{U 23}$ increasing with the increase of $|M^2_{U 23}|$ 
as expected from decreasing $m_{\su_1}$ for increasing $|M^2_{U 23}|$, 
and it is large ($0.06 \lsim \DEV(\gamma/g) \lsim 0.09$) in the allowed region. 
As for the contours of $\DEV(\gamma/g)$ around P1 in the planes of down-type squark 
parameters $T_{D23}$, $T_{D32}$, $T_{D33}$ and $M^2_{D23}$, we have found 
that $\DEV(\gamma/g)$ is insensitive to these parameters as expected. \\
As shown in Table \ref{table_DEVerror_LC}, the expected absolute 1$\sigma$ error 
of $\DEV(\gamma/g)$ measured at ILC is given by $\Delta \DEV(\gamma/g)$ = 
(3.3\%, 2.8\%, 2.3\%) at (ILC250, ILC250+500, ILC250+500+1000) and similar 
results are obtained for the future lepton colliders other than ILC. 
Therefore, such large deviation $\DEV(\gamma/g)$ ($\sim$ +6\% to +9\%) in the sizable 
region allowed by all the constraints (including the expected sparticle mass limits 
from HL-LHC) as shown in Fig.~\ref{fig_DEVgam2g} can be observed with fairly high 
significance at the future lepton colliders such as ILC. \\

%

\subsection{Theoretical errors of the MSSM prediction for DEVs}
\label{subsec:Theoretical errors}
%
Before closing this section we comment briefly on the theoretical error of 
the MSSM prediction $\DEV(X)_{MSSM}$. 
We find that, in general, the theoretical error of the MSSM prediction $\DEV(X)_{MSSM}$ 
tends to be significantly small compared with the expected experimental error of the 
DEV(X) measured at the future lepton colliders \cite{Bartl:h2cc, h2gagagg}. 
The theoretical error of the MSSM prediction $\DEV(X)_{MSSM}$ comes from two sources. 
One is the parametric uncertainty, and the other is the renormalization scale-dependence 
uncertainty. The former is due to the errors of the SM input parameters such as 
$\alpha_s^{\overline{MS}}(m_Z)$ and the latter is due to unknown higher order corrections. 
The main reason of this tendency is as follows:
the theoretical errors (the parametric errors and the scale-dependence errors) of 
$\Gamma(X)_{MSSM}$ and $\Gamma(X)_{SM}$ significantly cancel out in the ratio 
$\Gamma(X)_{MSSM}/\Gamma(X)_{SM}$ in the MSSM prediction 
$\DEV(X)_{MSSM} = \Gamma(X)_{MSSM}/\Gamma(X)_{SM} - 1$, 
resulting in a significantly small theoretical error of $\DEV(X)_{MSSM}$. 
Here, note that we compute the SM widths $\Gamma(X)_{SM}$ by taking 
the decoupling SUSY/Higgs limit of the MSSM width $\Gamma(X)_{MSSM}$. \\
For example, as for the MSSM prediction $\DEV(g)_{MSSM}$ at the benchmark point P1, 
by using the theoretical error estimation method described in \cite{h2gagagg}, we find that 
the parametric absolute 1$\sigma$ error is 0.073\% 
and the scale-dependence absolute error is 0.080\% .
Hence, we find the total theoretical absolute 1$\sigma$ error of the MSSM prediction of DEV(g) at P1 
is 0.153\% (= 0.073\% + 0.080\%) by adding the parametric absolute 1$\sigma$ error to the 
scale-dependence absolute error linearly.
On the other hand, the expected experimental absolute 1$\sigma$ error of DEV(g) is 
1.8\%, 1.4\%, 1.1\% at ILC250, ILC250+500, ILC250+500+1000 together with HL-LHC, 
respectively (see Table \ref{table_DEVerror_LC}). 
Therefore, we find that indeed the theoretical error of the MSSM prediction of DEV(g) 
is significantly small compared with the expected experimental error of DEV(g) 
measured at the future lepton colliders such as ILC. 

\section{Summary and Conclusion}
\label{sec:concl}
We have studied the decays $h^0(125)\to c \bar c, b \bar b, b \bar s, 
\gamma \gamma, g g$ in the MSSM with general QFV due to squark generation mixings. 
{\it In strong contrast to} the usual studies in the MSSM with Minimal Flavor Violation, 
we have found that the deviations of these MSSM decay widths from the SM values can 
be quite sizable. 
{\it For the first time}, we have performed a systematic MSSM parameter scan 
for these decay widths respecting all relevant theoretical and experimental 
constraints. 
From the parameter scan, we have found the following: 
\begin{itemize}
\item DEV(c) and DEV(b) can be very large simultaneously:
      DEV(c) can be as large as $\sim \pm60\%$, and 
      DEV(b) can be as large as $\sim \pm20\%$
      (see Fig.~\ref{DEVc_DEVb})

\item DEV(b/c) can exceed $\sim +100\%$ 
      (see Fig.~\ref{DEVb2c_DEVbc})

\item $B(h^0 \to b s)$ can be as sizable as $\sim 0.15\%$ exceeding  
      the ILC250+500+1000 sensitivity of $\sim 0.1\%$ at 4$\sigma$ 
      signal significance 
      (see Fig.~\ref{BRbs_TU32tanb})

\item DEV($\gamma$) and DEV(g) can be sizable simultaneously:
      DEV($\gamma$) can become $\sim \pm 1\%$, and 
      DEV(g) can be as large as $\sim +4\%$ and $\sim -7\%$
      (see Fig.~\ref{DEVgam_DEVg} and Fig.~\ref{DEVg_TU32M2U23})      

\item DEV($\gamma$/g) can be as large as $\sim +9\%$ 
      (see Fig.~\ref{DEVgam2g_TU32M2U23})    

\item There are significant correlations among these DEVs and $B(h^0 \to b s)$:
  \begin{itemize}
  
  \item There is a very strong correlation between DEV(b/c) and DEV(c)
        (see Fig.~\ref{DEVb2c_DEVc})

  \item There is a significant correlation between $B(h^0 \to b s)$ and DEV(b)
        (see Fig.~\ref{BRbs_DEVb})

  \item There is a very strong correlation between DEV($\gamma$) and DEV(g). 
        This correlation is due to the fact that the $\su_{1,2,3}$-loop 
        contributions dominate in the two DEVs 
        (see Fig.~\ref{DEVgam_DEVg})

  \end{itemize}

\item We have pointed out that the experimental measurement uncertainties
      as well as the MSSM prediction uncertainties tend to cancel 
      out significantly in the width ratios, making the measurement 
      of these width ratios a very sensitive probe of virtual SUSY 
      loop effects in these $h^0$ decays at future lepton colliders. 
      
\item All of these sizable deviations in the $h^0$ decays are due to 
      (i) large scharm-stop mixing and large scharm/stop involved trilinear couplings 
      $T_{U23}, T_{U32}, T_{U33}$, (ii) large sstrange-sbottom mixing and large 
      sstrange/sbottom involved trilinear couplings $T_{D23}, T_{D32}, T_{D33}$, 
      and (iii) large bottom Yukawa coupling $Y_b$ for large $\tan\beta$ and 
      large top Yukawa coupling $Y_t$. 
\end{itemize}

Such sizable deviations from the SM can be observed at high signal 
significance in future lepton colliders such as ILC, CLIC, CEPC, FCC-ee, 
and MuC {\it even after} the failure of SUSY particle discovery at the HL-LHC.
In case the deviation pattern shown here is really observed at the lepton 
colliders, then it would strongly suggest the discovery of QFV SUSY (the 
MSSM with general QFV). 

%
\section*{Acknowledgments}

We would like to thank Werner Porod for helpful discussions, especially for the 
permanent support concerning SPheno. We also thank Junping Tian and Jorge de Blas 
for supplying us helpful information on the experiments at the future lepton colliders, 
especially at ILC. We also thank Sven Heinemeyer for helpful discussions.

\newpage
\begin{appendix}

\section{Theoretical and experimental constraints}
\label{sec:constr}

The experimental and theoretical constraints taken into account in the 
present work are discussed in detail in~\cite{Eberl_17}. 
Here we list the updated constraints from $K$- and B-physics and those 
on the Higgs boson mass and couplings in Table~\ref{TabConstraints}.
For the mass of the Higgs boson $h^0$, taking the combination of the ATLAS and 
CMS measurements  $m_{h^0} = 125.09 \pm 0.24~\GeV$ \cite{Higgs_mass_ATLAS_CMS} and 
adding the theoretical uncertainty of $\sim \pm 3~\GeV$ ~\cite{Higgs_mass_Heinemeyer} 
linearly to the experimental uncertainty at 2$\sigma$, 
we take $m_{h^0} = 125.09 \pm 3.48 ~\GeV$. \\
Here, we remark that the uncertainty of $\sim \pm 3~\GeV$ is the intrinsic theoretical 
error due to unknown higher-order corrections (i.e., the error due to renormalization 
scheme/scale dependence). 
It is well known that the parametric uncertainties due to the experimental errors of 
the SM input parameters (such as $m_t$, $\alpha_s(m_Z)$, ...) are also important 
\cite{Djouadi_Porod, NLL_Heinemeyer}. The dominant source of the parametric uncertainties 
is the experimental error of the top quark mass $m_t$ \cite{NLL_Heinemeyer} since the 
radiative corrected $m_{h^0}$ is very sensitive to $m_t$ in the MSSM. 
Therefore, the total theoretical error given by the sum of the intrinsic theoretical 
error and the parametric errors should be significantly larger than $\pm 3~\GeV$. 
Here, we take $\pm 3~\GeV$ as the total theoretical error conservatively. 
It is important to note that the parametric errors, especially that due to 
experimental error of the top quark mass, remain to be significant 
even if the intrinsic theoretical error is improved to be very small. \\
The $h^0$ couplings that receive SUSY QFV effects significantly are 
$g(hbb)$ ~\cite{Eberl:h2bb}, $g(hcc)$ ~\cite{Bartl:h2cc}, 
$g(hgg)$ and $g(h\gamma\gamma)$ ~\cite{h2gagagg}. 
\footnote{
Precisely speaking, in principle, $g(htt)$ coupling could also 
receive SUSY QFV effects significantly. 
However, predicting the effective coupling $g(htt)$ at loop levels in the MSSM 
is very difficult since its theoretical definition in the context of $t\bar{t}h$ 
production at LHC is unclear ~\cite{tth@LHC}.
} 
The measurement of $g(hcc)$ is very difficult 
due to the huge QCD background at LHC; there is no significant experimental data 
on $g(hcc)$ at this time. Hence, the relevant $h^0$ couplings to be compared 
with the LHC observations are $g(hbb)$, $g(hgg)$ and $g(h\gamma\gamma)$. 
Therefore, we list the LHC data on $g(hbb)$ ($\kappa_b$), $g(hgg)$ ($\kappa_g$) 
and $g(h\gamma\gamma)$ ($\kappa_\gamma$) in Table~\ref{TabConstraints}.\\
 
As the constraints from the decays $B\to D^{(*)}\,\tau\,\nu$ are unclear due to 
large theoretical uncertainties \cite{Bartl:h2cc}, 
\footnote{
As pointed out in \cite{Nierste:2008qe}, the theoretical 
predictions (in the SM and MSSM) on B$(B \to D\, l\, \nu)$ and B$(B \to D^*\, l\, \nu)$~$
(l = \tau, \mu, e)$ have potentially large theoretical uncertainties due to the 
theoretical assumptions on the hadronic form factors at the $B\,D\,W^+$ and $B\,D^*\,W^+$ 
vertices (also at the $B\,D\,H^+$  and $B\,D^*\,H^+$ vertices in the MSSM). Hence, 
the constraints from these decays are unclear. 
}
we do not take these constraints into account in our paper.
As the issues of possible anomalies of $R(D^{(*)}) = B(B\to D^{(*)}\,\tau\,\nu)/B(B\to D^{(*)}\,\ell\,\nu)$ 
with $\ell = e \ \mbox{or} \ \mu$ are not yet settled \cite{B@EPS-HEP2023}, we do not take 
the constraints from these ratios into account either. Note that the possible related anomaly of 
$R_{K^{(*)}} = B(B\to K^{(*)}\,e^+\, e^-)/B(B\to K^{(*)}\,\mu^+\,\mu^-)$ is gone away now \cite{LHCb_RK}.\\
In \cite{Dedes}, the QFV decays $t \to q h^0$ with $q = u, c$, have 
been studied in the MSSM with general QFV. It is found that these decays cannot 
be visible at the current and high luminosity LHC runs due to the very small 
decay branching ratios B($t \to q h^0$), giving no significant constraint on the 
$\tilde c - \tilde t$ and $\tilde s - \tilde b$ mixings.\\ 
In \cite{Soni}, the QFV decay branching ratio $B(Z^0 \to b \, s) \equiv 
B(Z^0 \to b \, \bar s) + B(Z^0 \to \bar b \, s)$ was studied in the 
MSSM with general QFV. It was shown that the current best experimental 
upper limit on $B(Z^0 \to b \, s)$ gives no significant constraint 
on the $\ti{s} - \ti{b}$ and $\ti{c} - \ti{t}$ mixings.
\footnote{
Note that no experimental upper limit on $B(Z^0 \to b \, s)$ is listed in PDG2022 
\cite{PDG2022}.
}\\
We comment on the recent data on the anomalous magnetic moment of the muon $a_\mu$ 
from the Fermilab experiment \cite{a_muon_Fermilab}. The Fermilab data result 
in 5.0$\sigma$ discrepancy between the experimental data and the SM prediction 
\cite{a_muon_Theory_Initiative}.
\footnote{
On the other hand, Refs. \cite{BMW_collab, CDM-3, a_muon_EPS-HEP2023} 
point to agreement of the SM prediction with the current experimental data.
}
In our scenario with heavy sleptons/sneutrinos with masses of about 1.5 TeV, 
the MSSM loop contributions to $a_\mu$ are so small that they can not 
explain the discrepancy between the new data and the SM prediction.
Therefore, in the context of our scenario, this discrepancy should be 
explained by the loop contributions of another new physics coexisting 
with SUSY.\\
We also comment on the recent W boson mass data from CDF II \cite{CDF_II}, 
which is about +7$\sigma$ away from the SM prediction. However, the CDF II 
data disagrees significantly with the world average of the $m_W$ data from 
the other experiments \cite{PDG2022}.
\footnote{
We remark the caveats from Refs. \cite{Wilson_mW, Heinemeyer_mW}: According to the PDG 
prescription, the new "scale-factored" world average of $m_W$ data including the 
CDF II data is +3.2$\sigma$ off the SM value used by CDF II. 
We also note that the new improved ATLAS $m_W$ data \cite{ATLAS_mW_2023} is consistent with 
their old data \cite{ATLAS_mW_2017} and the previous world average \cite{PDG2022}.
} 
This issue of the $m_W$ data is not yet settled. Hence, we do not take 
into account this $m_W$ constraint on the MSSM parameters in our analysis. \\

In addition to these, we also require our scenarios to be 
consistent with the following experimental constraints:
%
\begin{table*}[t!]
\footnotesize{
\caption{
Constraints on the MSSM parameters from the $K$- and $B$-meson data 
relevant mainly for the mixing between the second and the third generations of 
squarks and from the data on the $h^0$ mass and couplings $\kappa_b$, $\kappa_g$, 
$\kappa_\gamma$. The fourth column shows constraints at $95 \%$ CL obtained by 
combining the experimental error quadratically with the theoretical uncertainty, 
except for $B(K^0_L \to \pi^0 \nu \bar{\nu})$, $m_{h^0}$ and $\kappa_{b,g,\gamma}$.
}
\begin{center}
%
\setlength{\tabcolsep}{3pt} 
\begin{tabular}{|c|c|c|c|}
    \hline
    Observable & Exp.\ data & Theor.\ uncertainty & \ Constr.\ (95$\%$CL) \\
    \hline\hline
    &&&\\
    $10^3\times|\epsilon_K|$ & $2.228 \pm 0.011$ (68$\%$ CL)~\cite{PDG2020} 
    & $\pm 0.28$ (68$\%$ CL)~\cite{epsK_DMK_SM} &
    $2.228 \pm 0.549$\\
    $10^{15}\times\Delta M_K$ [\GeV] & $3.484\pm 0.006$ (68$\%$ CL)~\cite{PDG2020} 
    & $\pm 1.2 $ (68$\%$ CL)~\cite{epsK_DMK_SM} &
    $3.484 \pm 2.352$\\
    $10^{9}\times$B($K^0_L \to \pi^0 \nu \bar{\nu}$) & $< 3.0$ (90$\%$ CL)~\cite{PDG2020} 
    & $\pm 0.002 $ (68$\%$ CL)~\cite{PDG2020} &
    $< 3.0$ (90$\%$ CL)\\
    $10^{10}\times$B($K^+ \to \pi^+ \nu \bar{\nu}$) & $1.7 \pm 1.1$ (68$\%$ CL)~\cite{PDG2020} 
    & $\pm 0.04 $ (68$\%$ CL)~\cite{PDG2020} &
    $1.7^{+2.16}_{-1.70}$\\
    $\Delta M_{B_s}$ [ps$^{-1}$] & $17.757 \pm 0.021$ (68$\%$ CL)~\cite{HFAG2019, PDG2020} 
    & $\pm 2.7$ (68$\%$ CL)~\cite{DeltaMBs_SM} &
    $17.757 \pm 5.29$\\
    $10^4\times$B($b \to s \gamma)$ & $3.32 \pm 0.15$ (68$\%$ CL)~\cite{HFAG2019, PDG2020} 
    & $\pm 0.23$ (68$\%$ CL)~\cite{Misiak_2015} &  $3.32\pm 0.54$\\
    $10^6\times$B($b \to s~l^+ l^-$)& $1.60 ~ ^{+0.48}_{-0.45}$ (68$\%$ CL)~\cite{bsll_BABAR_2014}
    & $\pm 0.11$ (68$\%$ CL)~\cite{Huber_2008} & $1.60 ~ ^{+0.97}_{-0.91}$\\
    $(l=e~{\rm or}~\mu)$ &&&\\
    $10^9\times$B($B_s\to \mu^+\mu^-$) & $2.69~^{+0.37}_{-0.35}$ (68$\%$CL)~\cite{B@ICHEP2020}
    & $\pm0.23$  (68$\%$ CL)~\cite{Bsmumu_SM_Bobeth_2014} 
    & $2.69~^{+0.85}_{-0.82}$ \\
    $10^4\times$B($B^+ \to \tau^+ \nu $) & $1.06 \pm 0.19$ (68$\%$CL)
    ~\cite{HFAG2019}
    &$\pm0.29$  (68$\%$ CL)~\cite{Btotaunu_LP2013} & $1.06 \pm 0.69$\\
    $ m_{h^0}$ [\GeV] & $125.09 \pm 0.24~(68\%~ \rm{CL})$ \cite{Higgs_mass_ATLAS_CMS}
    & $\pm 3$~\cite{Higgs_mass_Heinemeyer} & $125.09 \pm 3.48$ \\
    $\kappa_b$ & $0.89^{+0.22}_{-0.21}~(95\%~ \rm{CL})$ \cite{kappa_bgamg_ATLAS}
    &  & $0.89^{+0.22}_{-0.21}$ (ATLAS)\\
    & $0.99^{+0.33}_{-0.31}~(95\%~ \rm{CL})$ \cite{kappa_bgamg_CMS}
    &  & $0.99^{+0.33}_{-0.31}$ (CMS)\\
    $\kappa_g$ & $0.95^{+0.14}_{-0.13}~(95\%~ \rm{CL})$ \cite{kappa_bgamg_ATLAS}
    &  & $0.95^{+0.14}_{-0.13}$ (ATLAS)\\
    & $0.92 \pm 0.16~(95\%~ \rm{CL})$ \cite{kappa_bgamg_CMS}
    &  & $0.92 \pm 0.16$ (CMS)\\
    $\kappa_\gamma$ & $1.01 \pm 0.12~(95\%~ \rm{CL})$ \cite{kappa_bgamg_ATLAS}
    &  & $1.01 \pm 0.12$ (ATLAS)\\
    & $1.10 \pm 0.16~(95\%~ \rm{CL})$ \cite{kappa_bgamg_CMS}
    &  & $1.10 \pm 0.16$ (CMS)\\
&&&\\
    \hline
\end{tabular}
\end{center}
\label{TabConstraints}}
\end{table*}
%

\begin{itemize}

\item
The LHC limits on sparticle masses (at 95\% CL)~\cite{SUSY@LP2019, SUSY@ICHEP2020, 
SUSY_summary_plot@ATLAS, SUSY_summary_plot@CMS, SUSY@ATLAS2020}:

We impose conservative limits for safety though actual limits are 
somewhat weaker than those shown here.
In the context of simplified models, gluino masses $\msg \lesssim 2.35~\TeV$ are 
excluded for $\mnt1 < 1.55~\TeV$. There is no gluino mass limit for $\mnt1 > 1.55~\TeV$.
The eightfold degenerate first two generation squark masses are excluded below 1.92~TeV for $\mnt1 < 0.9~\TeV$.
There is no limit on the masses for $\mnt1 > 0.9~\TeV$. 
We impose this squark mass limit on $m_{\su_3}$ and $m_{\sd_3}$. 
Bottom-squark masses are excluded below 1.26~TeV for $\mnt1 < 0.73~\TeV$. 
There is no bottom-squark mass limit for $\mnt1 > 0.73~\TeV$.
Here the bottom-squark mass means the lighter sbottom mass $m_{\sb_1}$.
We impose this limit on $m_{\sd_1}$ since $\sd_1 \sim \sb_R$ (see Table \ref{flavourdecomp}).
A typical top-squark mass lower limit is $\sim$ 1.26~TeV for $m_{\nt_1} < 0.62$ TeV. 
There is no top-squark mass limit for $m_{\nt_1} > 0.62$ TeV. 
Here the top-squark mass means the lighter stop mass $m_{\st_1}$.
We impose this limit on $m_{\su_1}$ since $\su_1 \sim \st_R$ (see Table \ref{flavourdecomp}).
For sleptons/sneutrinos heavier than the lighter chargino $\ch_1$ and the second neutralino $\nt_2$, 
the mass limits are $m_{\ch_1}, m_{\nt_2} > 0.74$ TeV for $m_{\nt_1} \lesssim 0.3$ TeV, and 
there is no $m_{\ch_1}$, $m_{\nt_2}$ limits for $m_{\nt_1} > 0.3$ TeV; 
for sleptons/sneutrinos lighter than $\ch_1$ and $\nt_2$, 
the mass limits are $m_{\ch_1}, m_{\nt_2} > 1.15$ TeV for $m_{\nt_1} \lesssim 0.72$ TeV, and 
there is no $m_{\ch_1}$, $m_{\nt_2}$ limits for $m_{\nt_1} > 0.72$ TeV.
For mass degenerate selectrons $\ti{e}_{L,R}$ and smuons $\ti{\mu}_{L,R}$, masses below 
0.7 TeV are excluded for $m_{\nt_1} < 0.41$ TeV. For mass degenerate staus $\stau_L$ and 
$\stau_R$, masses below 0.39 TeV are excluded for $m_{\nt_1} < 0.14$ TeV. 
There is no sneutrino $\ti{\nu}$ mass limit from LHC yet.
Sneutrino masses below 94 GeV are excluded by LEP200 experiment \cite{PDG2020}.

\item
The constraint on ($m_{A^0, H^+}, \tan\beta$) (at 95\% CL) from the negative searches for 
the MSSM Higgs bosons $H^0$, $A^0$ and $H^+$ at LHC ~\cite{H_to_tautau@ATLAS, H_to_tautau@CMS, 
H_tb@ATLAS, H_tb@ATLAS_2021, H_tau_nu@ATLAS, H_tb@CMS, H_tau_nu@CMS, ATLAS_2402, 
ATLAS_Note_2024_008, H_tau_nu@ATLAS_2412, H_tautau@CMS_2208}, where $H^0$ is the heavier 
$CP$-even neutral Higgs boson and $H^+$ is the charged Higgs boson: \\
The ($m_{A^0, H^+}, \tan\beta$) limit from the negative search for the heavier MSSM 
Higgs bosons $H^0$, $A^0$ and $H^+$ at LHC depends on the MSSM scenarios (such as 
hMSSM scenario, $M_h^{125}$ scenario, ...). 
However, from ~\cite{H_to_tautau@ATLAS, H_to_tautau@CMS, 
H_tb@ATLAS, H_tb@ATLAS_2021, H_tau_nu@ATLAS, H_tb@CMS, H_tau_nu@CMS, ATLAS_2402, 
ATLAS_Note_2024_008, H_tau_nu@ATLAS_2412, H_tautau@CMS_2208}, 
we find that in general the ($m_{A^0, H^+}, \tan\beta$) 
limits of ATLAS/CMS are rather insensitive to the choice of the MSSM scenarios.
Here, note that $m_{H^0} \simeq m_{A^0} \simeq m_{H^+}$ in the decoupling Higgs 
scenarios as ours. 
From ATLAS/CMS data ~\cite{H_to_tautau@ATLAS, H_to_tautau@CMS, 
H_tb@ATLAS, H_tb@ATLAS_2021, H_tau_nu@ATLAS, H_tb@CMS, H_tau_nu@CMS, ATLAS_2402, 
ATLAS_Note_2024_008, H_tau_nu@ATLAS_2412, H_tautau@CMS_2208}, especially 
~\cite{ATLAS_Note_2024_008}, we find that the ($m_{A^0}, \tan\beta$) limit from 
the negative search for the decay $H^0/A^0 \to \tau^+ \tau^-$ (as shown in Fig. 2(c) 
of \cite{H_to_tautau@ATLAS}) is most important for $\tan\beta > 10$. 
The ($m_{A^0}, \tan\beta$) limit (at 95\% CL) shown in Fig.2(c) of \cite{H_to_tautau@ATLAS} 
is the strongest for $\tan\beta > 10$ among those obtained in ~\cite{H_to_tautau@ATLAS, 
H_to_tautau@CMS, H_tb@ATLAS, H_tb@ATLAS_2021, H_tau_nu@ATLAS, H_tb@CMS, H_tau_nu@CMS, ATLAS_2402, 
ATLAS_Note_2024_008, H_tau_nu@ATLAS_2412, H_tautau@CMS_2208}. 
Therefore, we take and respect this strongest ($m_{A^0}, \tan\beta$) limit in our 
parameter scan analysis for $\tan\beta > 10$.

\item
The experimental limit on SUSY contributions on the electroweak
$\rho$ parameter ~\cite{Altarelli:1997et}: $\Delta \rho~ (\rm SUSY) < 0.0012.$

\end{itemize}

Furthermore, we impose the following theoretical constraints from the vacuum 
stability conditions for the trilinear coupling matrices~\cite{Casas}: 
\begin{eqnarray}
|T_{U\alpha\alpha}|^2 &<&
3~Y^2_{U\alpha}~(M^2_{Q \alpha\alpha}+M^2_{U\alpha\alpha}+m^2_2)~,
\label{eq:CCBfcU}\\[2mm]
|T_{D\alpha\alpha}|^2 &<&
3~Y^2_{D\alpha}~(M^2_{Q\alpha\alpha}+M^2_{D\alpha\alpha}+m^2_1)~,
\label{eq:CCBfcD}\\[2mm]
|T_{U\alpha\beta}|^2 &<&
Y^2_{U\gamma}~(M^2_{Q \beta\beta}+M^2_{U\alpha\alpha}+m^2_2)~,
\label{eq:CCBfvU}\\[2mm]
|T_{D\alpha\beta}|^2 &<&
Y^2_{D\gamma}~(M^2_{Q \beta\beta}+M^2_{D\alpha\alpha}+m^2_1)~,
\label{eq:CCBfvD}
\end{eqnarray}
where
$\a,\b=1,2,3,~\a\neq\b;~\gamma={\rm Max}(\a,\b)$, and
$m^2_1=(m^2_{H^+}+m^2_Z\sin^2\theta_W)\sin^2\b-\frac{1}{2}m_Z^2$,
$m^2_2=(m^2_{H^+}+$  
$m^2_Z\sin^2\theta_W)$ $\cos^2\beta-\frac{1}{2}m_Z^2$.
The Yukawa couplings of the up-type and down-type quarks are
$Y_{U\alpha}=\sqrt{2}m_{u_\alpha}/v_2=\frac{g}{\sqrt{2}}\frac{m_{u_\alpha}}{m_W
\sin\beta}$
$(u_\a=u,c,t)$ and
$Y_{D\alpha}=\sqrt{2}m_{d_\alpha}/v_1=\frac{g}{\sqrt{2}}\frac{m_{d_\alpha}}{m_W
\cos\beta}$
$(d_\a=d,s,b)$,
with $m_{u_\a}$ and $m_{d_\a}$ being the running quark masses at the 
scale $\rm Q=1$~TeV and $g$ being the SU(2) gauge coupling. All soft SUSY-breaking parameters 
are given at $\rm Q=1$~TeV. As SM parameters we take $m_Z=91.2~\GeV$ and
the on-shell top-quark mass $m_t=172.9~\GeV$ \cite{PDG2020}.

\section{Expected experimental errors of the deviations\\
DEVs and the effective Higgs couplings at future lepton colliders} 
\label{sec:error}

Here we summarize expected experimental {\it absolute} 1$\sigma$ errors 
of the deviations DEV(X) and DEV(X/Y) measured at future lepton colliders.

\noindent According to Eq. (\ref{DEV_Coup_kappa}), the expected {\it absolute} 1$\sigma$ error 
of the relative deviation DEV(X) denoted by $\Delta \DEV(X)$ is related with 
the expected {\it relative} 1$\sigma$ error of the effective coupling $g(h^0 X X)$ 
denoted by $\delta g(h^0 X X)$ as follows:
\be
\Delta \DEV(X) \simeq 2 \delta g(h^0 X X). 
  \label{DDEV_dCoup}
\ee
In Table \ref{table_DEVerror_LC}, we show the expected {\it absolute} 1$\sigma$ 
errors of the measured deviations DEV(X) and DEV(X/Y) denoted by $\Delta \DEV(X)$ and 
$\Delta \DEV(X/Y)$ at future lepton colliders. The errors $\Delta \DEV(X)$ are obtained 
by using Eq.(\ref{DDEV_dCoup}) and the expected {\it relative} 1$\sigma$ error of the 
effective coupling $\delta g(h^0 X X)$ given in Table 29 of Ref. \cite{Snowmass2021_Rep}. 
\footnote{
The expected {\it relative} 1$\sigma$ errors of the effective couplings $\delta g(h^0 X X)$ 
given in Table 7 of Ref. \cite{ESU2020_Rep} are similar to those given in Table 29 
of Ref. \cite{Snowmass2021_Rep}.
}
The errors $\Delta \DEV(X/Y)$ are obtained according to Ref. \cite{Mateo}.

\begin{table}[h!]
\caption{
The expected {\it absolute} 1$\sigma$ error of the deviations DEV(X) and 
DEV(X/Y) (denoted by $\Delta \DEV(X)$ and $\Delta \DEV(X/Y)$) measured at future 
lepton colliders: 
ILC-I = ILC250 + Giga-Z, ILC-II = ILC250+500 + Giga-Z, ILC-III = ILC250+500+1000 + Giga-Z; 
CLIC-I = CLIC380, CLIC-II = CLIC380+1500, CLIC-III = CLIC380+1500+3000;
FCC-ee I = FCC-ee240 + Z/WW, FCC-ee II = FCC-ee240+365 + Z/WW;
CEPC-I = CEPC240 + Z/WW, CEPC-II = CEPC240+360 + Z/WW;
MuC-I = MuC3TeV, MuC-II = MuC10TeV, MuC-III = MuC10TeV+125\GeV.
As for ILC, the results without Giga-Z run are almost identical to 
those with Giga-Z one.
The Z/WW denote Z-pole and WW threshold runs.
All results except for MuC-I and MuC-II are those from the 
free-$\Gamma_H$ fit and the results for MuC-I and MuC-II are 
those from the constrained-$\Gamma_H$ fit, where $\Gamma_H$ 
is the total width of the Higgs boson $h^0$. 
The details of run scenarios of the lepton colliders are explained in 
Ref. \cite{Snowmass2021_Rep}. The HL-LHC and LEP/SLD measurements are 
combined with all lepton collider run scenarios.
} 
\begin{center}
\begin{tabular}{|c|c|c|c|c|c|c|c|}
  \hline
                   & ILC-I & ILC-II & ILC-III & CLIC-I & CLIC-II & CLIC-III \\
  \hline
  $\Delta \DEV(b)$  & 1.7\% & 1.1\%  & 0.9\% & 2.2\% & 1.2\% & 1.1\% \\
  \hline 
  $\Delta \DEV(c)$  & 3.6\% & 2.4\% & 1.8\% & 8.6\% & 3.8\% & 3.0\% \\
  \hline 
  $\Delta \DEV(\gamma)$  & 2.4\% & 2.2\% & 2.0\% & 2.6\% & 2.4\% & 2.2\% \\
  \hline
  $\Delta \DEV(g)$ & 1.8\% & 1.4\% & 1.1\% & 2.2\% & 1.6\% & 1.4\% \\
  \hline  \hline
  $\Delta \DEV(b/c)$  & 3.1\% & 2.1\% & 1.3\% & 8.2\% & 3.5\% & 2.5\% \\
  \hline 
  $\Delta \DEV(\gamma/g)$  & 3.3\% & 2.8\% & 2.3\% & 3.4\% & 3.1\% & 2.6\% \\
  \hline
\end{tabular}
\vskip 0.4cm
\begin{tabular}{|c|c|c|c|c|c|c|c|}
  \hline
                   & FCC-ee I & FCC-ee II & CEPC-I & CEPC-II & MuC-I & MuC-II & MuC-III \\
  \hline
  $\Delta \DEV(b)$  & 1.3\% & 1.2\% & 0.86\% & 0.84\% & 1.8\% & 0.92\% & 1.1\% \\
  \hline
  $\Delta \DEV(c)$  & 2.8\% &  2.6\% & 2.4\% & 2.2\% & 12.4\% & 3.8\% & 3.6\% \\
  \hline 
  $\Delta \DEV(\gamma)$  & 2.4\% &  2.2\% & 1.8\% & 1.8\% & 2.4\% & 1.4\% & 1.5\% \\
  \hline
  $\Delta \DEV(g)$ & 1.5\% & 1.4\% & 0.9\% & 0.88\% & 1.7\% & 0.92\% & 1.0\% \\
  \hline  \hline
  $\Delta \DEV(b/c)$  & 2.2\% & 2.1\% & 2.0\% & 1.9\% & 11.9\%  & 3.5\% & 3.4\% \\
  \hline 
  $\Delta \DEV(\gamma/g)$  & 3.0\% & 2.9\% & 2.1\% & 2.0\% & 3.2\% & 1.6\% & 1.6\% \\
  \hline
\end{tabular}
\end{center}
\label{table_DEVerror_LC}
\end{table}

\newpage
\section{ILC sensitivity to the branching ratio $B(h^0 \to b \, s)$} 
\label{sec:ILC_sensitivity_to_BRbs}

The setup conditions for this ILC sensitivity are as follows \cite{Tian}:  
\begin{itemize}

\item The signal is the QFV decay $h^0 \to b \bar{q}$ and $h^0 \to \bar{b} q$ 
    with the light quark $q = d/s$. 
    The dominant background is the QFC decay $h^0 \to b \bar{b}$, where either 
    $b$-jet or $\bar{b}$-jet is misidentified as a light quark-jet (a q-jet).
    We consider the case $B(h^0 \to b q) \equiv B(h^0 \to b \bar{q}) + 
    B(h^0 \to \bar{b} q) = 0.1\%$ and take $B(h^0 \to b \bar{b}) = 58\%$. 
    
\item As for the efficiencies, the followings are assumed: \\
    The optimal b-tagging efficiency for a b-jet, $\epsilon_{b/b}$ is $\sim 94\%$.\\
    The q-tagging efficiency for a q-jet, $\epsilon_{q/q}$ is $\sim 90\%$.\\
    The q-tagging efficiency for a b-jet, $\epsilon_{q/b}$ is $\sim 6\%$.\\
    The b-tagging efficiency for a q-jet, $\epsilon_{b/q}$ is $\sim 10\%$.\\
    The efficiency of other selection cuts, $\epsilon_{other}$ is assumed to be 
    around 70\%, such as that from Higgs mass cut, Z mass cut, angular cuts, etc., 
    which would be needed to suppress background events other than 
    $h^0 \to b \bar{b}$. This 70\% efficiency would be common for the 
    signal $h^0 \to b q$ and the dominant background $h^0 \to b \bar{b}$.

\item Regarding the amount of accumulated data at ILC, the specifications
    of the H-20 scenario \cite{Barklow} (see also \cite{Snowmass2021_Rep, ILC_white_paper}) 
    are adopted basically; 

  \begin{itemize}
  
   \item ILC250 stage;\\
        The total number of $h^0$ production events $N_{tot}^{h^0}$ is assumed 
        to be $\sim 0.5 \cdot 10^6$, for which the expected significance of 
        the signal $h^0 \to b q$, $\sigma_{sig}$ is $\sim 2\sigma$ in case $B(h^0 \to b q) = 0.1\%$:\\
        $\sigma_{sig} = N_{sig}/\sqrt{N_{sig} + N_{bg}} \sim 2$, where \\
        $N_{sig} = N_{tot}^{h^0} \cdot B(h^0 \to b q) \cdot \epsilon_{b/b} \cdot \epsilon_{q/q} \cdot \epsilon_{other} \sim 300$ and \\
        $N_{bg} = N_{tot}^{h^0} \cdot B(h^0 \to b \bar{b}) \cdot 2 \cdot \epsilon_{b/b} \cdot \epsilon_{q/b} \cdot \epsilon_{other} \sim 23000$.\\
        Alternatively, one can say that the expected upper bound on the branching ratio of the QFV decay, 
        $h^0 \to b \, d/s$, at 95\% CL, is 0.1\%: $B(h^0 \to b \, d/s) < 0.1\%$ (95\% CL) at ILC250.

  \item ILC250+500 stage;\\
        The total number of $h^0$ production events $N_{tot}^{h^0}$ is increased 
        from $\sim 0.5 \cdot 10^6$ to $\sim 1.1 \cdot 10^6$, for which the expected 
        significance of the signal $h^0 \to b q$, $\sigma_{sig}$ is $\sim 3\sigma$ in case 
        $B(h^0 \to b q) = 0.1\%$.

  \item ILC250+500+1000 stage;\\
        The total number of $h^0$ production events $N_{tot}^{h^0}$ is 
        further increased from $\sim 1.1 \cdot 10^6$ to $\sim 2.3 \cdot 10^6$, 
        for which the expected significance of the signal $h^0 \to b q$, $\sigma_{sig}$ is 
        $\sim 4\sigma$ in case $B(h^0 \to b q) = 0.1\%$.

  \end{itemize}

\end{itemize}

\section{Coupling modifiers $\kappa_X$, $B_{inv}$ and $B_{und}$} 
\label{sec:kappas}
%
\noindent
Here we show that the leading-order (LO) coupling modifiers $\kappa_X$ (X = W, Z, t, 
$\tau$, $\mu$, Z$\gamma$), $B_{inv}$ and $B_{und}$ at the benchmark scenario P1 satisfy 
the corresponding LHC data \cite{kappa_bgamg_ATLAS, kappa_bgamg_CMS}. 
In Table \ref{LOkappaTab} we show the LO coupling modifiers 
$\kappa_X$ (X = W, Z, t, $\tau$, $\mu$, Z$\gamma$), $B_{inv}$ and $B_{und}$ at P1 
and the corresponding ATLAS/CMS data at 95\% CL \cite{kappa_bgamg_ATLAS, kappa_bgamg_CMS}. 
\newpage
\noindent We compute the LO $\kappa_X$ (X = W, Z, t, $\tau$, $\mu$) using the LO formulas given 
in \cite{Djouadi_PhysRep, Djouadi_PhysRev}, in which we input $\alpha$ and $\beta$ values 
at P1: $\a = -0.03034721$ and $\b = tan^{-1}(33) = 1.54050257$; e.g., 
$\kappa_W = \kappa_Z = \sin(\b - \a) = \sin(1.54050257 - (-0.03034721)) = 0.99999999857$.  
We compute the LO $\kappa_{Z\gamma}$ at P1 using our own Fortran code \cite{Eberl_k_Zgam}.
\footnote{
We compute the LO MSSM width $\Gamma(h^0 \to Z \gamma)_{MSSM}$ at the full 1-loop level at P1 
in the MSSM with general QFV and the LO SM width $\Gamma(h^0 \to Z \gamma)_{SM}$ at the full 
1-loop level in the SM using our own Fortran code \cite{Eberl_k_Zgam}. 
We obtain the LO $\kappa_{Z\gamma} = 0.98351169$ using $\kappa_{Z\gamma}^2 = 
\Gamma(h^0 \to Z \gamma)_{MSSM}/\Gamma(h^0 \to Z \gamma)_{SM}$.
}
We compute the LO $B_{inv}$ and $B_{und}$ at P1 using the public code 
{\tt SPheno} v3.3.8~\cite{SPheno1, SPheno2}.
\footnote{
We compute $B_{inv}$ and $B_{und}$ at P1 using {\tt SPheno} v3.3.8 as follows: 
$B_{inv} = B(h^0 \to Z \nu \bar\nu) B(Z \to \nu \bar\nu) = 0.01413 \cdot 0.20 = 0.00283$. 
$B_{und}= B(h^0 \to \mbox{undetected New Physics particles}) = B(h^0 \to \mbox{sparticles}) = 0$ 
as the LSP neutralino mass $\mnt{1} = 781 \GeV$ at P1.
} 
From Table \ref{LOkappaTab}, we find that the LO coupling modifiers 
$\kappa_X$ (X = W, Z, t, $\tau$, $\mu$, Z$\gamma$), $B_{inv}$ and $B_{und}$ at P1 satisfy 
the corresponding LHC data at 95\% CL \cite{kappa_bgamg_ATLAS, kappa_bgamg_CMS}.\\
\indent Similarly, we have confirmed that the LO coupling modifiers $\kappa_X$ (X = W, Z, t, $\tau$, 
$\mu$, Z$\gamma$), $B_{inv}$ and $B_{und}$ at all the survival points in our MSSM parameter 
scan satisfy the corresponding LHC data at 95\% CL \cite{kappa_bgamg_ATLAS, kappa_bgamg_CMS}.
%
\begin{table*}[t!]
\footnotesize{
\caption{
In this Table we show the LO coupling modifiers $\kappa_X$ (X = W, Z, 
t, $\tau$, $\mu$, Z$\gamma$), $B_{inv}$ and $B_{und}$ in the benchmark scenario P1 and 
the corresponding LHC data at 95\% CL \cite{kappa_bgamg_ATLAS, kappa_bgamg_CMS}. 
}
\begin{center}
%
\setlength{\tabcolsep}{3pt} 
\begin{tabular}{|c|c|c|c|}
    \hline
    X & $\kappa_X$ at P1 & $\kappa_X$ (95\% CL) (ATLAS) & $\kappa_X$ (95\% CL) (CMS) \\
    \hline\hline
    &&&\\
    W 
    & 0.9999999986
    & $1.054^{+0.117}_{-0.116}$
    & $1.02 \pm 0.16$ \\
    &&&\\
    Z 
    & 0.9999999986
    & $0.993 \pm 0.111$ 
    & $1.04 \pm 0.14$ \\
    &&&\\
    t 
    & $0.9999983786$
    & $0.944^{+0.218}_{-0.214}$
    & $1.01^{+0.22}_{-0.20}$ \\
    &&&\\
    $\tau$ 
    & $1.0017641128$
    & $0.929^{+0.143}_{-0.137}$
    & $0.92 \pm 0.16$ \\
    &&&\\
    $\mu$ 
    & $1.0017641128$
    & $1.063^{+0.482}_{-0.595}$
    & $1.12^{+0.41}_{-0.43}$ \\
    &&&\\
    $Z \gamma$ 
    & $0.98351169$
    & $1.377^{+0.607}_{-0.720}$
    & $1.65^{+0.67}_{-0.73}$ \\
    &&&\\
    \hline \hline
    $B_X$ 
    & $B_X$ at P1 
    & $B_X$ upper bound (95\% CL) (ATLAS)
    & $B_X$ upper bound (95\% CL) (CMS) \\
    \hline\hline
    &&&\\
    $B_{inv}$
    & 0.00283
    & $B_{inv} < 0.13$ 
    & $B_{inv} < 0.17$ \\
    &&&\\
    $B_{und}$
    & 0
    & $B_{und} < 0.12$ 
    & $B_{und} < 0.16$ \\
    &&&\\
    \hline    
\end{tabular}
\end{center}
\label{LOkappaTab}}
\end{table*}
%

%
\end{appendix}

\newpage
%
%

\end{document}